\documentclass[hyperref,amsmath,amssymb,showpacs,floatfix,journal=ancac3,manuscript=article]{achemso}
\usepackage{graphicx}	
\usepackage{amsmath}	
\usepackage{amssymb}	
\usepackage{extarrows}
\usepackage{morefloats}
\mciteErrorOnUnknownfalse 
\usepackage[version=4]{mhchem} 
\setkeys{acs}{etalmode=truncate, maxauthors=200}
\setkeys{acs}{keywords = true}
\setkeys{acs}{usetitle=true}
\usepackage{epstopdf} 
\usepackage[perpage]{footmisc}
\renewcommand{\thefootnote}{\fnsymbol{footnote}}
\usepackage{changes}

\newcommand{\bad}[1]{{\color{red}#1}}
\DeclareGraphicsExtensions{.eps, .ps, .pdf}
\graphicspath{{./}{/home/ioan/floppy/astro/mnras/}{/home/ioan/floppy/astro/mnras/data/}{/home/ioan/floppy/astro/mnras/data_c4n-/}}

\newcommand{\figname}{Figure~} 
\newcommand{\figsname}{Figures~}
\newcommand{\secname}{Section~}

\newcommand{\si}{SI}
\usepackage{mfirstuc}
\MFUnocap{of}
\MFUnocap{that}
\MFUnocap{by}
\MFUnocap{to}
\MFUnocap{on}
\MFUnocap{and}
\MFUnocap{in}
\MFUnocap{the}
\MFUnocap{for}
\MFUnocap{with}
\SectionNumbersOff 
\author{Ioan B\^aldea}
\email{ioan.baldea@pci.uni-heidelberg.de}
\affiliation{Theoretische Chemie, Universit\"at Heidelberg, Im Neuenheimer Feld 229, D-69120 Heidelberg, Germany}
\alsoaffiliation{Institute of Space Sciences, National Institute of Lasers, Plasma and Radiation Physics,
RO 077125, Bucharest-M\u{a}gurele, Romania}
\title[Extensive Quantum Chemistry Study of Neutral and Charged \ce{C4N} Chains] 
      {Extensive Quantum Chemistry Study of Neutral and Charged \ce{C4N} Chains.
        An Attempt to Aid Astronomical Observations}

\begin{document}

\begin{abstract}
  Many molecular species can presumably still be observed in space if they are adequately characterized chemically. In this paper, we suggest that this could be the case of the neutral (\ce{C4N^0}) and anion (\ce{C4N-}) cyanopropynylidene chains, which were not yet identified in space although both the neutral (\ce{C3N^0} and \ce{C5N^0}) and anion (\ce{C3N-} and \ce{C5N-}) neighboring members of the homologous series were observed. Extensive data obtained from quantum chemical calculations using density functional theory (DFT), coupled cluster
  (CC), and quadratic configuration interaction (QCI) methods for all charge and spin states of interest for space science (doublet and quartet neutrals, triplet and singlet anions, and singlet and triplet cations) are reported: e.g., bond metric and natural bond order data, enthalpies of formation, dissociation and reaction energies, spin gaps, rotational constants, vibrational properties, dipole and quadrupole momenta, electron attachment energies ($EA$) and ionization potentials ($IP$). The fact that (not only for \ce{C4N} but also for 
\ce{C2N}  and \ce{C6N}) the quantum chemical methods utilized here are able to excellently reproduce the experimental $EA$ value --- which is often a challenge for theory --- is particularly encouraging, since this indicates that theoretical estimates of chemical reactivity indices (which are key input parameters for modeling astrochemical evolution) can be trusted. The presently calculated enthalpies of formation and dissociation energies do not substantiate any reason to assume that \ce{C4N} is absent in space. To further support this idea, we analyze potential chemical pathways of formation of both \ce{C4N^0} and \ce{C4N-}, which include association and exchange reactions.In view of the substantially larger dipole moment ($D_{anion} \gg D_{neutral}$), we suggest that astronomical detection should first focus on \ce{C4N-} chains rather than on neutral \ce{C4N^0} chains.
\end{abstract}
\noindent \textbf{Keywords:}
          {astrochemistry;
            interstellar medium;
            carbon chains;
            cyanopropynylidenes \ce{C4N};
            ab initio calculations;
            singlet-triplet interplay;
            ionization and electron attachment energies;
reaction and dissociation enthalpies;
            chemical reactivity indices;
            chemical pathways of \ce{C4N} formation
          }
\section{Introduction}
Carbon chains or obtained by adding heteroatoms at their ends represent a
continuing important topic in space sciences
\cite{Cernicharo:96,Cernicharo:01,Herbst:93,Guelin:98,McCarthy:99a,Maier:01b,Thaddeus:08,Cernicharo:08,Stanton:09,Agundez:10}.
Thanks to intensive and extensive efforts numerous carbon-based chains 
could be detected in the last decades. Given the fact that many of these molecular species
have a rather minor importance for terrestrial applications, information on many
molecular species of this kind needed to properly interpret data acquired (or to be acquired)
in astronomical observations is often very scarce. This state of affair may at least partially
explain the puzzling fact that, across a given homologous series, certain members could not yet
be observed although longer molecules were already detected.

The cyanopropynylidene (\ce{C4N}) chains 
investigated theoretically in the present paper,
which are expected to be relevant for interstellar chemistry \cite{Herbst:93,Doty:98},
belong to this category.
Although the presence in space of neutral \ce{C3N^0} and \ce{C5N^0} chains \cite{Friberg:80,Guelin:98}
was reported, \ce{C4N^0} chains could not be detected so far.
The situation of the corresponding anions is similar.
\ce{C3N-} and \ce{C5N-} chains were astronomically observed \cite{Thaddeus:08,Agundez:10}
but \ce{C4N-} could not yet be reported.

Cyanopropynylidenes made the object of several publications.
The neutral \ce{C4N^0} radical has been previously investigated
experimentally by microwave spectroscopy \cite{McCarthy:03} and
theoretically at Hartree-Fock (HF) level \cite{Pauzat:91} and
within density functional theory (DFT) \cite{Ding:01,Belbruno:01}.
Previous theoretically studies to \ce{C4N-} anion
reported results of second order M{\o}ller-Plesset (MP2) \cite{Zhan:96} and
DFT \cite{Wang:95,Pascoli:99,Neumark:09} calculations.
\ce{C4N-} anions produced by laser ablation \cite{Huang:95,Wang:95}
or sputtering \cite{Gupta:01} 
were studied experimentally via mass spectroscopy.
The experimental study using slow photoelectron velocity-map imaging spectroscopy (SEVI)
\cite{Neumark:09} is of particular interest in the context of the present paper.
To anticipate, the very accurate experimental electron affinity ($EA$) reported there
is very well reproduced by our theoretical calculations.

By reporting extensive data on doublet (spin $S=1/2$) and quartet ($S=3/2$) neutral (\ce{C4N^0}) chains, 
as well as on singlet ($S=0$) and triplet ($S=1$) anions (\ce{C4N-}) and cations (\ce{C4N+})
obtained within standard quantum chemical approaches
the present paper aims at filling a gap in the literature and at assisting ongoing efforts
in astronomic observation.\label{page:uccsd_t}

\section{Methods}
\label{sec:methods}

The results reported below were obtained by performing quantum chemical calculations on the
bwHPC platform \cite{bwHPC} using the GAUSSIAN 
\cite{g16,g09} and CFOUR \cite{cfour} packages. They are 
based both on the density functional theory (DFT)
and on ab initio methods. The latter comprise
coupled-cluster (CC) \cite{Bartlett:78,Bartlett:82,Cizek:07,Schirmer:09}
and quadratic configuration interaction (QCI) \label{page:qci} 
\cite{Bartlett:82,Head-Gordon:89,Head-Gordon:94}
approaches
including single and double excitations (CCSD, QCISD) 
also augmented with perturbative corrections due to triple excitations (CCSD(T), QCISD(T)).

All molecular geometries were optimized at the DFT level of theory using the B3LYP
hybrid exchange-correlation functional
\cite{Becke:88,Becke:93a,Frisch:94}
and the largest Pople 
6-311++G(3df, 3pd) 
basis sets \cite{Petersson:88,Petersson:91}.
For comparison purposes, 
the hybrid parameter free PBE0 \cite{Adamo:99},
M06-2X \cite{Truhlar:08},
and double-hybrid B2GP-PLYP \cite{g16} functionals 
were also used for DFT geometry optimization.
In all cases, we checked that all vibrational frequencies were real.

Similar to the spin gaps of all charge species ($\Delta^{q}; q=0, \pm$) 
the lowest electronic attachment energies ($EA$) and ionization potentials ($IP$)
can be and have been computed by using
``$\Delta$'' methods \cite{Gunnarson:89,Baldea:2014c}, i.e., by taking differences
between the total energies $\mathcal{E}_{X}\left(\mathbf{R}\right)$ of the corresponding molecular species
(neutral doublet (D$^0$) and quartet (Q$^0$), anion singlet (S$^{-}$) and triplet (T$^{-}$), and
cation singlet (S$^{+}$) and triplet (T$^{+}$)
at the appropriate geometries ($\mathbf{R}=\mathbf{R}_{S^{\pm}}, \mathbf{R}_{T^{\pm}}, \mathbf{R}_{D^{0}}, \mathbf{R}_{Q^{0}}$)
optimized as described above.

While not appearing to affect unrestricted DFT calculations
\cite{Pascoli:99,Baldea:2019e},
spin contamination becomes important and raises serious
doubt on results obtained within ab initio approaches like CCSD/CCSD(T) 
applied on top of unrestricted Hartree-Fock (UHF) wave functions.
For completeness and for comparison with previous studies
employing unrestricted methods \cite{Zhan:96,Pascoli:99},
along with estimates obtained within more reliable restricted open shell
(ROCCSD/ROCCD(T)) 
calculations, 
we also present properties obtained from unrestricted
coupled-cluster (UCCSD/UCCSD(T)) calculations.
The inspection of the various tables indicates \label{page:uhf}
that, without an adequate elimination of spin contamination
  (a task beyond the scope of the present paper), UCCSD/UCCSD(T)-based values 
cannot be trusted.

The quantities $EA$ and $IP$ were also computed by CC-based
equation-of-motion (EOM) methods (EA-EOM-ROCCSD and IP-EOM-ROCCSD) \cite{Nooijen:95a,Stanton:93,Stanton:94}.
To check whether long-range corrections improve
the DFT-based estimates, $EA$ and $IP$ values were also computed by using the
long-range corrected exchange-correlations
LC-BLYP \cite{Iikura:01} and LC-$\omega$PBE \cite{Scuseria:06} functionals.

Due to some numerical issues with the 6-311++G(3df, 3pd) basis sets, 
aug-cc-pVTZ \cite{Dunning:89,Woon:93} basis sets were used in
the EOM-ROCCSD calculations with CFOUR \cite{cfour}
and the natural bond orbital (NBO) analysis
\cite{NBO:5.9} carried out on top of RCCSD(T) and ROCCSD(T) calculations
with GAUSSIAN 09 \cite{g09}.

Thermochemistry data presented in the main text 
were obtained in the standard way \cite{Ochterski:00} using the CBS-QB3 protocol
as implemented in GAUSSIAN 16 \cite{g16}. 
Additional results based on the CBS-APNO and CBS-4M protocols 
\cite{Ochterski:96,Montgomery:00} are included in the {\si}.\label{page:apno-1}

To end this section, we note that at the small molecular size considered, geometry optimization and 
numerical frequency calculations at the more computationally demanding 
(RO)CCSD(T) level can and have also been done (cf.~Table~S8).
However, we do not show
single-point results for those geometries because 
the ROCCSD(T)-based values of the
rotational constant of the neutral doublet 
\label{page:neutral-doublet} species with cc-PVTZ
and aug-cc-pVTZ basis sets ($B=2.39187$\,GHz and $B=2.39093$\,GHz, respectively) 
have larger deviation from experiment than the DFT-based values
(cf.~Table~\ref{table:B} and
S15). 

\section{Results and Discussion}
\label{sec:results}

\subsection{Electronic Structure and Chemical Bonding}
\label{sec:structure}

In this section we present detailed results on properties of
interest for all charge species and relevant spin states:
neutral doublet ($\tilde{X}^2\Pi$) and quartet ($\tilde{a}^4 \Sigma^{-}$),
anion linear and bent ($^1 A^\prime$) singlet, and anion triplet ($^3\Sigma^{-}$)
as well as cation singlet ($^1\Sigma^{+}$) and triplet ($^3\Sigma^{-}$).

\begin{figure*}
  \centerline{
    \includegraphics[width=0.4\textwidth]{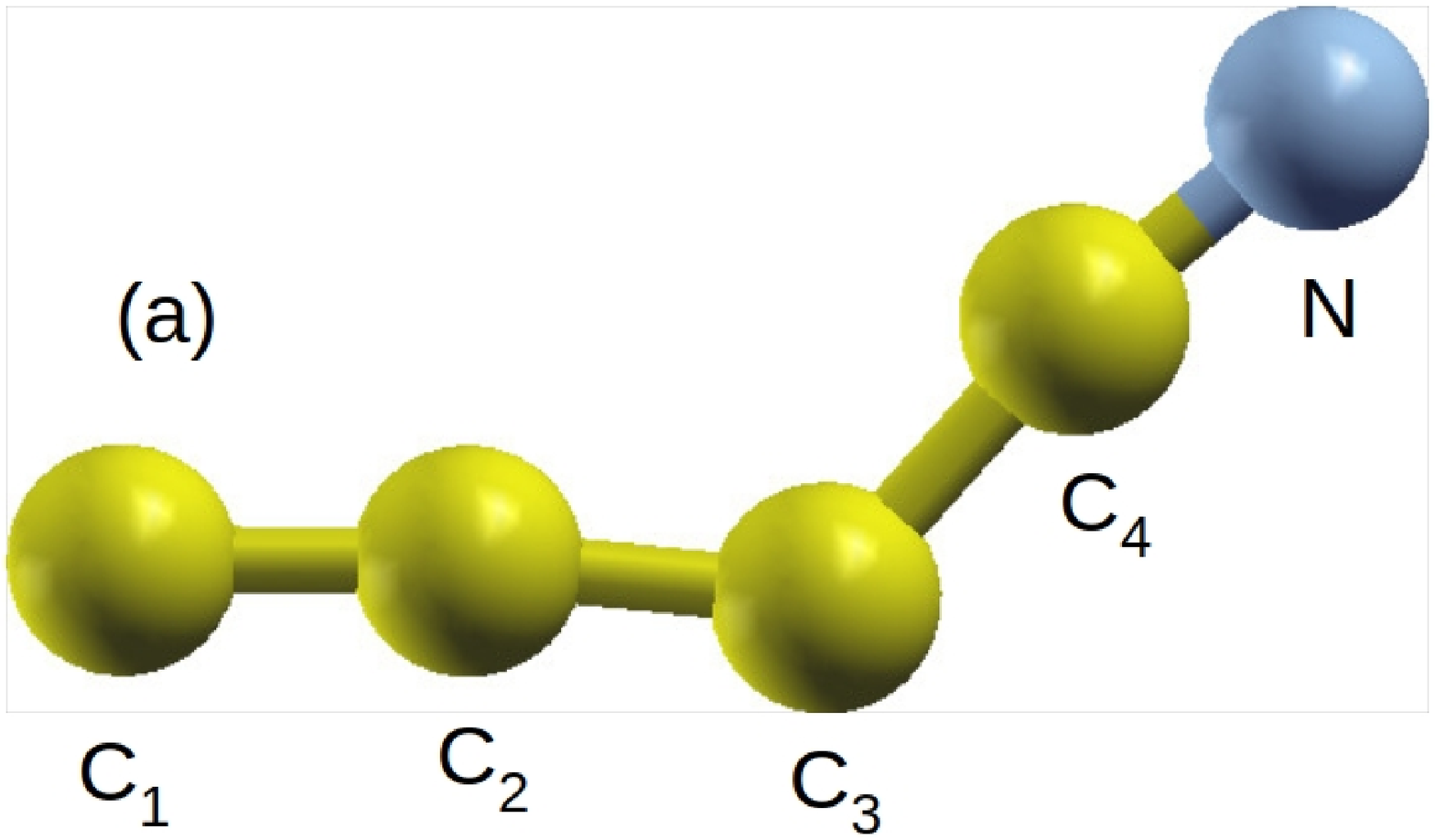}\hspace*{10ex}
    \includegraphics[width=0.4\textwidth]{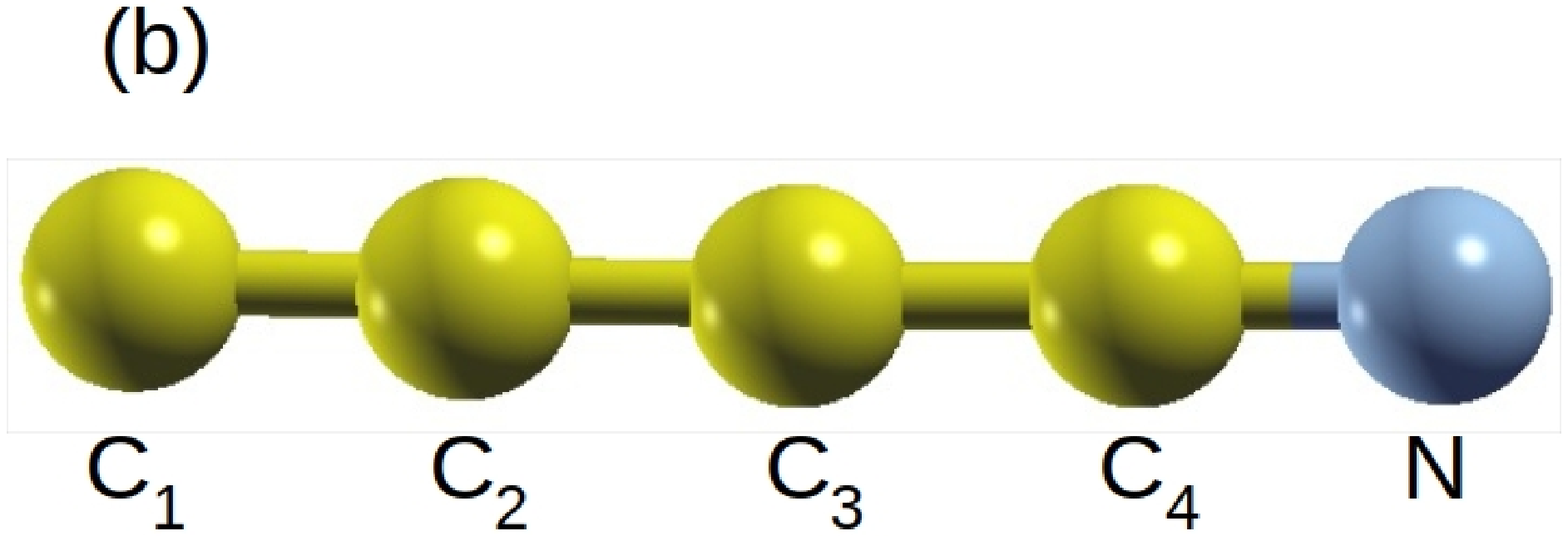}
  }
  \caption{Geometries of singlet ($^1 A^\prime$) and triplet  ($^3\,\Sigma^{-}$) 
    \ce{C4N-} anions (left and right panels, respectively)
    investigated in the present paper.
    Similar to the triplet \ce{C4N-} chain shown here, doublet and quartet \ce{C4N^0} neutral
    chains as well as
    singlet and triplet cation \ce{C4N+} chains are also linear.}
    \label{fig:geometries-c4n}
\end{figure*}

Except for the anion singlet
--- whose most stable conformation ($^1 A^\prime$) turned out to be bent
(cf.~\figname\ref{fig:geometries-c4n}) ---, 
geometry optimization (conducted without imposing symmetry constraints)
yielded (within numerical accuracy) linear chains.
To emphasize this aspect, in Table~\ref{table:geometries-c4n} we also included values of angles between adjacent atoms;
within numerical noise inherent in optimization without symmetry constraints,
they cannot be distinguished from the ideal value ($ 180^{\circ} $) characterizing strictly linear chains.
\begin{table*}
  \small 
  \centering
  \caption{Bond lengths $l$ between atoms XY (in angstrom),
    angles $\alpha$ between atoms $\widehat{\mbox{\ce{XYZ}}}$ (in degrees) and Wiberg bond order indices $\mathcal{N}$.
  Results of B3LYP/6-311++G(3df, 3pd) geometry optimization without imposing symmetry constraints.}
  \label{table:geometries-c4n}
  \begin{tabular}{ccccccccccccc} 
    \hline
    Species & Property & \ce{C1C2} & $\widehat{\mbox{\ce{C1C2C3}}}$ & \ce{C2C3} & $\widehat{\mbox{\ce{C2C3C4}}}$ & \ce{C3C4} & $\widehat{\mbox{\ce{C3C4N}}}$ & \ce{C4N} \\
    \hline
       metastable linear anion & $l$, $\alpha$ & 1.2892 & 179.97 & 1.2926 & 179.96 & 1.3178 & 179.98 & 1.1884 \\  
  singlet $^1 \Sigma^{+}$      & $\mathcal{N}$ & 2.0579 &       & 1.7873 &       & 1.4279 &       & 2.4787 \\
                               &               &        &       &        &       &        &       &        \\ 
   stable bent anion & $l$, $\alpha$ & 1.2780 & 174.3 & 1.3295 & 125.7 & 1.3846 & 171.4 & 1.1702 \\
  singlet $^1 A^{\prime}$      & $\mathcal{N}$ & 2.1785 &       & 1.6978 &       & 1.2309 &       & 2.6995 \\      
                               &               &        &       &        &       &        &       &        \\ 
     linear anion              & $l$, $\alpha$ & 1.2912 & 179.8 & 1.2917 & 178.7 & 1.3193 & 180.0 & 1.1874 \\
  triplet $^3 \Sigma^{-}$      & $\mathcal{N}$ & 1.8987 &       & 1.8251 &       & 1.3182 &       & 2.5775 \\
                               &               &        &       &        &       &        &       &        \\ 
     neutral                   & $l$, $\alpha$ & 1.3165 & 179.8 & 1.2536 & 178.8 & 1.3371 & 180.0 & 1.1670 \\
     doublet $^2 \Pi$          & $\mathcal{N}$ & 1.4710 &       & 2.2142 &       & 1.2069 &       & 2.7350 \\
                               &               &        &       &        &       &        &       &        \\ 
     neutral                   & $l$, $\alpha$ & 1.2585 & 179.9 & 1.2776 & 178.8 & 1.3257 & 179.9 & 1.1742 \\
quartet $\tilde{a} ^4 \Sigma^{-}$ & $\mathcal{N}$ & 1.4896 &       & 2.2643 &       & 1.2001 &       & 2.7540 \\
                               &               &        &       &        &       &        &       &        \\ 
          cation               & $l$, $\alpha$ & 1.3343 & 178.8 & 1.2383 & 179.7 & 1.3413 & 179.6 & 1.1648 \\
    singlet  $^1 \Sigma^{+}$   & $\mathcal{N}$ & 1.3932 &       & 2.3383 &       & 1.2060 &       & 2.7517 \\
                               &               &        &       &        &       &        &       &        \\ 
     cation                    & $l$, $\alpha$ & 1.2531 & 179.8 & 1.2747 & 178.9 & 1.3204 & 179.7 & 1.1751 \\
  triplet  $^3 \Sigma^{-}$     & $\mathcal{N}$ & 1.7742 &       & 1.9414 &       & 1.3077 &       & 2.6154 \\
    \hline
  \end{tabular}
\end{table*}

For the pertaining optimized geometries, we present detailed molecular properties:
Cartesian coordinates (adjusted to linearity where appropriate),
atomic NBO valencies and charges (Tables~S1 to S7), 
Results for bond lengths and Wiberg \cite{Wiberg:68} bond indices are collected in
Table~\ref{table:geometries-c4n},
\figsname\ref{fig:lengths_bonds_c4n_anion} and \ref{fig:lengths_bonds_c4n_cation}.
Changes of these quantities
with reference to the (most stable) neutral doublet are depicted in
\figsname S7 
and S8. 

In principle, chemical bonds of linear carbon species can be of cumulene type or of polyyne type.
Our NBO calculations
(see, e.g., \figsname\ref{fig:lengths_bonds_c4n_anion}b and \ref{fig:lengths_bonds_c4n_cation}b)
indicate that none of these structures (which are incompatible with standard rules of valence)
is present in the \ce{C4N} species investigated here. They do not support claims
that neutral clusters favor a cumulenic bonding while anionic species
prefer polyyne-like bonding \cite{Bartlett:92,Hutter:94,Huang:95,Wang:95}.
The results presented in \figname S20 
also reveal that --- contrary to straightforward
chemical intuition --- there is no simple relationship
between bond order indices and bond lengths. 

\begin{figure*}
  \centerline{
    \includegraphics[width=0.4\textwidth]{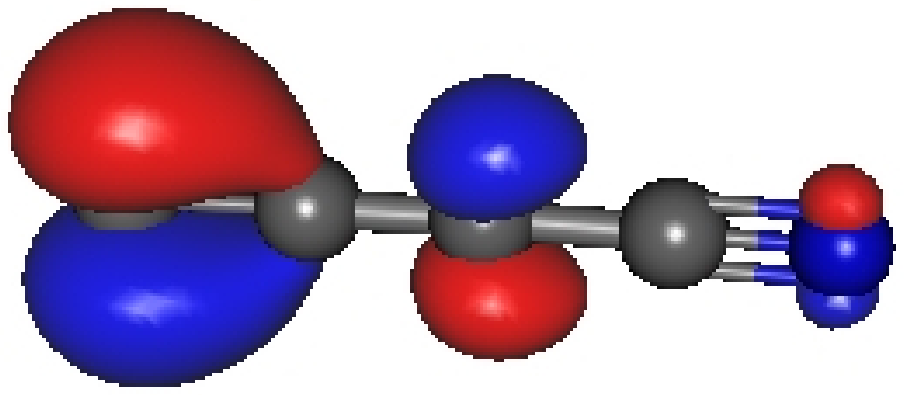}\hspace*{10ex}
    \includegraphics[width=0.4\textwidth]{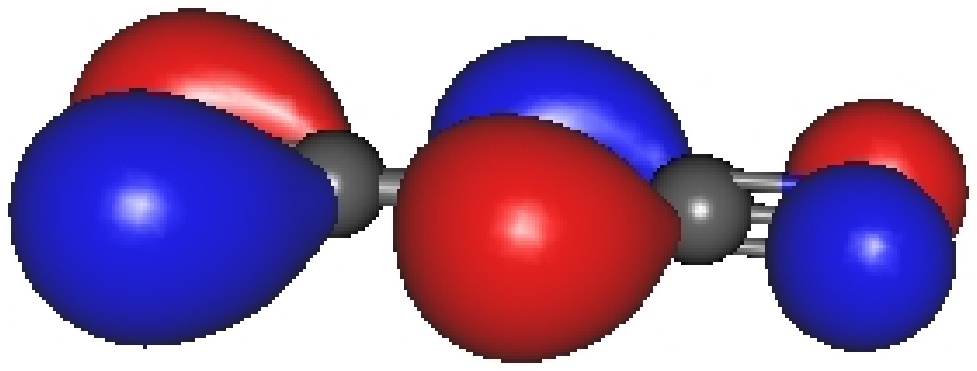}\hspace*{10ex}
  }
  \caption{HOMO and LUMO (left and right panel, respectively) of the \ce{C4N^0} doublet ($\tilde{X}^2\Pi$).}
    \label{fig:homo-lumo-c4n-doublet}
\end{figure*}

\begin{figure*}
  \centerline{
    \includegraphics[width=0.4\textwidth]{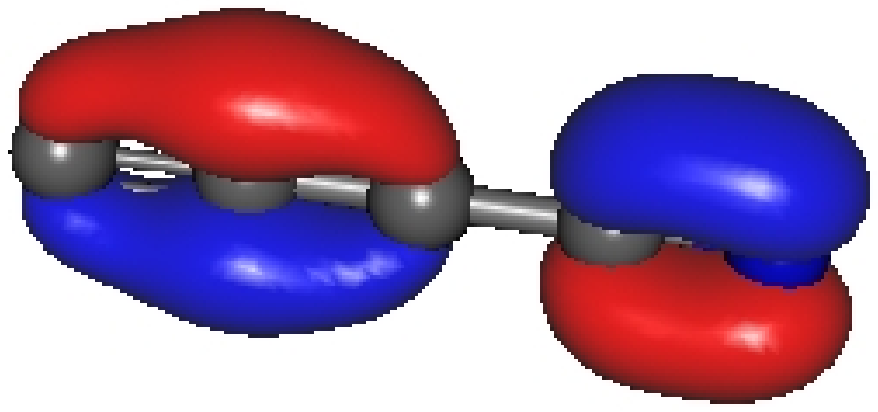}\hspace*{10ex}
    \includegraphics[width=0.4\textwidth]{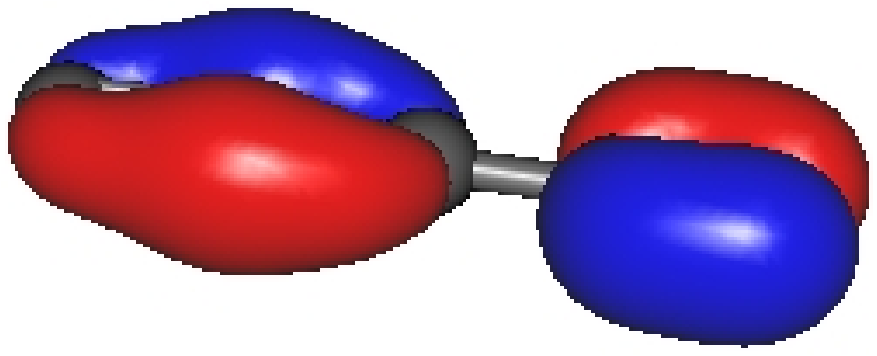}
  }
  \centerline{
    \includegraphics[width=0.4\textwidth]{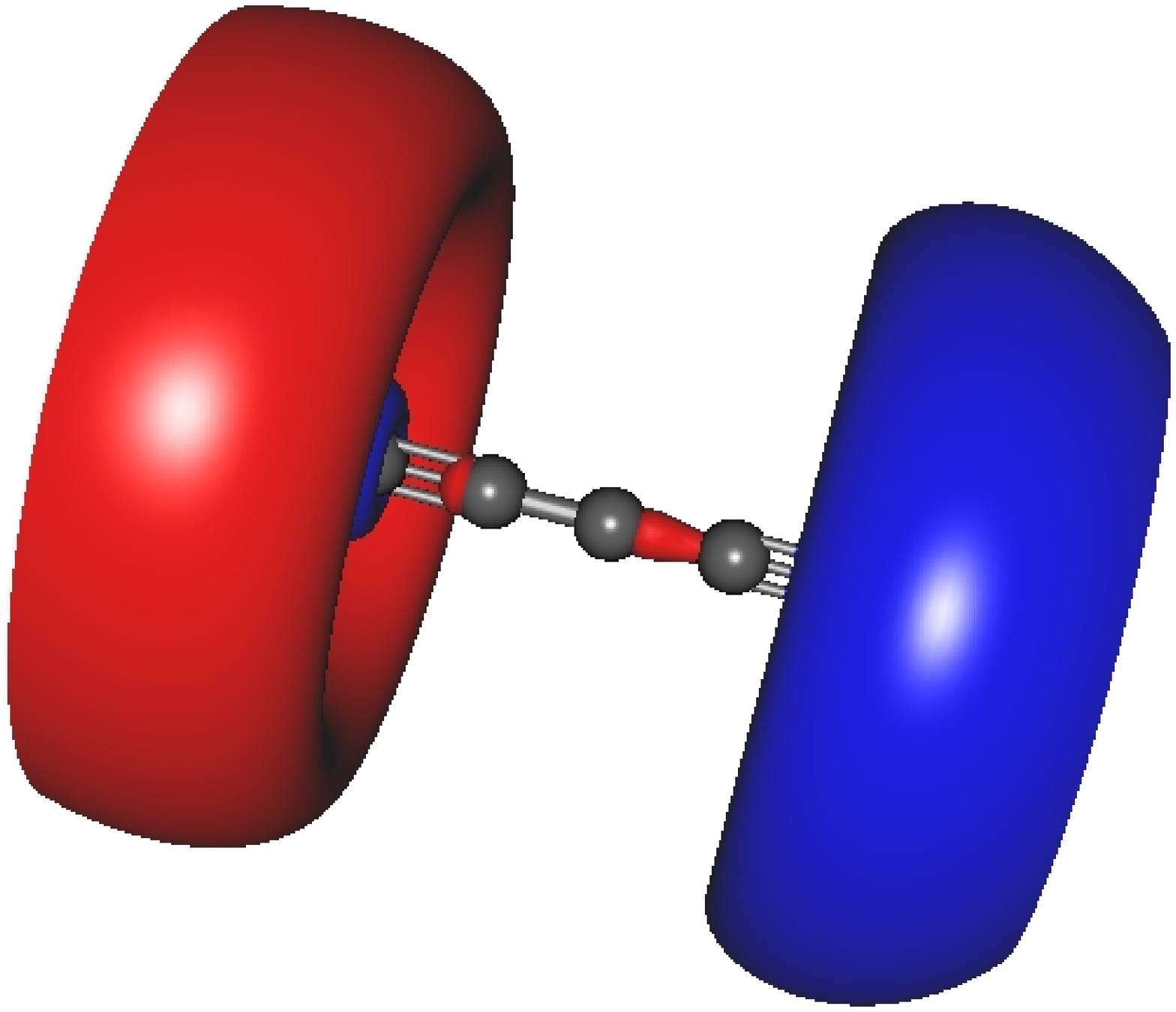}\hspace*{10ex}
    \includegraphics[width=0.4\textwidth]{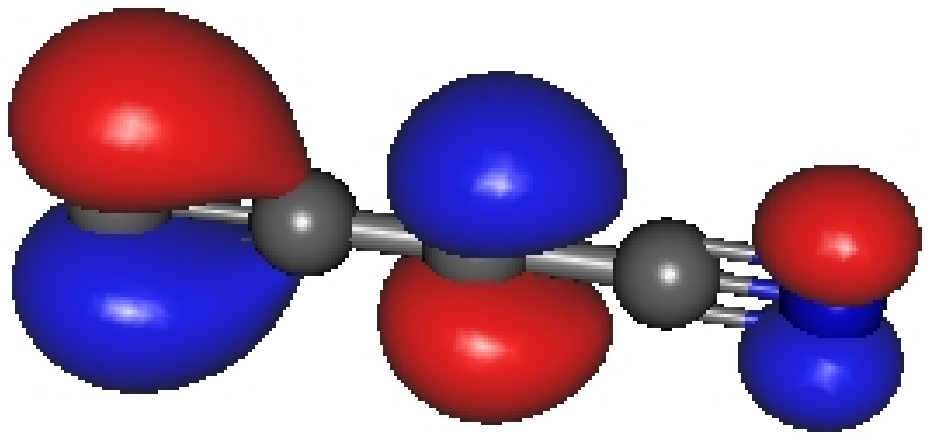}
  }
  \caption{Degenerate HOMO and HOMO-1 (upper left and right panel, respectively)
    and nearly degenerate LUMO and LUMO+1 (lower left and right panel, respectively)
    of the \ce{C4N+} singlet ($^1\Sigma^{+}$).}
    \label{fig:homo-homo-1-deg-lumo-lumo+1-deg-c4n+_singlet}
\end{figure*}

Spatial distributions of the frontier molecular orbitals (highest occupied HOMO and
lowest unoccupied LUMO) are depicted in
\figsname\ref{fig:homo-deg-lumo-c4n-_triplet},
S17, 
S18, 
\ref{fig:homo-lumo-c4n-doublet},
S16, 
\ref{fig:homo-homo-1-deg-lumo-lumo+1-deg-c4n+_singlet} and
S19. 
Because Kohn-Sham orbitals utilized in the DFT are mathematical rather than physical objects
\cite{Parr:89,Baldea:2014c} and ubiquitously utilized HF molecular orbitals rely on a very crude
description, the MO spatial distributions depicted in the aforementioned figures have been obtained
from the natural orbital expansion of the reduced density matrices at the EOM-CCSD level \cite{Baldea:2014c}.
For open-shell cases these results were obtained via restricted open-shell (ROHF-based) approaches.
The inspection of the MOs is useful also because it provides insight into issues under debate
in the past (see \secname''Negatively Charged C$_4$N$^{-}$ Chains''). 

\subsubsection{Neutral {C$_4$N$^0$} Chains}
\label{sec:neutrals}

Within an MO-based picture, having an unpaired $\pi$ electron,
the neutral ground state is a spin doublet with the electronic configuration
\begin{equation}
  \label{eq-X2Pi_neutral}
  \tilde{X}\, ^2\Pi = \mbox{[core]}\,
  6 \sigma^2 \, 7 \sigma^2 \,
  8 \sigma^2 \, 9 \sigma^2 \,
  1\pi^4 \, 10 \sigma^2\,
  2 \pi^4\, 11 \sigma^2 \,
  3\pi^1
\end{equation}

The lowest excited state of the neutral chain, obtained by promoting
a $\sigma$ electron into a $\pi$ orbital, is a spin quartet possessing the following
electronic configuration
\begin{equation}
  \label{eq-a4Sigma-_neutral}
  \tilde{a}\, ^4\Sigma^{-} = \mbox{[core]}\,
  6 \sigma^2 \, 7 \sigma^2 \,
  8 \sigma^2 \, 9 \sigma^2 \,
  1\pi^4 \, 10 \sigma^2\,
  2 \pi^4\, 11 \sigma^1 \,
  3\pi^2
\end{equation}

In agreement with these intuitive considerations, our calculations found that the quartet state
lies higher in energy than the doublet state.
By inspecting the values of the doublet-quartet splitting
collected in Table~\ref{table:Delta-doublet-quadruplet}, a significant difference between the estimates obtained
within the various quantum chemical methods utilized can be concluded. This behavior confirms the
fact noted recently \cite{Baldea:2019e,Baldea:2019g} that electron correlations
(implying by definition that departures from the above MO-based picture are substantial)
in carbon-based chains are strong, which represents a challenge for theory. Still, the large values ($\sim 1$\,eV)
clearly demonstrate that the neutral quartet lies considerably higher in energy than the doublet.

The inspection of Table~\ref{table:geometries-c4n}
and \figsname\ref{fig:lengths_bonds_c4n_cation} and S8 
reveals that the strongest impact of the spin state
on the neutral chain is on the moiety
opposite to the N atoms. The \ce{C1C2} bond of the quartet is longer
than that of the doublet, while \ce{C2C3} bond of the quartet is shorter than that of the doublet
(cf.~Table~\ref{table:geometries-c4n} and \figsname\ref{fig:lengths_bonds_c4n_cation}).
Electronic charge from the \ce{C3} atom in the middle of the chain moves toward the
end opposite to the N atom
(cf.~TablesS1 and S2, 
and \figname S7). 
This renders the dipole moment of the quartet
substantially larger than that of the doublet
(cf.~Table~\ref{table:D} and S16). 
    
\begin{table*}
  \centering
  \caption{Vertical $\Delta_{DQ}^{0}\left(\mathbf{R}_{D,Q}^{0}\right) \equiv \mathcal{E}^{0}_{Q}\left(\mathbf{R}_{D,Q}^{0}\right) - \mathcal{E}^{0}_{D}\left(\mathbf{R}_{D,Q}^{0}\right)$
    and adiabatic
    $\Delta_{DQ}^{0, ad} \equiv \mathcal{E}^{0}_{Q}\left(\mathbf{R}_{Q}^{0}\right) - \mathcal{E}^{0}_{D}\left(\mathbf{R}_{D}^{0}\right)$
    of the doublet-quartet splitting 
    computed by using the total energies $\mathcal{E}^{0}$ of the neutral (\ce{C4N^0}) chains taken at
    B3LYP/6-311++G(3df, 3pd) optimized geometries $\mathbf{R}_{D,Q}^{0}$ of the neutral doublet (label $D$) and quartet (label $Q$)
    without and with corrections due to zero point motion.}
  \label{table:Delta-doublet-quadruplet}
  \begin{tabular}{ccccccc}
    \hline
       & & B3LYP & UCCSD & UCCSD(T) & ROCCSD & ROCCSD(T) \\
    \hline
    $\Delta_{DQ}^{0}\left(\mathbf{R}_{D}^{0}\right)$ & uncorrected     & 1.167  & 0.891 &    1.071 &  0.740 &     0.826 \\    
    &  corrected     & 1.182  & 0.906 &    1.085 &  0.755 &     0.840 \\
    &                &        &       &          &        &           \\
    $\Delta_{DQ}^{0}\left(\mathbf{R}_{Q}^{0}\right)$ & uncorrected & 1.062  & 0.714 & 0.898    &  0.557 &    0.679  \\    
    & corrected & 1.076  & 0.729 & 0.914    &  0.571 &    0.693  \\
        &                &        &       &          &        &           \\
    $\Delta_{DQ}^{0, ad}$ &  uncorrected & 1.167  & 0.891 & 1.071    & 0.740  & 0.826     \\    
                          &    corrected & 1.182  & 0.906 & 1.085    & 0.755  & 0.840     \\    
    \hline
  \end{tabular}
\end{table*}

\subsubsection{Negatively Charged C$_4$N$^{-}$ Chains}
\label{sec:anions}

%
  %
%
Whether the ground state of the \ce{C4H-} chain is a spin singlet or triplet
was an issue of debate in the past.
Based on earlier time-of-flight mass spectroscopy measurements, 
it was \cite{Wang:95}
concluded that the anion possesses a linear triplet ground state. 
This conclusion was challenged by subsequent MP2-based calculations \cite{Zhan:96}
which suggested a bent singlet conformer. In contrast to them,
DFT/B3LYP geometry optimization using aug-cc-pVTZ and 6-311G$^\ast$ basis sets
\cite{Pascoli:99} yielded a linear triplet structure, a result
supported by slow photoelectron velocity-map imaging spectroscopy (SEVI) data \cite{Neumark:09}.

In view of the aforementioned, we paid particular attention to this aspect
and conducted geometry optimization using several exchange-correlation functionals
--- B3LYP, PBE0 and M06-2X --- which are among the most successful
in correctly predicting the lowest energy conformers of
a variety of molecules. In addition, we performed single-point calculations using 
CCSD, CCSD(T), QCISD and QCISD(T). Confirming the conclusion of ref.~\citenum{Pascoli:99},
all these results, which are collected in Table~\ref{table:Delta-anion}
and S12, 
indicate that the most stable conformer is a linear triplet
with the electronic configuration
\begin{equation}
  \label{eq-Sigma-_anion}
  ^3 \Sigma^{-} = \mbox{[core]}\,
  6 \sigma^2 \, 7 \sigma^2 \,
  8 \sigma^2 \, 9 \sigma^2 \,
  1\pi^4 \, 10 \sigma^2\,
  2 \pi^4\, 11 \sigma^2 \,
  3\pi^2
\end{equation}

At the linear triplet optimum geometry (cf.~\figname\ref{fig:geometries-c4n}b)
of $\mathbf{R}_{T}^{-}$, the \ce{C4N-} anion singlet lies substantially higher than the triplet; the energy difference
amounts to $\Delta_{ST}^{-}\left(\mathbf{R}_{T}^{-}\right) \sim 0.7$\,eV (cf.~Table~\ref{table:Delta-anion}).

\begin{figure*}
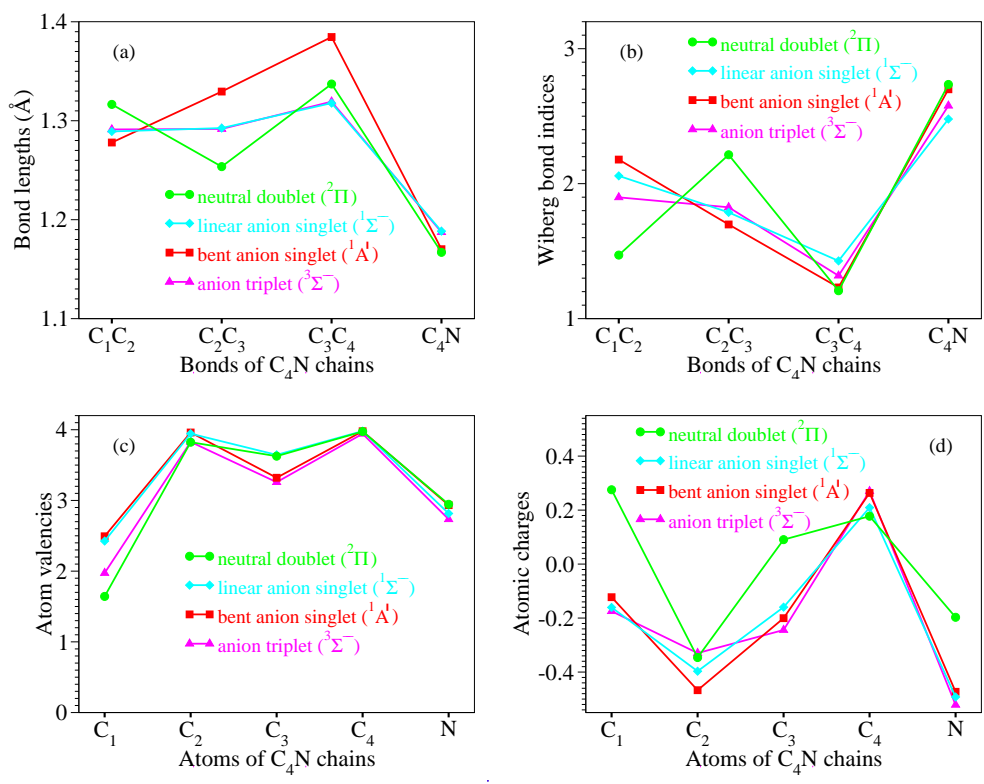

  \includegraphics[width=0.4\textwidth]{fig_bond_lengths_c4n_neutral_anion_b3lyp_MaxPople.eps}
  \includegraphics[width=0.4\textwidth]{fig_wiberg_indices_c4n_neutral_anion_roccsd_t_augccpvtz.eps}
  \includegraphics[width=0.4\textwidth]{fig_valencies_c4n_neutral_anion_roccsd_t_augccpvtz.eps}
  \includegraphics[width=0.4\textwidth]{fig_charges_c4n_neutral_anion_roccsd_t_augccpvtz.eps}
  \caption{(a) Bond lengths (in angstrom), (b) Wiberg bond indices, (c) Wiberg valencies
    and (d) atomic charges of \ce{C4N} chains considered in this paper.}
    \label{fig:lengths_bonds_c4n_anion}
\end{figure*}

Our DFT calculations (not only with the B3LYP exchange-correlation functional but also with
the PBE0 and M06\replaced{-2X}{\bad{2x}} functionals) yielded an energy minimum 
(all vibrational frequencies were real) of an
anion singlet chain possessing a linear (at list within numerical accuracy)
geometry $\mathbf{R}_{lS}^{-} \approx \mathcal{R}_{T}^{-}$
very similar to that of the \ce{C4N-} triplet; compare the bond lengths
(Table~\ref{table:geometries-c4n} and \figname\ref{fig:lengths_bonds_c4n_anion})
and the values of the corresponding reorganization energies
(Table~S26). 

However, rather than a linear conformer, the most stable form of the anion singlet
turned out to be a chain which is bent at the \ce{C3} position
by an angle of $\sim 54^{\circ}$ (\figname\ref{fig:geometries-c4n}a).
This stable bent singlet isomer lies significantly above the metastable linear singlet 
isomer; depending on the method utilized
the energy separation amounts to $\Delta_{lS,bS}^{-} \sim 0.3 - 0.5$\,eV (cf.~Table~\ref{table:Delta-anion}).

\begin{table*}
  \centering
  \caption{Values of the adiabatic (label $ad$) and vertical 
    singlet-triplet splitting
    $\Delta_{bS,T}^{-}\left(\mathbf{R}_{T}^{-}\right) \equiv
    \mathcal{E}^{-}_{T}\left(\mathbf{R}_{T}^{-}\right) - \mathcal{E}^{-}_{bS}\left(\mathbf{R}_{T}^{-}\right)$
    $\Delta_{bS,T}\left(\mathbf{R}_{bS}^{-}\right) \equiv
    \mathcal{E}^{-}_{T}\left(\mathbf{R}_{bS}^{-}\right) - \mathcal{E}^{-}_{bS}\left(\mathbf{R}_{bS}^{-}\right)$
    $\Delta_{lS,T}^{-}\left(\mathbf{R}_{lS}^{-}\right) \equiv
    \mathcal{E}^{-}_{T}\left(\mathbf{R}_{lS}^{-}\right) - \mathcal{E}^{-}_{lS}\left(\mathbf{R}_{lS}^{-}\right)$
    without or with corrections due to zero point motion.
    The geometries $\mathbf{R}^{-}_x$ of the triplet,
    bent singlet and linear singlet ($x=T, bS, lS$ were optimized at the B3LYP/6-311++G(3df, 3pd)
    level of theory.}
  \label{table:Delta-anion}
  \begin{tabular}{ccccccc}
    \hline
    & & B3LYP & UCCSD & UCCSD(T) & ROCCSD & ROCCSD(T) \\
    \hline
    $ -\Delta_{bS,T}^{-}\left(\mathbf{R}_{T}^{-}\right)$ & uncorrected & 0.785  & 0.815 & 0.646    & 0.743  & 0.711     \\    
    & corrected  & 0.791  & 0.821 & 0.652    & 0.749  & 0.717     \\    
     &                &        &       &          &        &           \\
    $ -\Delta_{bS,T}^{-}\left(\mathbf{R}_{bS}^{-}\right)$ & uncorrected & 0.103  & 0.080 & -0.046    & 0.016  & -0.002    \\
    &   corrected & 0.109  & 0.086 & -0.040    & 0.022  &  0.005    \\
    &                &        &       &          &        &           \\
    $ -\Delta_{bS,T}^{-, ad}$ & uncorrected & 0.527  & 0.361 & 0.228    & 0.289  & 0.292     \\    
                         &  corrected  & 0.533  & 0.367 & 0.234    & 0.295  & 0.299     \\    
            &                &        &       &          &        &           \\
    $ -\Delta_{lS,T}^{-}\left(\mathbf{R}_{lS}^{-}\right)$ & uncorrected &  0.785 & 0.816 & 0.647    &  0.744 & 0.712     \\    
    &  corrected  &  0.791 & 0.822 & 0.653    &  0.750 & 0.718     \\    
    \hline
  \end{tabular}
      \begin{tabular}{cccc}
    \hline
    $ - \Delta_{lS,bS}^{-, ad}$ & RB3LYP & RCCSD & RCCSD(T) \\
    \hline
    uncorrected & 0.258  & 0.457 & 0.421 \\
     corrected  & 0.264  & 0.463 & 0.427 \\
    \hline
  \end{tabular}
\end{table*}

To reiterate, the stable \ce{C4N-} triplet is a linear chain (\figname\ref{fig:geometries-c4n}b).
The bonds \ce{C1C2} and \ce{C3C4} of the \ce{C4N-} anion triplet are shorter than in the
neutral doublet while the bonds \ce{C2C3} and \ce{C4N} are longer (Table~\ref{table:geometries-c4n},
and \figsname\ref{fig:lengths_bonds_c4n_anion}a and S7a) 
As expected, bond order indices exhibit opposite changes: shorter bonds have
larger bond indices and vice versa
(\figsname\ref{fig:lengths_bonds_c4n_anion}b and
S7b). 
The calculated values of the atomic charges
(Tables~S1 and S5 
and \figsname\ref{fig:lengths_bonds_c4n_anion}d and
S7d) 
reveal that the 
excess electron of the \ce{C4N-} anion triplet is democratically ($\approx$1/3) shared by the \ce{C1}, \ce{C3} and N
atoms. Positively charged in the neutral chain, \ce{C1} and \ce{C3} atoms become negatively charged in the 
\ce{C4N-} anion triplet. By contrast, electron attachment has
little impact on the charge of the atoms \ce{C2} and \ce{C4}. Interestingly, changes in
the valence state of the atoms do not follow changes in the atomic charges in a simple intuitive way.
To exemplify, although both \ce{C1} and \ce{C3} atoms acquire negative charge,
the (fractional) valence of the former increases while that of the latter decreases
(\figsname\ref{fig:lengths_bonds_c4n_anion}c and S7c). 

Spin singlet appears to enhance delocalization of the excess electron.
Even at the triplet geometry, the singlet state favors delocalization of the excess
electronic charge, which also
involves the \ce{C2} atom. The (most stable) bent singlet geometry
(\figname\ref{fig:geometries-c4n}a and Table~\ref{table:geometries-c4n})
further enhances this delocalization; from their excess electronic charge in the triplet state,
atoms \ce{C1} and \ce{C3} pour electrons into the \ce{C2} atom, which becomes more negatively charged
in the stable bent singlet state.

\begin{figure*}
  \centerline{
    \includegraphics[width=0.4\textwidth]{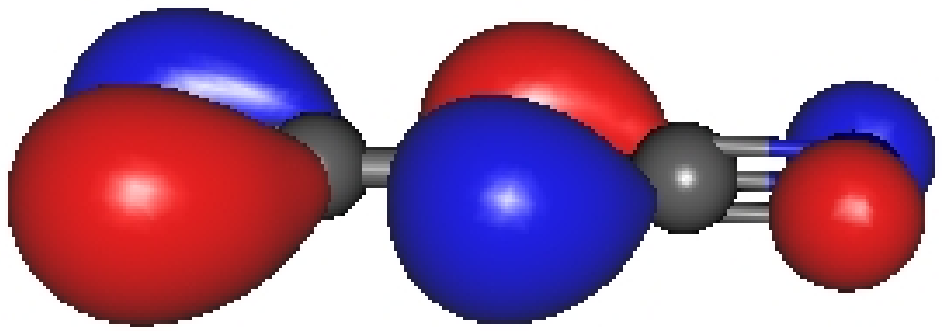}\hspace*{10ex}
    \includegraphics[width=0.4\textwidth]{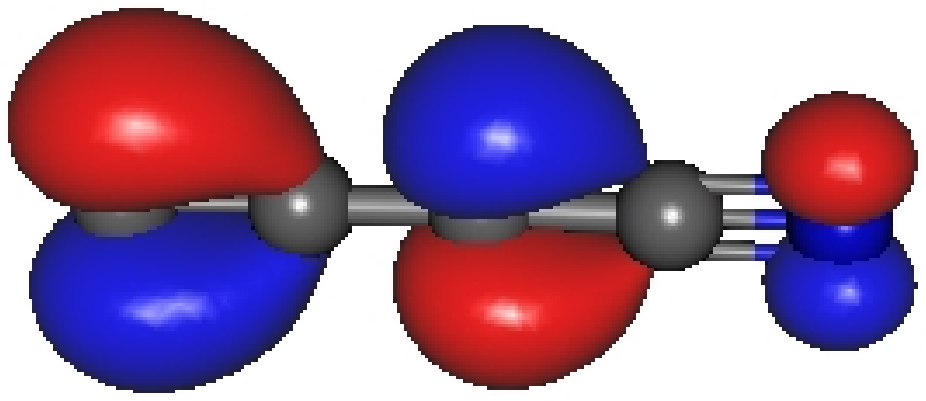}
  }
  \centerline{\includegraphics[width=0.4\textwidth]{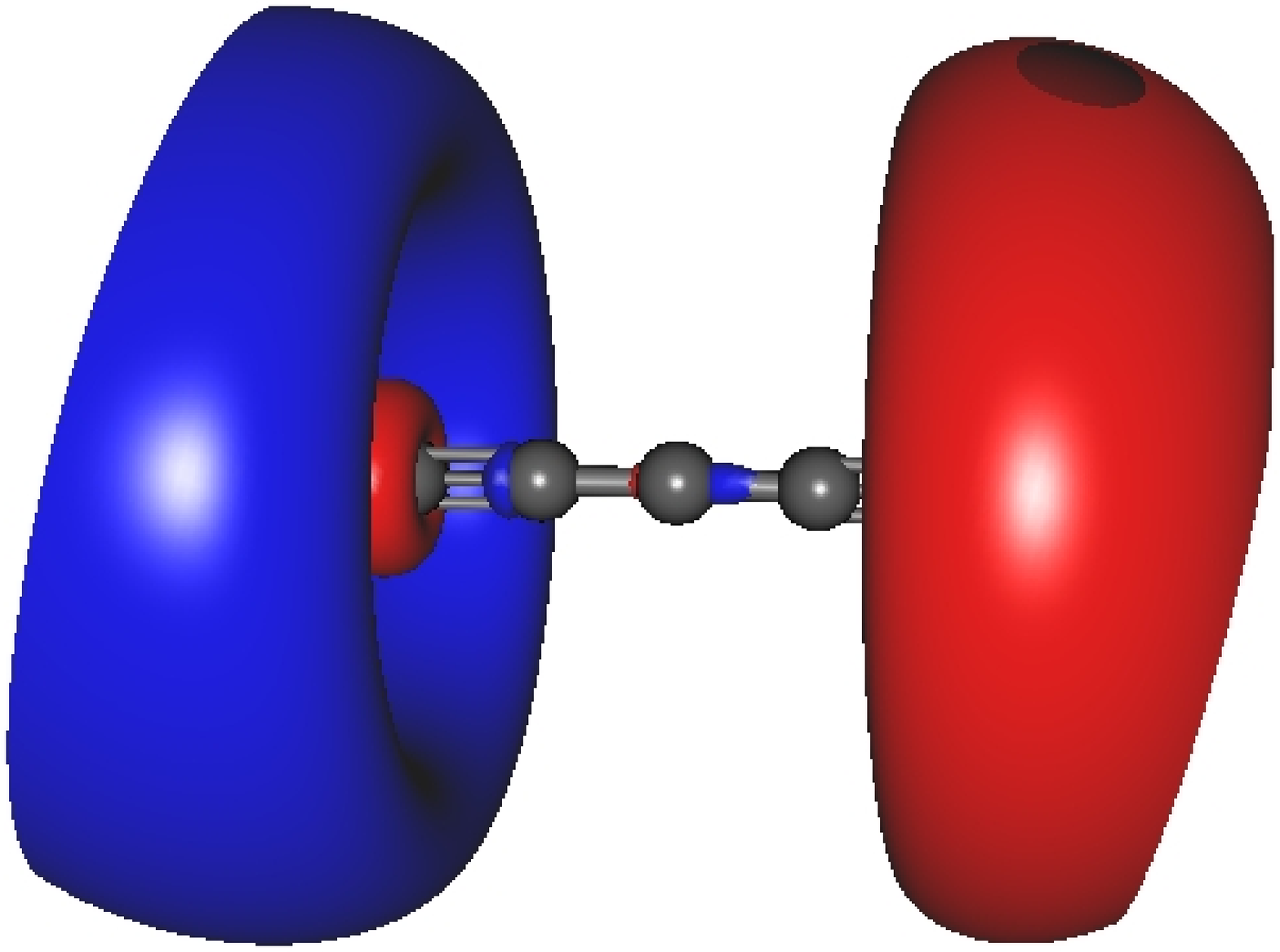}}  
  \caption{Degenerate HOMO and HOMO-1 (upper left and right panel, respectively)
    and LUMO (lower panel) of the \ce{C4N-} triplet ($^3\Sigma^{-}$).}
    \label{fig:homo-deg-lumo-c4n-_triplet}
\end{figure*}

Returning to the controversial aspect noted in the beginning of this section
--- whether the most stable \ce{C4N-} anion is a singlet (as claimed in ref.~\citenum{Zhan:96})
or a triplet (as emerged from subsequent work \cite{Pascoli:99,Neumark:09} and solidified by
our results) ---, it is worth emphasizing the overall strong delocalization
revealed by our NBO results for structures investigated.
This contrasts with the picture \label{page:ump2}
of ref.~\citenum{Zhan:96} claiming that the highest occupied molecular orbital
of the \ce{C4N-} anion is mostly localized at the chain ends.
The inspection of the HOMO spatial distributions of \emph{all} anion species 
(\figsname\ref{fig:homo-deg-lumo-c4n-_triplet},
S17 
and S18) 
reveals that this is in reality not the case.
This incorrect claim may be one reason why ref.~\citenum{Zhan:96} incorrectly ascribed
the most stable anion conformer to be a bent spin singlet.

\subsubsection{Positively Charged C$_4$N$^{+}$ Chains}
\label{sec:cations}

Less surprising that in the case of anion,
calculations indicate that the most stable cation \ce{C4N+} chain is a spin singlet
possessing the following electronic configuration

\begin{equation}
  \label{eq-Sigma+_cation}
  ^1 \Sigma^{+} = \mbox{[core]}\,
  6 \sigma^2 \, 7 \sigma^2 \,
  8 \sigma^2 \, 9 \sigma^2 \,
  1\pi^4 \, 10 \sigma^2\,
  2 \pi^4\, 11 \sigma^2 \,
  3\pi^0
\end{equation}

The lowest triplet state of the cation obtained by promoting a $\sigma$ electron into a $\pi$ orbital
\begin{equation}
  \label{eq-Sigma-_cation}
  ^3 \Sigma^{-} = \mbox{[core]}\,
  6 \sigma^2 \, 7 \sigma^2 \,
  8 \sigma^2 \, 9 \sigma^2 \,
  1\pi^4 \, 10 \sigma^2\,
  2 \pi^4\, 11 \sigma^1 \,
  3\pi^1
\end{equation}
lies above the cation singlet. 

The cationic triplet state possesses a considerably higher energy
(Table~\ref{table:Delta-cation}); the ROCCSD(T) approach predicts a value
$\Delta_{ST}^{+, ad} \approx 1.5$\,eV for the adiabatic singlet-triplet splitting.

\begin{figure*}
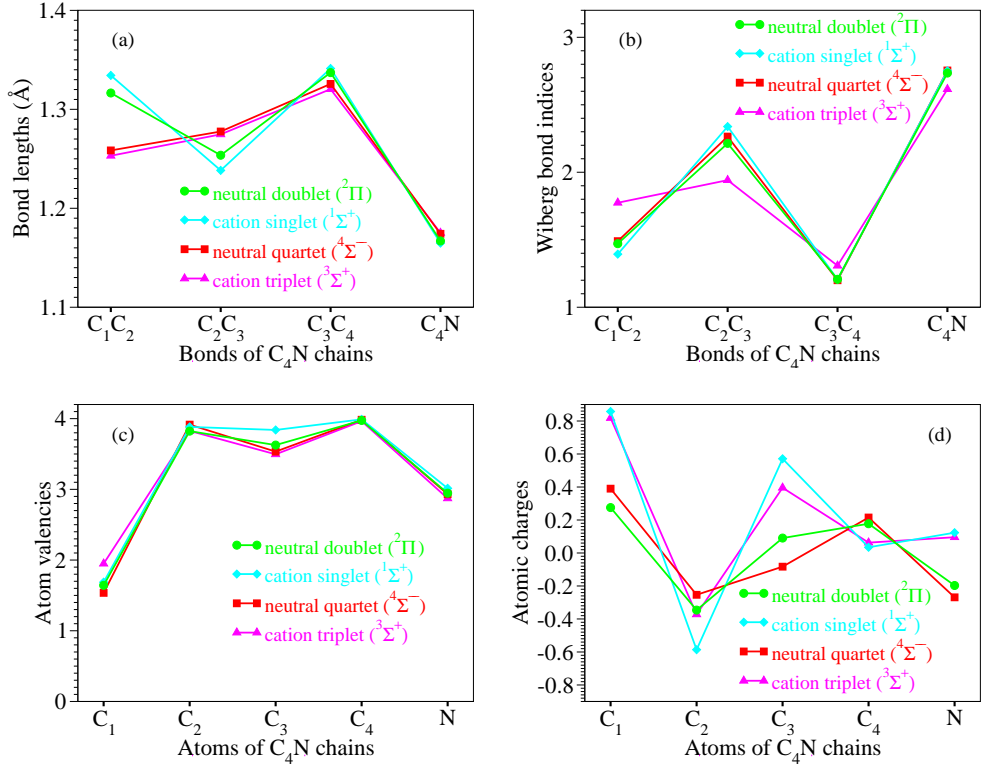

  \includegraphics[width=0.4\textwidth]{fig_bond_lengths_c4n_neutral_cation_b3lyp_MaxPople.eps}
  \includegraphics[width=0.4\textwidth]{fig_wiberg_indices_c4n_neutral_cation_roccsd_t_augccpvtz.eps}
  \includegraphics[width=0.4\textwidth]{fig_valencies_c4n_neutral_cation_roccsd_t_augccpvtz.eps}
  \includegraphics[width=0.4\textwidth]{fig_charges_c4n_neutral_cation_roccsd_t_augccpvtz.eps}
  \caption{(a) Bond lengths (in angstrom), (b) Wiberg bond indices, (c) Wiberg valencies
    and (d) atomic charges of \ce{C4N} chains considered in this paper.}
    \label{fig:lengths_bonds_c4n_cation}
\end{figure*}

Whether in a triplet or a singlet state of the cation \ce{C4N+} chain, electron removal mainly
affects (in decreasing order)
the atomic charges of \ce{C1} ($\sim$0.6\,e both for singlet and triplet),
\ce{C3} ($\sim$0.5\,e for singlet and $\sim$0.3\,e for triplet)
and N ($\sim$0.3\,e both for singlet and triplet); see
Table~S6, 
S7, 
and S1, 
and \figsname\ref{fig:lengths_bonds_c4n_cation}d and S8d. 
Substantial electron removal at the \ce{C1} site has a strong impact on the \ce{C1C2} bond,
which becomes longer
(\figsname\ref{fig:lengths_bonds_c4n_cation}a and S8a). 
Noteworthily, the impact on the bond order index is different: the \ce{C1C2} bond order
becomes significantly weaker in the cation triplet while remaining almost unaffected
in the cation singlet
(\figsname\ref{fig:lengths_bonds_c4n_cation}b and S8b). 

Charge redistribution upon ionization renders
in both cases the \ce{C4} atom more negative
(electronic charge excess $\sim$0.15\,e for singlet and $\sim$0.12\,e for triplet).
Interestingly, the charge of the \ce{C2} atom remains unaffected in the cation triplet;
this behavior is similar to that encountered above in the case of the anion triplet.

\begin{table*}
  \centering
   \caption{Values of the vertical 
     $\Delta_{ST}^{+}\left(\mathbf{R}_{S,T}^{+}\right) \equiv
     \mathcal{E}^{+}_{T}\left(\mathbf{R}_{S,T}^{+}\right) - \mathcal{E}^{+}_{S}\left(\mathbf{R}_{S,T}^{+}\right)$
     and adiabatic $\Delta_{ST}^{+, ad} \equiv \mathcal{E}^{+}_{T}\left(\mathbf{R}_{T}^{+}\right) - \mathcal{E}^{+}_{S}\left(\mathbf{R}_{S}^{+}\right)$
     singlet-triplet cation splitting computed
     without and with corrections due to zero point motion
     using the cation singlet (triplet) geometries
     $\mathbf{R}_{S}^{+}$ ($\mathbf{R}_{T}^{+}$) optimized at the B3LYP/6-311++G(3df, 3pd) level of theory.}
  \label{table:Delta-cation} 
  \begin{tabular}{ccccccc}
    \hline
    &  & B3LYP & UCCSD & UCCSD(T) & ROCCSD & ROCCSD(T) \\
    \hline
    $\Delta_{ST}^{+}\left(\mathbf{R}_{S}^{+}\right)$ & uncorrected & 1.517  & 1.489  & 1.774  & 1.527  &  1.678    \\    
    &  corrected & 1.489  & 1.461  & 1.746  & 1.499  &  1.650    \\    
        &                &        &       &          &        &           \\
     $\Delta_{ST}^{+}\left(\mathbf{R}_{T}^{+}\right)$ & uncorrected & 1.046  & 1.013  & 1,334 & 1.061  & 1.245     \\    
    &  corrected & 1.018  & 0.985  & 1.306 & 1.033  & 1.217     \\
        &                &        &       &          &        &           \\
    $\Delta_{ST}^{+, ad}$ & uncorrected & 1.311  & 1.350  & 1.626 & 1.398  & 1.538    \\
                        &  corrected & 1.283  & 1.322  & 1.598 & 1.370  & 1.510    \\
    \hline
  \end{tabular}
\end{table*}

\subsection{Rotational Constants}
\label{sec:A}

For the linear neutral doublet, the theoretical estimates of
the rotational constant $B$ 
significantly differ from the experimental value $B_{exp} = 2.4226963$\,GHz \cite{McCarthy:03}.
Based on these values for the neutral doublet,
a scaling factor of 0.991937 can be deduced to make more reliable B3LYP/6-311++G(3df, 3pd)-based
estimates for the \ce{C4N} species not investigated so far.

Table~\ref{table:B} collects values of the rotational constants $\mathbf{B}$ computed at
the B3LYP/6-311++G(3df, 3pd) level of theory along with the values scaled as indicated above. 

Excepting the non-linear anion singlet, the linear conformation is responsible to the fact that
the rotational constants $\mathbf{B}$ of all the other species have values close to each other.
\begin{table*}
  \centering
  \caption{Rotational constants of the \ce{C4N} chains investigated in this paper computed at the
    B3LYP/6-311++G(3df, 3pd) level of theory and scaled as described in the main text.
    The value given for the bent anion represents the average of the unscaled values 
    $B=2.82435$\,GHz and $C=2.69945$\,GHz.}
  \label{table:B}
  \begin{tabular}{rrr}
    \hline
    Species         &                 &       $B$\,(GHz) \\
    \hline
    neutral doublet &        computed &         2.44239  \\
                    & \emph{scaled}   &  \emph{2.42270}  \\[0.7ex]
    neutral quartet &        computed &         2.46635  \\    
                    & \emph{scaled}   &  \emph{2.44646}  \\[0.7ex]
    anion triplet   &        computed &         2.42267  \\
                    & \emph{scaled}   &  \emph{2.40313}  \\[0.7ex]
bent anion singlet  &        computed &         2.73467  \\
                    & \emph{scaled}   &  \emph{2.80158}  \\[0.7ex]
linear anion singlet&        computed &         2.42362  \\
                    & \emph{scaled}   &  \emph{2.40408}  \\[0.7ex]
cation singlet      &        computed &         2.44330  \\
                    & \emph{scaled}   &  \emph{2.42360}  \\[0.7ex] 
    cation triplet  &        computed &         2.47931  \\
                    & \emph{scaled}   &  \emph{2.45932}  \\
    \hline
  \end{tabular}
\end{table*}
As known from studies on many other molecular species,
differences in $\mathbf{B}$-values computed using different theoretical approaches visible in
Table~S15 
exceed the typical experimental accuracy ($\sim 10$\,kHz).
Nevertheless, the differences ($> 20$\,MHz)
between the various \ce{C4N} chains are in all cases sufficiently larger than the
measurement accuracy to not impede unambiguous assignment of a certain species from
data (to be) acquired in experiment.

\subsection{Vibrational Properties}
\label{sec:vibrations-c4n}

\begin{figure*}
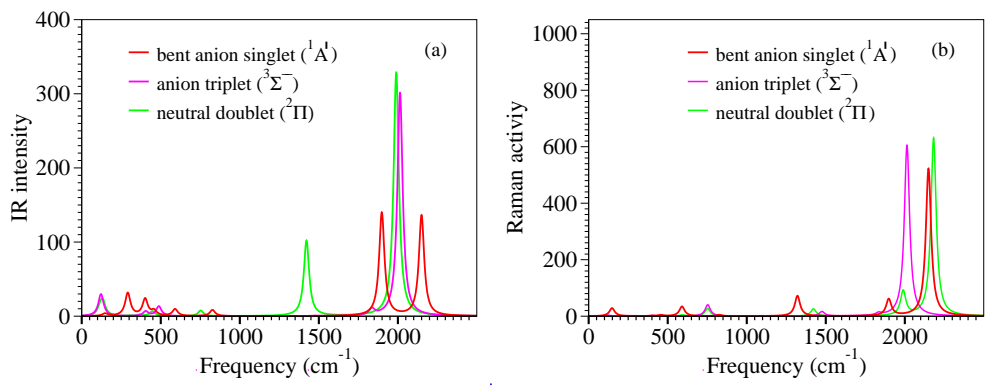

  \includegraphics[width=0.4\textwidth]{fig_ir_c4n_doublet_anion.eps}
  \includegraphics[width=0.4\textwidth]{fig_raman_c4n_doublet_anion.eps}
  \caption{(a) Infrared and (b) Raman spectra of \ce{C4N} neutral ($\tilde{X} ^2 \Pi$) and anion 
   ($^3 \Sigma^{-}$ and $^1 A^\prime$) chains investigated in the present paper.
  }
    \label{fig:vibr-neutrals-anions}
\end{figure*}

\begin{figure*}
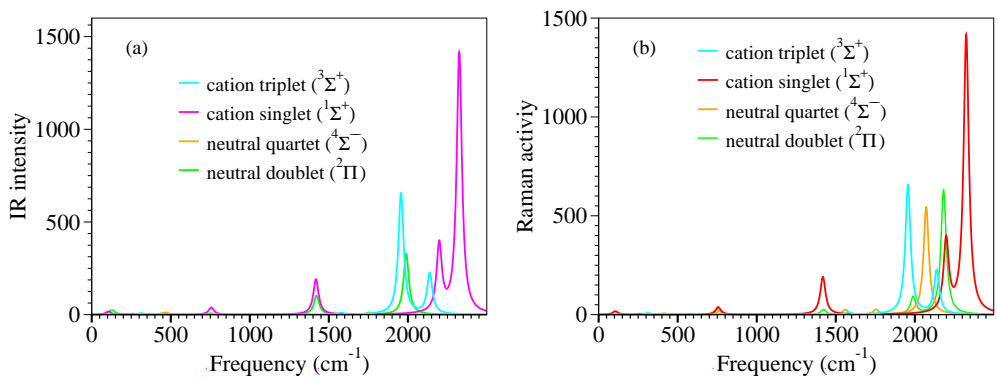

  \includegraphics[width=0.4\textwidth]{fig_ir_c4n_neutral_cation.eps}
  \includegraphics[width=0.4\textwidth]{fig_raman_c4n_neutral_cation.eps}
  \caption{(a) Infrared and (b) Raman spectra of \ce{C4N} neutral ($\tilde{X} ^2 \Pi$ and ($\tilde{a} ^4 \Sigma^{-}$) 
      and cation ($^3 \Sigma^{+}$ and $^1 \Sigma^{+}$ ) chains investigated in the present paper.
  }
    \label{fig:vibr-neutrals-cations}
\end{figure*}

The infrared and Raman spectra of the neutral and anionic species of \ce{C4N}
can be compared with each other by inspecting \figname\ref{fig:vibr-neutrals-anions}. 
The comparison between the neutral and cationic species of \ce{C4N} can be made based on 
\figname\ref{fig:vibr-neutrals-cations}.

Let us start our discussion with the highest vibrational frequency $\nu_{CN}$, which corresponds
to the nitrile radical \ce{C\bond{3}N} stretching, a mode known to be significantly
influenced by the adjacent atoms \cite{Baldea:2019e}.

For the various species investigated, the $\nu_{CN}$-values
vary within a range
$\Delta \nu_{\ce{CN}} < 312\,\mbox{cm}^{-1}$ ($ 2013\,\mbox{cm}^{-1} < \nu_{\ce{CN}} < 2325\,\mbox{cm}^{-1}$,
cf.~Table~S18). 
For a more quantitative comparison with experiment, the aforementioned values should be corrected by means of
appropriate scaling factors \cite{Scott:96,Baldea:2013a}. We do not discuss these details at length
here but still mention that the corresponding values are comparable to those estimated recently for
\ce{HC_nN} chains \cite{Baldea:2019e}.
Based on various experimental data
\cite{Shimanouchi:72,Stephany:74,Ball:00,Maki:01,Baldea:2018a},
a scaling factor of 0.945 appears appropriate for the \ce{C\bond{3}N} stretching mode.

Moving to lower frequencies, the vibration ($\nu_{\ce{C1C2}}$)
can be approximately described as a stretching mode of the \ce{C1C2} bond.
Within the various \ce{C4N} species considered, this vibrational frequency lies within a range
$\Delta \nu_{\ce{C1C2}} < 445\,\mbox{cm}^{-1}$ ($ 1753\,\mbox{cm}^{-1} < \nu_{\ce{C1C2}} < 2198\,\mbox{cm}^{-1}$)
broader than that for $\nu_{\ce{CN}}$.

The next mode can be approximately described as an out-of-phase combination of
\ce{C1C2} and \ce{C4N} stretchings whose frequency varies within a more narrow range
($\Delta \nu < 258\,\mbox{cm}^{-1}$, $ 1321\,\mbox{cm}^{-1} < \nu < 1579\,\mbox{cm}^{-1}$.
The next lower frequency corresponds to a symmetric stretching (``breathing'') mode
($\Delta \nu_{breath} < 75\,\mbox{cm}^{-1}$, $ 753\,\mbox{cm}^{-1} < \nu_{breath} < 827\,\mbox{cm}^{-1}$).
The lowest frequency is associated to a \ce{C1\bond{1}C3\bond{1}N} bending mode
corresponding to oscillations of the angle whose vertex is the \ce{C3} atom, its arms being determined by
\ce{C1C2C3} and \ce{C3C4N}: 
$\Delta \nu_{bent} < 52\,\mbox{cm}^{-1}$, $ 107\,\mbox{cm}^{-1} < \nu_{bent} < 159\,\mbox{cm}^{-1}$.

Modes lying between the aforementioned bending mode and the breathing mode 
are \ce{C3C4N} ($ 369\,\mbox{cm}^{-1} < \nu_{\ce{C3C4N}} < 636\,\mbox{cm}^{-1}$)
and \ce{C1C2C3} bending modes ($ 171\,\mbox{cm}^{-1} < \nu_{\ce{C1C2C3}} < 417\,\mbox{cm}^{-1}$)
Due to the bent shape, \ce{C3C4N} and \ce{C1C2C3} bending modes in the stable \ce{C4N-} singlet
yield in-phase and out-of-phase normal mode combinations. Their
frequencies are $\nu_{in-phase} \approx 591\,\mbox{cm}^{-1}$
$\nu_{out-of-phase} \approx 457\,\mbox{cm}^{-1}$, respectively.

\begin{figure*}
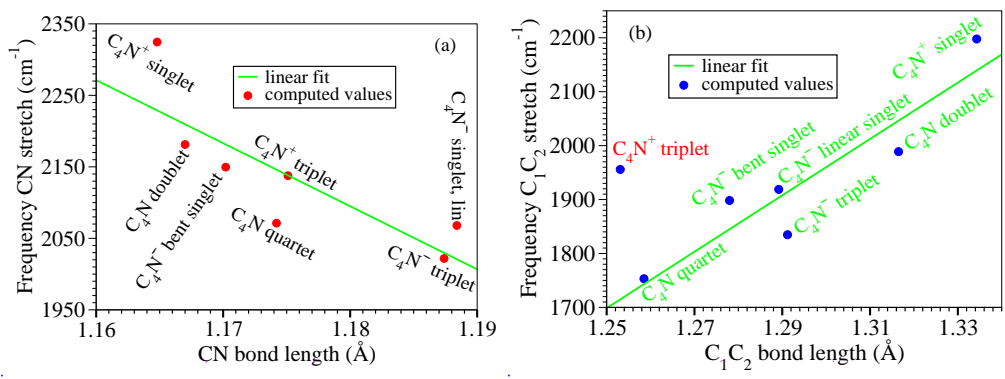

  \includegraphics[width=0.4\textwidth]{fig_CN_omega_vs_length.eps}
  \includegraphics[width=0.4\textwidth]{fig_C1C2_omega_vs_length.eps}
  \caption{Correlation between the vibrational frequency of the nitrile group CN stretching mode
    and the CN and \ce{C1C2} bond lengths (panels a and b, respectively).}
    \label{fig:frequency-length-correlation}
\end{figure*}

Notwithstanding the differences in charge and spin of the various species investigated, the $\nu_{CN}$-values
are reasonably well correlated with the length of the \ce{C\bond{3}N} bond
(\figname\ref{fig:frequency-length-correlation}a) in spite of the fact that the bond lengths
and the bond indices are not so good correlated with each other
(\figname S20a). 

In contrast to this, the $\nu_{\ce{C1C2}}$-values can be simply correlated neither with the
length $l_{\ce{C1C2}}$ nor with the index
$\mathcal{N}_{\ce{C1C2}}$ of this bond (\figname\ref{fig:frequency-length-correlation}b). In fact, even
establishing a correlation between lengths $l_{XY}$ and bond order indices $\mathcal{N}_{XY}$
across molecular species with different charge and/or spin appears to be problematic.
This is visualized in \figname S20, 
where possible $\mathcal{N}$-$l$ correlations are depicted by green lines.

\subsection{Dipole and Quadrupole Moments}
\label{sec:D-Q}

Values of electric dipole momentum $\mathbf{D}$ are collected in
Tables~\ref{table:D} and S16. 
Quadrupole moments $\mathbf{Q}$ are also presented
(Table~S17). 
  Attention should be paid in such calculations to the fact that, for charged species,
  these quantities depend on the coordinate system employed. By default,
  GAUSSIAN rotates/translates
  the molecule to the so-called ``Standard Orientation'', which puts
  the center of nuclear charge (for the most abundant isotopes $^{12}_{\ 6} C$ and $^{14}_{\ 7} N$ 
this coincides with the center of mass) at the origin of the Cartesian axes.
  Reported values refer to this ``Standard Orientation''. For dipole calculations,
  the GAUSSIAN keyword ``NoSymm''
  (used for geometry optimizations in order to unbiasedly search for absolute minima, cf.~\secname''Electronic Structure and Chemical Bonding'')
  should \emph{not} be given in the section route; otherwise the geometry utilized is that
  of the input file, wherein the center of charge is not necessarily
  at the origin of the Cartesian axes.\label{page:dipole}
\begin{table*}
  \centering
  \caption{Values of the dipole momentum $\mathbf{D}$ (field independent basis, debye)
    of the \ce{C4N} chains investigated in this paper at various levels of theory using B3LYP/6-311++G(3df, 3pd) optimized
    geometries.}
  \label{table:D}
  \begin{tabular}{cccccc}
    \hline
    Species              &          Method              & $D_{X}$ & $D_{Y}$ & $D_{Z}$ & $D_{total}$ \\
    \hline                            
    neutral doublet      &     B3LYP/6-311++G(3df, 3pd) &  0.0000 &  0.0000 &  0.3347 & 0.3347 \\
                         &  UCCSD(T)/6-311++G(3df, 3pd) &  0.0000 &  0.0000 &  0.0907 & 0.0907 \\
                         & ROCCSD(T)/6-311++G(3df, 3pd) &  0.0000 &  0.0000 &  0.4512 & 0.4512 \\ 
                         &                              &         &         &   \\
    neutral quartet      &     B3LYP/6-311++G(3df, 3pd) &  0.0000 &  0.0000 & 3.4628 & 3.4628 \\    
                         & UCCSD(T)/6-311++G(3df, 3pd)  &  0.0000 &  0.0000 & 3.2558 & 3.2558 \\
                         & ROCCSD(T)/6-311++G(3df, 3pd) &  0.0000 &  0.0000 & 4.5003 & 4.5003 \\
                         &                              &         &         &   \\
    anion triplet        &     B3LYP/6-311++G(3df, 3pd) &  0.0000 &  0.0000 & 2.9398 & 2.9398  \\
                         &  UCCSD(T)/6-311++G(3df, 3pd) &  0.0000 &  0.0000 & 4.4930 & 4.4930  \\
                         & ROCCSD(T)/6-311++G(3df, 3pd) &  0.0000 &  0.0000 & 2.2379 & 2.2379  \\
                         &                              &         &         &         &         \\ 
    bent anion singlet   &     B3LYP/6-311++G(3df, 3pd) &  0.0000 &  0.5970  &  -2.6678 & 2.7338 \\
                         &  RCCSD(T)/6-311++G(3df, 3pd) &  0.0000 &  0.4227  &  -2.5695 & 2.6040 \\
                         &                              &         &          &         &        \\                            
    linear anion singlet &     B3LYP/6-311++G(3df, 3pd) &  0.0000 &  0.0000  &  2.5344 & 2.5344 \\  
                         &  RCCSD(T)/6-311++G(3df, 3pd) &  0.0000 &  0.0000  &  2.4302 & 2.4302 \\
                         &                              &         &          &         &        \\
    cation singlet       &     B3LYP/6-311++G(3df, 3pd) &  0.0000 &  0.0000  &  2.1890 & 2.1890 \\  
                         &  RCCSD(T)/6-311++G(3df, 3pd) &  0.0000 &  0.0000  &  2.1898 & 2.1898 \\
                         &                              &         &          &         \\                          
    cation triplet       &     B3LYP/6-311++G(3df, 3pd) &  0.0000 &  0.0000  &  4.8254 & 4.8254 \\
                         &  UCCSD(T)/6-311++G(3df, 3pd) &  0.0000 &  0.0000  &  4.1447 & 4.1447 \\
                         & ROCCSD(T)/6-311++G(3df, 3pd) &  0.0000 &  0.0000  &  6.0065 & 6.0065 \\
    \hline
  \end{tabular}
\end{table*}

From computational perspective, the inspection of
Table~\ref{table:D} and S16 
reveals that spin contamination is an important challenge for unrestricted calculations to open-shell chains;
the UHF-based values substantially differ from the ROHF-based values. Noteworthily, the UHF-based estimates 
are not uniformly affected. For the neutral \ce{C4N} doublet, values computed within UHF --- like 
those reported earlier \cite{Pauzat:91}
(cf.~Table~S16) 
--- or on top of UHF
are drastically underestimated with respect to the DFT-based values, which are not notably 
affected by spin contamination. Albeit less pronounced, this is also the case of the cation \ce{C4N+} triplet.
By contrast, UHF-based estimates of $D$ for the neutral quartet and anion triplet 
are smaller than those based on ROHF, although on 
the other side they are closer to the DFT-based values.

The fact that all \ce{C4N} species possess a non-vanishing dipole momentum $D$ is
particularly appealing for astronomical detection because
the intensities of rotational lines scales as $D^2$.
According to Table~\ref{table:D}, the most stable anion \ce{C4H-} triplet chain possesses
a dipole momentum roughly ten times larger than the most stable natural \ce{C4H} doublet.
For this reason, at comparable abundances, spectral lines of \ce{C4H-}
anion chains should be about hundred times stronger than those of their neutral counterparts.
Moreover, in view of the fact that anions have an enthalpy of formation lower than
neutral chains (cf.~Table~\ref{table:enthalpies})
the aforementioned factor ($\sim 100$) is likely underestimated.
By and large, based on these arguments we expect that
the astronomical detection of anion \ce{C4N-} chains will
be easier than that of neutral \ce{C4N^0} chains.

\subsection{Chemical Reactivity Indices}
\label{sec:ip-ea}
Anionic \cite{Millar:17,Vuitton:09} and
cationic \cite{Herbst:78,Weilmuenster:99} species are important constituents
of extraterrestrial environment.
Therefore, the lowest electron attachment energies $EA$
(Tables~\ref{table:EA-triplet@doublet},
S19, 
S20, 
and S21) 
and ionization potentials $IP$ 
(Tables~\ref{table:IP-singlet@doublet},
S22, 
S23, 
and S24) 
examined in this section are quantities of central interest in astrochemistry.

Table~\ref{table:EA-triplet@doublet} and \ref{table:IP-singlet@doublet} include values pertaining
both to vertical and to adiabatic processes. Vertical quantities
correspond to electron addition (or removal) at a given geometry,
e.g.~$EA_{TD}^{vert}\left(\mathbf{R}_{D}^{0}\right) \equiv
\mathcal{E}_{D}^{0}\left(\mathbf{R}_{D}^{0}\right) - \mathcal{E}_{T}^{-}\left(\mathbf{R}_{D}^{0}\right)$
at the energy minimum ($\mathbf{R}_{D}^{0}$) of the (most stable) neutral doublet.
Adiabatic values correspond to molecular energy differences
of anionic (or cationic) and neutral species computed at their own optimum geometries,
e.g.~$EA_{TD}^{ad} \equiv
\mathcal{E}_{D}^{0}\left(\mathbf{R}_{D}^{0}\right) - \mathcal{E}_{T}^{-}\left(\mathbf{R}_{T}^{-}\right)$.
The inspection of Table~\ref{table:EA-triplet@doublet} and \ref{table:IP-singlet@doublet} reveals
that differences between the vertical and adiabatic $EA$ and $IP$ values are very small.
Accordingly, most reorganization energies $\lambda$ are small
(cf.~Tables~S26 
and S27). 
This is the consequence of the fact that most molecular isomers are linear. 
The only notable exception is the (most stable) bent (non-linear) anion singlet ($bS$)
(\figname\ref{fig:geometries-c4n}).

Importantly, our CC-based estimate for the adiabatic electron affinity $EA^{ad} \simeq 3.1$\,eV
(Table~\ref{table:EA-triplet@doublet}) excellently agrees with the experimental value
$EA_{exp}^{ad} = 3.1113 \pm 0.0010$\,eV deduced by means of
high-resolution SEVI spectroscopy \cite{Neumark:09}. One could note in this context that
inaccuracies of up to $\sim 50$\,meV of CC-based $EA$-estimates are unavoidable for
present state-of-the-art of theory. 

Such inaccuracies result because 
various single-point ab initio calculations ($\Delta$-UCCSD, $\Delta$-UCCSD(T), $\Delta$-ROCCSD, $\Delta$-ROCCSD(T),
$\Delta$-QCISD, $\Delta$-QCISD(T) and EA-EOM-CCSD) at a certain (optimum) geometry yield values 
slightly differ from each other (cf.~Table~\ref{table:EA-triplet@doublet}).
Another source of inaccuracy
is the geometry utilized in single-point calculations, which is also slightly 
affected by the optimization procedure (B3LYP/6-311++G(3df, 3pd), PBE0/6-311++G(3df, 3pd),
or M06-2X/6-311++G(3df, 3pd), 
cf.~Tables~S19, 
S20, 
and S21). 

Confirming previous report \cite{Neumark:09}, the DFT/B3LYP-based estimate
(Table~\ref{table:EA-triplet@doublet}) departs from the experimental value by about 0.2\,eV.
This DFT-based inaccuracy
--- which is comparable to that found for other molecular species \cite{Baldea:2014c} ---
may not be sufficient to derive accurate chemical reactivity indices (see below) needed 
for reliable astrochemical modeling.   

In addition to electron affinities, ionization energies $IP$ are also needed for
modeling extraterrestrial environments. $IP$-values of vertical and adiabatic ionization energies with and
without corrections due to zero-point motion were also computed by means of the same methods utilized
for $EA$'s. They are reported in
Tables~\ref{table:IP-singlet@doublet},
S22, 
S23, 
and S24. 

Once the $IP$ and $EA$ values are known, other basic chemical reactivity indices can be estimated,
which are important because they serve as input information for modeling astrochemical evolution
of a given environment. 
As an example, results for the chemical hardness $\eta \equiv IP - EA$ are reported in
Table~\ref{table:eta-doublet} and \ref{table:eta-quadruplet}. These tables include values of
both ``global'' ($\eta^{vert}$) and ``combined'' hardness ($\eta^{ad}$);
the former are computed using the vertical $IP^{vert}$ and $EA^{vert}$ values, the latter
are obtained from the adiabatic  $IP^{ad}$ and $EA^{ad}$ values.

We chose to show this quantity ($\eta$)
also because it reveals that, along with the recently examined \ce{HC10N} chain
\cite{Baldea:2019e},
\ce{C4N} is another carbon-based chain of astrochemical interest agreeing with
Pearson's conjecture \cite{Pearson:87}. Being more stable than the neutral
quartet (cf.~Table~\ref{table:enthalpies} and \figname\ref{fig:H}),
the neutral doublet is chemically harder 
($\eta_{D} > \eta_{Q}$, cf.~Table~\ref{table:eta-doublet} and \ref{table:eta-quadruplet}).

\begin{table*}
  \small
  \centering
  \caption{Values of the vertical and adiabatic doublet-triplet electron attachment
    ($EA_{TD}^{vert}\left(\mathbf{R}\right) \equiv \mathcal{E}^{0}_{D}\left(\mathbf{R}\right) - \mathcal{E}^{-}_{T}\left(\mathbf{R}\right)$
    and
    $EA_{TD}^{ad} \equiv
    \mathcal{E}^{0}_{D}\left(\mathbf{R}_{D}^{0}\right) - \mathcal{E}^{-}_{T}\left(\mathbf{R}_{T}^{-}\right)$, respectively)
    computed using the neutral doublet $\mathbf{R}_{D}^{0}$ and anion triplet $\mathbf{R}_{T}^{-}$
    B3LYP/6-311++G(3df, 3pd) optimum geometries without or with corrections due to zero point motion.
    The present B3LYP/6-311++G(3df, 3pd)-based adiabatic value ($EA_{TD}^{ad} = 3.274$\,eV)
    agrees with the B3LYP/aug-cc-pVTZ estimate (3.29\,eV \cite{Neumark:09}). The vertical
    uncorrected value $EA_{TD}^{vert}\left(\mathbf{R}_{T}^{-}\right) = 1.91$\,eV deduced via Koopmans
    theorem at RHF/3-21G level \cite{Wang:95} is drastically underestimated.    
  }
  \label{table:EA-triplet@doublet}
  \begin{tabular}{cccccccc}
    \hline
                                                &             & B3LYP & UCCSD & UCCSD(T)& ROCCSD & ROCCSD(T) & EOM-ROCCSD \\
    \hline
 $EA_{TD}^{vert}\left(\mathbf{R}_{D}^{0}\right)$    & uncorrected & 3.217  & 3.077 &  3.038  & 3.003  & 2.983     & 3.027      \\    
                                                 &  corrected  & 3.207  & 3.066 &  3.029  & 2.993  & 2.973     & 3.017     \\    
                                                 &             &        &       &         &        &           &         \\
 $EA_{TD}^{vert}\left(\mathbf{R}_{T}^{-}\right)$ & uncorrected & 3.360  & 3.262 &  3.218  & 3.182  & 3.135     & 3.199       \\    
                                                 &  corrected  & 3.350  & 3.252 &  3.208  & 3.172  & 3.124     & 3.189    \\
                                                 &             &        &       &         &        &           &          \\
                                  $EA_{TD}^{ad}$ & uncorrected & 3.285  & 3.121 &  3.100  & 3.034  & 3.059     & 3.109      \\    
                                                 &  corrected  & 3.274  & 3.111 &  3.090  & 3.024  & 3.048     & 3.099    \\
    \hline
  \end{tabular}
\end{table*}

\begin{table*}
  \small
  \centering
   \caption{Values of the vertical and adiabatic doublet-singlet ionization energy
    ($IP_{SD}^{vert}\left(\mathbf{R}\right) \equiv \mathcal{E}^{+}_{S}\left(\mathbf{R}\right) - \mathcal{E}^{0}_{D}\left(\mathbf{R}\right)$
     and
     $IP_{SD}^{ad} \equiv \mathcal{E}^{+}_{S}\left(\mathbf{R}_{S}^{+}\right) - \mathcal{E}^{0}_{D}\left(\mathbf{R}_{D}^{0}\right)$,
     respectively) computed using 
     the neutral doublet $\mathbf{R}_{D}^{0}$ and cation singlet $\mathbf{R}_{S}^{+}$
     B3LYP/6-311++G(3df, 3pd) optimum geometries
     without and with corrections due to zero point motion.}
  \label{table:IP-singlet@doublet} 
  \begin{tabular}{cccccccc}
    \hline
                                                    &             & B3LYP&  UCCSD &UCCSD(T)& ROCCSD &ROCCSD(T)& EOM-ROCCSD \\
    \hline
    $IP_{SD}^{vert}\left(\mathbf{R}_{D}^{0}\right)$ & uncorrected & 9.812 & 9.666  & 9.408  & 9.681  & 9.514   & 9.802      \\    
                                                    &  corrected  & 9.852 & 9.705  & 9.448  & 9.721  & 9.554   & 9.842      \\    
                                                    &             &       &        &        &        &         &            \\
   $IP_{SD}^{vert}\left(\mathbf{R}_{S}^{+}\right)$  & uncorrected & 9.780 & 9.639  & 9.396  & 9.663  & 9.493   & 9.797      \\    
                                                    & corrected   & 9.819 & 9.678  & 9.436  & 9.703  & 9.533   & 9.836      \\    
                                                    &             &       &        &        &        &         &            \\
                                     $IP_{S}^{ad}$  & uncorrected & 9.794 & 9.631  & 9.392  & 9.646  & 9.497   & 9.783      \\    
                                                    &  corrected  & 9.833 & 9.670  & 9.431  & 9.686  & 9.537   & 9.823      \\    
    \hline
  \end{tabular}
\end{table*}

\begin{table*}
  \centering
  \caption{Values of the ``global'' (vertical) $\eta_{D}^{vert}$ and ``combined'' (adiabatic)
    $\eta_{D}^{ad}$ chemical hardness of the neutral doublet without and with corrections due to zero point motion.
    All geometries were optimized at the B3LYP/6-311++G(3df, 3pd) level of theory.}
  \label{table:eta-doublet} 
  \begin{tabular}{cccccccc}
    \hline
    & & B3LYP & UCCSD & UCCSD(T) & ROCCSD & ROCCSD(T) & EOM-ROCCSD \\
    \hline
    $\eta_{D}^{vert}$  & uncorrected & 6.595  &  6.589 & 6.370  & 6.678  & 6.531 & 6.775 \\    
                     &   corrected & 6.645  &  6.639 & 6.420  & 6.728  & 6.581 & 6.825 \\    
                     &             &        &        &        &        &       & \\
      $\eta_{D}^{ad}$ & uncorrected & 6.509  &  6.509 & 6.292  & 6.612  & 6.438 & 6.675 \\    
                     &  corrected & 6.559  &  6.559 & 6.342  & 6.662  & 6.489 & 6.724 \\    
    \hline
  \end{tabular}
\end{table*}

\begin{table*}
  \centering
  \caption{Values of the ``global'' (vertical) $\eta_{Q}^{vert}$ and ``combined'' (adiabatic)
    $\eta_{Q}^{ad}$ chemical hardness of the neutral quartet 
    without and with corrections due to zero point motion.
    All geometries were optimized at the B3LYP/6-311++G(3df, 3pd) level of theory.}
  \label{table:eta-quadruplet} 
  \begin{tabular}{ccccccc}
    \hline
    & & B3LYP & UCCSD & UCCSD(T) & ROCCSD & ROCCSD(T) \\
    \hline
    $\eta_{Q}^{vert}$ & uncorrected & 4.467  & 5.093  &  4.491 &  4.952 & 4.558 \\    
                    &   corrected & 4.488  & 5.113  &  4.512 &  4.973 & 4.579 \\    
                    &             &        &        &        &        &       \\
      $\eta_{Q}^{ad}$ & uncorrected & 4.175  & 4.725  &  4.148 &  4.590 & 4.208 \\    
                    &  corrected  & 4.195  & 4.746  &  4.169 &  4.611 & 4.229 \\    
    \hline 
  \end{tabular}
\end{table*}

We do not want to end this section without making two technical remarks.

First, to improve
the agreement with experiment, DFT approaches to estimate $EA$ and $IP$ often
use long-range corrected functionals. To check whether this is the case of \ce{C4N} chains,
we also conducted DFT calculations using two such functionals
(LC-BLYP and LC-$\omega$PBE) embodying long-range corrections.
The results presented in Tables~S19 
and
S22 
do not substantiate this expectation.
The long-range corrected ($lrc$) LC-BLYP and LC-$\omega$PBE functionals estimates
($EA_{lrc}^{ad} \sim 3.5$\,eV) for $EA$
yield larger deviations from the experimental value ($EA_{exp}^{ad} = 3.1113 \pm 0.001 $\,eV)
\cite{Neumark:09} than the values ($EA_{wo-lrc}^{ad} \sim 3.3$\,eV)
based on the non-corrected B3LYP and PBE0 functionals.
Double-hybrid functionals do not perform better;
with zero-point energy corrections DSD-PBEP86/6-311++G(3df, 3pd)
gives $EA^{ad} = 3.423$\,eV.\label{page:dsd-pbep86}

Second, we noted above that spin-splitting ($\Delta_{DQ}^{0}$, $\Delta_{ST}^{\pm}$) values
estimated within unrestricted CC-based approaches substantially differ from those
based on restricted open shell approaches. By contrast, similar to other cases,\cite{Baldea:2014c,Baldea:2020a}
spin contamination does not appear
to be an issue for $IP$ and $EA$; UCC-based values do not notably differ from those obtained within
ROCC approaches.

\subsection{C$_4$N$^{-}$ versus HC$_4$N$^0$}
\label{sec:c4n-_vs_hc4n}

The \ce{C4N-} and \ce{HC4N} chains are isoelectronic. For this reason, it may be not surprising that
they both have spin triplet ground states.
Likewise, their most stable singlet state is a bent conformer
(\figsname\ref{fig:geometries-c4n}
and S9). 

Surprisingly, notwithstanding these qualitative similarities, quantitative differences between their 
properties are significant. This holds for all properties:
structural (Table~\ref{table:bonds-c4n-_vs_hc4n}),
electronic
(Tables~\ref{table:D-c4n-_vs_hc4n},
S25, 
S9, 
and S10) 
and vibrational
(\figsname S14 
and S15). 

Let us refer to a few specific aspects. As visible in
\figsname S10a, 
S11a, 
S12a, 
and
S13a, 
bond lengths differences can amount up to
0.05\,{\AA}, as the case of the \ce{C1C2} bond at the molecular end opposite to the N atom.
This is associated with a substantial change in the (fractional) valence state of the \ce{C1} atom
(\figsname S10c, 
S11c, 
S12c, 
and
S13c). 
Noteworthily, the other molecular end is also
affected; see, for example, the charge of the N atom both in singlet
(\figsname S10d 
and
S12d) 
and triplet
(\figsname S11d 
and
S13d) 
states. Again, in spite of their isoelectronicity, both infrared and Raman spectra of \ce{HC4N^0}
are significantly different from those of \ce{C4N-}; compare
\figsname S14a and S14b 
and 
\figsname S15a and S15b, 
respectively.

\begin{table*}
  \centering
  \caption{Results for the isoelectronic molecular pair (\ce{C4N-}, \ce{HC4N})
    geometries optimized at the B3LYP/6-311++G(3df, 3pd) level of theory
    without imposing symmetry constraints.
    Bond lengths $l$ between atoms XY (in angstrom),
    angles $\alpha$ between atoms $\widehat{\mbox{\ce{XYZ}}}$ (in degrees) and Wiberg bond order indices $\mathcal{N}$.
  }
  \label{table:bonds-c4n-_vs_hc4n}
  \begin{tabular}{ccccccccccccc} 
    \hline
    Species & Property & \ce{C1C2} & $\widehat{\mbox{\ce{C1C2C3}}}$ & \ce{C2C3} & $\widehat{\mbox{\ce{C2C3C4}}}$ & \ce{C3C4} & $\widehat{\mbox{\ce{C3C4N}}}$ & \ce{C4N} \\
    \hline
     stable bent \ce{C4N-} singlet  & $l$, $\alpha$ & 1.2780 & 174.3 & 1.3295 & 125.7 & 1.3846 & 171.4 & 1.1702 \\
     & $\mathcal{N}$ & 2.1785 &       & 1.6978 &       & 1.2309 &       & 2.6995 \\
     &               &        &       &        &       &        &       &        \\
     \ce{HC4N} singlet            & $l$, $\alpha$ & 1.2267 & 172.5 & 1.3353 & 127.5 & 1.3645 & 173.2 & 1.1682 \\
                                  & $\mathcal{N}$ & 2.4495 &       & 1.4297 &       & 1.2551 &       & 2.6754 \\
    \hline
     linear \ce{C4N-} triplet   & $l$, $\alpha$ & 1.2912 & 179.8 & 1.2917 & 178.7 & 1.3193 & 180.0 & 1.1874 \\
                                & $\mathcal{N}$ & 1.8987 &       & 1.8251 &       & 1.3182 &       & 2.5775 \\
     &               &        &       &        &       &        &       &        \\
     \ce{HC4N} triplet          & $l$, $\alpha$ & 1.2406 & 179.8 & 1.2920 & 178.8 & 1.3181 & 179.9 & 1.1790 \\
                                & $\mathcal{N}$ & 2.4537 &       & 1.4572 &       & 1.2929 &       & 2.6509 \\      
     \hline
  \end{tabular} 
\end{table*}
\begin{table*}
  \centering
  \caption{Values of the dipole momentum $\mathbf{D}$ (field independent basis, debye) of the 
           isoelectronic \ce{C4N-} and \ce{HC4N} chains at various levels of theory.}
  \label{table:D-c4n-_vs_hc4n}
  \begin{tabular}{cccccc}
    \hline
    Species              &          Method              & $D_{X}$  & $D_{Y}$   & $D_{Z}$  & $D_{total}$ \\
    \hline                            
   \ce{C4N-} triplet     &     B3LYP/6-311++G(3df, 3pd) &  0.0000 &   0.0000  &  2.9398 & 2.9398  \\
                         &     B3LYP/aug-cc-pVTZ        &  0.0000 &   0.0000  &  2.9400 & 2.9400  \\  
                         &  UCCSD(T)/6-311++G(3df, 3pd) &  0.0000 &   0.0000  &  4.5215 & 4.5215  \\
                         &  UCCSD(T)/aug-cc-pVTZ        &  0.0000 &   0.0000  &  4.5002 & 4.5002  \\
                         & ROCCSD(T)/6-311++G(3df, 3pd) &  0.0000 &   0.0000  &  2.2379 & 2.2379  \\
                         & ROCCSD(T)/aug-cc-pVTZ        &  0.0000 &   0.0000  &  2.2447 & 2.2447  \\
                         &                              &         &           &         &         \\
   \ce{HC4N} triplet     &     B3LYP/6-311++G(3df, 3pd) &  0.0000 &   0.0000  &  4.3495 &  4.3495 \\  
                         &     B3LYP/aug-cc-pVTZ                            &  0.0000 &   0.0000  &  4.3460 &  4.3460 \\
                         &  UCCSD(T)/6-311++G(3df, 3pd) &  0.0000 &   0.0000  &  4.0705 &  4.0705 \\
                         &  UCCSD(T)/aug-cc-pvtz                            &  0.0000 &   0.0000  &  4.0656 &  4.0656 \\
                         & ROCCSD(T)/6-311++G(3df, 3pd) &  0.0000 &   0.0000  &  4.6327 &  6.6327 \\
                         & ROCCSD(T)/aug-cc-pVTZ                            &  0.0000 &   0.0000  &  4.6267 &  4.6267 \\ 
\hline
bent \ce{C4N-} singlet   &     B3LYP/6-311++G(3df, 3pd) &  0.0000 &   0.5970  & -2.6678 &  2.7338 \\
                         &     B3LYP/aug-cc-pVTZ                            &  0.0000 &   0.5970  & -2.6678 &  2.7338 \\
                         &  RCCSD(T)/6-311++G(3df, 3pd) &  0.0000 &   0.4227  & -2.5695 &  2.6040 \\
                         &  RCCSD(T)/aug-cc-pVTZ                            &  0.0000 &   0.4233  & -2.5714 &  2.6061 \\
                         &                                                  &         &           &         &         \\
\ce{HC4N}   singlet     &      B3LYP/6-311++G(3df, 3pd) &  0.0000 &  -1.0005  &   4.1944 &  4.3121 \\
                        &      B3LYP/aug-cc-pVTZ                            &  0.0000 &  -1.0020  &   4.1968 &  4.3148 \\
                        &  RCCSD(T)/6-311++G(3df, 3pd)  &  0.0000 &  -1.1285  &   4.4508 &  4.5916 \\
                        &      RCCSD(T)/aug-cc-pVTZ                         &  0.0000 &  -1.1300  &   4.4518 &  4.5929 \\
\hline
  \end{tabular}
\end{table*}

\subsection{Chemical Stability and Potential Chemical Pathways of Formation}
\label{sec:enthalpies}

\begin{figure*}
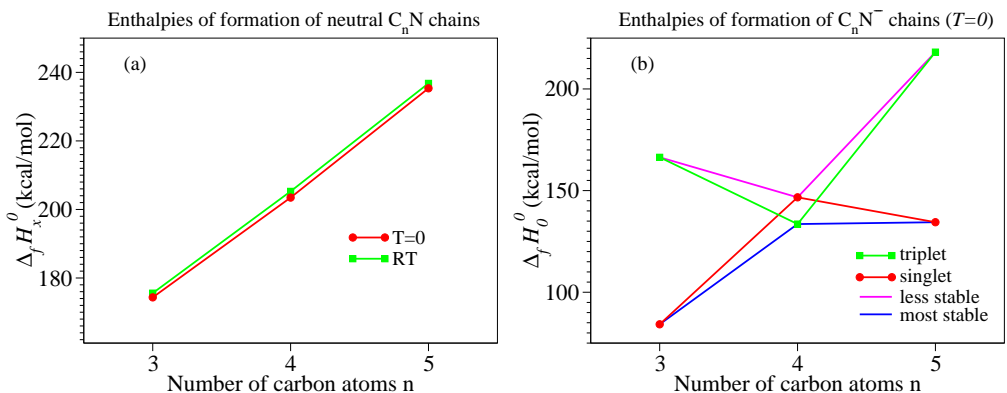

  \centerline{
    \includegraphics[width=0.4\textwidth]{fig_H_cxn-doublets.eps}
    \includegraphics[width=0.4\textwidth]{fig_H_cxn-.eps}
  }
  \caption{Enthalpies of formation of neutral doublet \ce{C_nN^0} and anion \ce{C_nN^-} chains
    (panel a and b, respectively).}
    \label{fig:H}
\end{figure*}

To address the problem of chemical stability,
we first calculated relevant enthalpies of formation 
$\Delta_f H_{x}^0$'s. Values at zero ($x=0$) and at room ($x=RT$) temperature
are presented in Table~\ref{table:enthalpies}
and \figname\ref{fig:H}. According to \figname\ref{fig:H}a, the enthalpies of formation for spin doublets
--- which are the most stable states of the neutral chains --- linearly increase with the chain size ($n$).

\begin{table}
\setcounter{table}{12}
  \centering
  \caption{Enthalpies of Formation of Carbon Chains Discussed in This Paper at zero (Subscript 0)
    and Room Temperature (Subscript RT) Computed by Using CBS-QB3 Protocol. All values in kcal/mol. 
    Values at RT for Anions Computed Using
    the ion convention (see, e.g., https://webbook.nist.gov/chemistry/ion/\#A.) Notice that the values in this table 
  are those corrected in ref.~\citenum{Baldea:2021e} which are somewhat different from those of ref.~\citenum{Baldea:2020b}.}
  \label{table:enthalpies}
  \begin{tabular}{rrr}
    \hline
    Species & $\Delta_{f} H_{0}^{0}$ & $\Delta_{f} H_{RT}^{0}$ \\
    \hline
    \ce{C3N^0} doublet       &  175.502 & 178.540  \\
    \ce{C4N^0} doublet       &  206.071 & 209.938  \\
    \ce{C5N^0} doublet       &  231.981 & 235.726  \\
                             &          &          \\
    \ce{C3N-} singlet        &   72.856 &  75.820  \\
    \ce{C4N-} triplet        &  133.775 & 137.288  \\
    \ce{C5N-} singlet        &  123.892 & 127.992  \\
    \hline
  \end{tabular}
\end{table}

A similar conclusion emerges from the inspection of the lower part of Table~\ref{table:enthalpies},
where the values for the most stable anionic species are presented.
Because the values for the \ce{C4N-} (triplet) chains are smaller than those for the longer (singlet)
\ce{C5N-} chains --- which were
already reported in space \cite{Agundez:10} --- one can still hope that they are observable. 
The results for enthalpies of formation (Table~\ref{table:enthalpies} and \figname\ref{fig:H}b) 
indicate again that
\ce{C4N-} triplet chains are more stable than \ce{C4N-} singlet chains, in contrast to \ce{C3N-}
and \ce{C5N-}, for which singlets are more stable than triplets, in agreement with the
odd-even singlet-triplet
alternation reported earlier for the \ce{C_nN-} homologous series \cite{Pascoli:99} as well as for
its \ce{HC_nN} isoelectronic counterpart \cite{Baldea:2019e,Baldea:2019g}.

So, the above results suggest that neutral \ce{C4N^0} chains can still be observed in space; the values for  
\ce{C4N} chains are smaller than for the longer \ce{C5N} chains already astronomically detected
\cite{Guelin:98}.
This result is supported by the fact that 
in laboratory molecular beam \ce{C4N} was indeed produced more abundantly than \ce{C5N} \cite{McCarthy:03}.
We next checked that 
neutral \ce{C4N} chains are stable against dissociation.
Our calculations indicated that all possible 
dissociation processes (reactions 1 to 4 in Tables~\ref{table:dissociation-c4n} and 
S28) 
are endoenergetic.  The same holds true for anion's dissociation;
reactions 5 to 8 in the same Tables~\ref{table:dissociation-c4n} and 
S28 
are also endoenergetic.
In particular, this rules out a(n a priori conceivable) fragmentation of the not yet detected
\ce{C4N^0} and \ce{C4N-} into already astronomically observed \ce{C3N^0} and \ce{C3N-} or \ce{C2N^0}.

\clearpage

\begin{table*}
  \centering
  \caption{Dissociation of neutral and anion \ce{C4N} chains. Enthalpies of reaction at zero (subscript 0)
    and room temperature (subscript RT) computed by using CBS-QB3 protocol. All values (in kcal/mol)
    refer to the electronic ground states.}
  \label{table:dissociation-c4n}
  \begin{tabular}{lclclclrr}
    \hline
        No. & Species  &               & Reaction&              &       &                  &  $\Delta_{r} H_{0}^{0}$ & $\Delta_{r} H_{RT}^{0}$ \\
    \hline
         1  &   \ce{C4N} & $ \ce{C4N} $ & $ \to $ & $ \ce{C} $ & $ + $ & $ \ce{C3N}     $   & 139.4 & 140.1 \\
         2  &            & $ \ce{C4N} $ & $ \to $ & $ \ce{C2}$ & $ + $ & $ \ce{C2N}    $   & 152.0 & 152.8 \\
         3  &            & $ \ce{C4N} $ & $ \to $ & $ \ce{C3}$ & $ + $ & $ \ce{CN}     $   &  95.3 &  96.4 \\
         4  &            & $ \ce{C4N} $ & $ \to $ & $ \ce{C4}$ & $ + $ & $ \ce{N}      $   & 159.6 & 160.5 \\
    \hline
         5a &  \ce{C4N-} & $ \ce{C4N-}$ & $ \to $ & $ \ce{C} $ & $ + $ & $ \ce{C3N-}   $   & 109.1 & 109.4 \\
         5b &            & $  \ce{C4N-} $ & $ \to $ & $\ce{C-}$ & $ + $ & $ \ce{C3N}  $   & 184.4 & 185.4 \\[1ex]
         6a &            & $ \ce{C4N-} $ & $ \to $ & $  \ce{C2} $ & $ + $ & $ \ce{C2N-} $  & 160.6 & 161.6 \\
         6b &            & $ \ce{C4N-} $ & $ \to $ & $  \ce{C2-}$ & $ + $ & $ \ce{C2N} $   & 151.0 & 152.1 \\[1ex]
         7a &            & $ \ce{C4N-} $ & $ \to $ & $ \ce{C3} $ & $ + $ & $ \ce{CN-}   $   &  77.6 &  79.0 \\
         7b &            & $ \ce{C4N-} $ & $ \to $ & $ \ce{C3-} $ & $ + $ & $ \ce{CN}   $   & 122.0 & 123.3 \\[1ex]
         8a &            & $ \ce{C4N-} $ & $ \to $ & $ \ce{C4} $ & $ + $ & $ \ce{N-}    $   & 238.1 & 239.4 \\
         8b &            & $ \ce{C4N-} $ & $ \to $ & $ \ce{C4-}$ & $ + $ & $ \ce{N}    $   & 141.8 & 142.8 \\
    \hline
  \end{tabular}
\end{table*}

Furthermore, our values for bond dissociation enthalpies do not support claims that \ce{C4N} is less stable than
the already detected \ce{C2N}, \ce{C3N} and \ce{C5N}. For illustration, let us consider the bond breaking at molecular ends.
The values presented in 
Tables~\ref{table:dissociation-c4n}, 
S28, 
\ref{table:dissociation-c3n-c5n},
and 
S29 
reveal that breaking the end \ce{C\bond{3}N} bond in \ce{C4N} requires a significantly
larger energy than in all (\ce{C2N}, \ce{C3N} and \ce{C5N}) already astronomically detected members of the 
homologous series \ce{C_nN} \cite{Anderson:14,Friberg:80,Guelin:98}.
Switching to the opposite molecular end, removing the terminal C atom of \ce{C4N^0} requires an energy 
larger than for \ce{C2N^0} and comparable to that for \ce{C5N^0} 
(cf.~reactions 9a and 12a in Tables~\ref{table:dissociation-c3n-c5n} and 
S29), 
respectively.
The same conclusion emerges from the direct comparison  
of the structural and chemical properties of \ce{C3N}, \ce{C4N}, and \ce{C5N}
presented in detail in  
\figsname S1, 
S2, 
S3, 
and
S4. 

\begin{table*}
  \centering
  \caption{Dissociation of neutral \ce{C2N}, \ce{C3N}, and \ce{C5N} chains already detected in space. 
    Enthalpies of reaction at zero (subscript 0)
    and room temperature (subscript RT) computed by using CBS-QB3 protocol. All values (in kcal/mol)
    refer to the electronic ground states.}
  \label{table:dissociation-c3n-c5n}
  \begin{tabular}{rclclclrr}
    \hline
        No. & Species  &               & Reaction&              &       &                  &  $\Delta_{r} H_{0}^{0}$ & $\Delta_{r} H_{RT}^{0}$ \\
    \hline
         9a  & \ce{C2N} & $ \ce{C2N} $ & $ \to $ & $  \ce{C} $ & $ + $ & $ \ce{CN}     $   & 113.4 & 114.4 \\
         9b  &          & $ \ce{C2N} $ & $ \to $ & $  \ce{C2}$ & $ + $ & $ \ce{N}     $   & 145.8 & 146.8 \\[1ex]
        10a  & \ce{C3N} & $ \ce{C3N} $ & $ \to $ & $  \ce{C} $ & $ + $ & $ \ce{C2N}    $   & 156.8 & 157.9 \\
        10b  &          & $ \ce{C3N} $ & $ \to $ & $  \ce{C2}$ & $ + $ & $ \ce{CN}    $   & 126.0 & 127.1 \\
        10c  &          & $ \ce{C3N} $ & $ \to $ & $  \ce{C3}$ & $ + $ & $ \ce{N}     $   & 132.6 & 134.0 \\[1ex]
        12a  & \ce{C5N} & $ \ce{C5N} $ & $ \to $ & $   \ce{C}$ & $ + $ & $ \ce{C4N}   $   & 144.1 & 145.7 \\
        12b  &          & $ \ce{C5N} $ & $ \to $ & $  \ce{C2}$ & $ + $ & $ \ce{C3N}   $   & 139.2 & 140.6 \\
        12c  &          & $ \ce{C5N} $ & $ \to $ & $  \ce{C3}$ & $ + $ & $ \ce{C2N}   $   & 126.0 & 127.8 \\
        12d  &          & $ \ce{C5N} $ & $ \to $ & $  \ce{C4}$ & $ + $ & $ \ce{CN}   $   &  126.9 & 128.6 \\
        12e  &          & $  \ce{C5N}$ & $ \to $ & $   \ce{C5}$ & $ + $ & $ \ce{N}   $   & 135.8 & 137.3 \\
     \hline
  \end{tabular}
\end{table*}

To sum up, the foregoing analysis indicates that \ce{C4N} nonobservability in space cannot be due 
to molecule's fragmentation; none of the above processs involves an unusually low dissociation energy. 

Putting differently, the stability against dissociation says that
all converse (association) reactions depicted in Tables~\ref{table:dissociation-c4n} and 
S28 
are exoenergetic. That is, they represent potential chemical pathways of \ce{C4N^0}/\ce{C4N-} formation
from precursors already reported in space
(\ce{CN},\cite{Jefferts:70} \ce{C2N},\cite{Anderson:14} \ce{C3N},\cite{Friberg:80} 
\ce{C2},\cite{Souza:77} \ce{C3} \cite{Hinkle:88,Cernicharo:00}).

In addition, a series of exchange reactions (wherein parents molecules 
\ce{C5},\cite{Hinkle:88,Cernicharo:00} \ce{CH},\cite{Swings:37} \ce{C2H},\cite{Tucker:74}
\ce{C2H-},\cite{Agundez:10} \ce{C3H},\cite{Thaddeus:85a} \ce{C4H},\cite{Guelin:78} and \ce{C4H-} \cite{Cernicharo:07}
were also astronomically observed) presented in
Tables~\ref{table:exchange},
S30, 
S31, 
and
S32 
are chemical pathways of production that come into question.
Along with exoenergetic exchange reactions (which may also be problematic 
without third party energy removal), we also included there several endoenergetic processes 
(e.g.,~reactions 13, 14a, 15e, 17d, and 17e) corresponding to small or moderate (positive) reaction 
energies; they are significant for reactants in electronic excited states.

\begin{table*}
  \centering
  \caption{Relevant exchange reactions.
    Enthalpies of reaction at zero (subscript 0)
    and room temperature (subscript RT) computed by using CBS-QB3 protocol. All values (in kcal/mol)
    refer to the electronic ground states.}
  \label{table:exchange}
  \begin{tabular}{llclclrlrr}
    \hline
        No. &  &          &           & Reaction  &       &        &          &  $\Delta_{r} H_{0}^{0}$ & $\Delta_{r} H_{RT}^{0}$ \\
    \hline
        13   & $ \ce{C5}$ & $ + $ & $  \ce{N} $  & $ \to $ & $  \ce{C}$ & $ + $ & $ \ce{C4N}        $   &    8.3 &    8.4 \\[1ex]
        14a  & $ \ce{N} $ & $ + $ & $  \ce{C4H-}$ & $ \to $ & $  \ce{C4N}$ & $ + $ & $  \ce{H-}   $   &   23.9 &   24.3 \\
        14b  & $ \ce{N} $ & $ + $ & $  \ce{C4H-} $ & $ \to $ & $  \ce{C4N-}$ & $ + $ & $  \ce{H}   $   &  -36.0 &  -35.9 \\ 
        14c  & $ \ce{N-}$ & $ + $ & $  \ce{C4H} $ & $ \to $ & $  \ce{C4N-}$ & $ + $ & $  \ce{H}   $   & -125.0 & -124.8 \\  
        14d  & $ \ce{N-}$ & $ + $ & $  \ce{C4H} $ & $ \to $ & $  \ce{C4N}$ & $ + $ & $  \ce{H-}   $   &  -65.1 &  -64.6 \\[1ex]
        15a  & $ \ce{CN}$ & $ + $ & $  \ce{C3H} $ & $ \to $ & $  \ce{H}$ & $ + $ & $  \ce{C4N}    $   &  -20.5 &  -20.4 \\
        15b  & $ \ce{CN-}$ & $ + $ & $  \ce{C3H}$ & $ \to $ & $  \ce{H}$ & $ + $ & $  \ce{C4N-}  $   &   -2.8 &   -3.0 \\ 
        15c  & $ \ce{CN-}$ & $ + $ & $  \ce{C3H}$ & $ \to $ & $  \ce{H-}$ & $ + $ & $  \ce{C4N}  $   &   57.1 &  57.2 \\
        15d  & $ \ce{CN} $ & $ + $ & $  \ce{C3H-}$ & $ \to $ & $  \ce{H}$ & $ + $ & $  \ce{C4N-}  $   &  -50.9 &  -50.8 \\
        15e  & $ \ce{CN} $ & $ + $ & $  \ce{C3H-}$ & $ \to $ & $  \ce{H-}$ & $ + $ & $  \ce{C4N}  $   &    9.0 &    9.4 \\[1ex]  
        16a  & $ \ce{CH} $ & $ + $ & $  \ce{C3N} $ & $ \to $ & $  \ce{H}$ & $ + $ & $  \ce{C4N}   $   &  -59.4 &  -59.2 \\
        16b  & $ \ce{CH-}$ & $ + $ & $  \ce{C3N}$ & $ \to $ & $  \ce{H}$ & $ + $ & $  \ce{C4N-}  $   & -105.6 & -105.8 \\
        16c  & $ \ce{CH-}$ & $ + $ & $  \ce{C3N}$ & $ \to $ & $  \ce{H-}$ & $ + $ & $  \ce{C4N}  $   &  -45.8 &  -45.5 \\
        16d  & $ \ce{CH} $ & $ + $ & $  \ce{C3N-}$ & $ \to $ & $  \ce{H}$ & $ + $ & $  \ce{C4N-}  $   &  -29.1 &  -29.1 \\[1ex]
        17a  & $ \ce{CH} $ & $ + $ & $  \ce{C3N-}$ & $ \to $ & $  \ce{H-}$ & $ + $ & $  \ce{C4N}  $   &   30.8 &  31.1 \\
        17b  & $ \ce{C2H}$ & $ + $ & $  \ce{C2N}$ & $ \to $ & $  \ce{H} $ & $ + $ & $  \ce{C4N}   $    & -40.5 &  -40.3  \\
        17c  & $ \ce{C2H-}$ & $ + $ & $  \ce{C2N}$ & $ \to $ & $ \ce{H} $ & $ + $ & $  \ce{C4N-} $   &  -44.6 &  -44.5 \\
        17d  & $ \ce{C2H-}$ & $ + $ & $  \ce{C2N}$ & $ \to $ & $  \ce{H-}$ & $ + $ & $  \ce{C4N} $   &   15.3  &  15.7 \\
        17e  & $ \ce{C2H} $ & $ + $ & $  \ce{C2N-}$ & $ \to $ & $  \ce{H-}$ & $ + $ & $  \ce{C4N} $   &   10.8  &  11.2 \\
        17f  & $ \ce{C2H} $ & $ + $ & $  \ce{C2N-}$ & $ \to $ & $  \ce{H} $ & $ + $ & $  \ce{C4N-} $   & -49.1  &  -49.0 \\[1ex] 
        18   & $ \ce{NC2N} $ & $ + $ & $   \ce{C2}  $ & $ \to $ & $  \ce{N} $ & $ + $ & $  \ce{C4N}  $   & 48.6 & 49.0 \\[1ex]
        19   & $ \ce{NC2N} $ & $ + $ & $   \ce{C2N}  $ & $ \to $ & $  \ce{N2} $ & $ + $ & $  \ce{C4N}  $   & -29.8 & -29.3 \\[1ex]
        20   & $ \ce{NC2N} $ & $ + $ & $   \ce{C2H}  $ & $ \to $ & $  \ce{NH} $ & $ + $ & $  \ce{C4N}  $   & 82.3 & 82.7 \\
     \hline
  \end{tabular}
\end{table*}
\subsection{Technical Remark on the Complete Basis Set (CBS) Approaches}
\label{sec:cbs-qb3}
To compute enthalpies of formation and reaction energies 
(cf.~{\secname}``Chemical Stability and Potential Chemical Pathways of Formation''), 
GAUSSIAN 16 allows choosing between several complete basis set protocols:
CBS-4M, CBS-APNO, and CBS-QB3.\cite{Ochterski:96,Montgomery:99,Montgomery:00,Ochterski:00}
The first is recommended for new studies.\cite{g16}
In some cases, the second attains a root mean square deviation
$RMSD_{CBS-APNO} = 1.16$\,kcal/mol, which is better than
$RMSD_{CBS-QB3} = 2.27$\,kcal/mol.\cite{Karton:09}
One may therefore wonder why we have presented in the main text 
numerical values obtained via the ``less'' performant 
CBS-QB3 protocol, and put CBS-4M- and CBS-APNO-based estimates in the SI.

To justify this preference, CBS-based values of the adiabatic electron affinity 
$EA_{CBS}$ 
for all even-members (\ce{C2N}, \ce{C4N}, and \ce{C6N}) of astrophysical interest
are shown in Table~\ref{table:ea_cbs-c2n-c4n-c6n}. 
In these calculations, $EA_{CBS}$ was estimated as an energy of reaction 
($\ce{C_n N-} \xlongrightarrow[\text{ }]{\text{$ EA \equiv \Delta_r H_{0}^0  $}} \ce{C_n N^0} + e^{-}$).
For completeness, adiabatic ionization potentials $IP_{CBS}$ estimated in a similar manner 
($\ce{C_n N^0} \xlongrightarrow[\text{ }]{\text{$ IP \equiv \Delta_r H_{0}^0  $}} \ce{C_n N+} + e^{-}$) 
are also shown (cf.~Table~S33). 

We focused on the adiabatic electron affinity because it is the only quantity 
that can be compared with high accuracy SEVI experimental data.\cite{Neumark:09}
The comparison reveals that for this quantity --- which is notoriously problematic even 
for small normal (nonradical) carbon-based chains \cite{Sommerfeld:05} ---
the CBS-QB3 protocol attains the best agreement with experiment.\label{page:apno}

One should still aid that values of the energy of reactions 
computed by means of CBS-APNO and CBS-4M 
(cf.~Tables~S28, 
S29, 
S30, 
S31, 
and
S32) 
do not notably differ from those based on CBS-QB3.
Most importantly, they do by no means change the above 
conclusions on \ce{C4N}'s stability and formation pathways.

\begin{table}
  \centering
  \caption{Adiabatic electron affinities of \ce{C2N}, \ce{C4N} and \ce{C6N} radicals measured in 
           high-resolution SEVI experiments \cite{Neumark:09} 
           and computed using several CBS protocols.\cite{g16} All values are in eV.
           Notice that the CBS-QB3 estimates are the closest to experiment. The root mean square deviations
           are RMSD$_{CBS-QB3}=0.862$\,kcal/mol, RMSD$_{CBS-APNO}=1.841$\,kcal/mol, and RMSD$_{CBS-4M}=4.538$\,kcal/mol.} 
  \label{table:ea_cbs-c2n-c4n-c6n}
  \begin{tabular}{lllll}
    \hline
  Method                    &    &      \ce{C2N}           &      \ce{C4N}           &              \ce{C6N}           \\
    \hline
  Experiment  & EA\,(eV) & $ 2.7489 \pm 0.0010 $   &  $ 3.1113 \pm 0.0010$   & $ 3.3715 \pm 0.0010 $  \\
  CBS-QB3         &         EA\,(eV)        &        2.7615           &         3.1351          &              3.4804   \\
                  & $\delta\,$EA\,(kcal/mol) &        0.291            &         0.549           &              2.511   \\
  CBS-APNO        &         EA\,(eV)        &        2.7728           &         3.2506          &               3.5648  \\
                  & $\delta\,$EA\,(kcal/mol) &        0.551            &         3.212           &               4.458  \\
  CBS-4M          &          EA\,(eV)        &        3.0115           &         3.4596          &               3.7693 \\
                  & $\delta\,$EA\,(kcal/mol) &        6.056            &         8.032           &               9.173  \\
    \hline
  \end{tabular}
\end{table}

\section{Conclusions}  

In closing, our results for electronic structure, chemical bonding, 
and chemical stability \label{page:enthalpies} do not substantiate claims
(made explicitly or implicitly in previous literature on similar molecular species)
that even \ce{C_nN} chains are less stable than odd members to justify why the
former were not detected in space; \figname\ref{fig:H} shows that the point for the neutral \ce{C4N}
chain lies exactly on the line joining the points for neutral \ce{C3N} and neutral \ce{C5N} chains
which were already astronomically observed.

The present investigation demonstrates that whether neutral or charged, all \ce{C4N} chains
possess strongly delocalized structures. In particular, in spite of the significantly
different atomic ionization energies ($IP_C = 11.26$\,eV versus $IP_N = 14.53$\,eV)
electron removal also affects the charge of the nitrogen atom
(\figsname\ref{fig:lengths_bonds_c4n_anion}d and
S7d). 
Furthermore, the excess electron attached to a neutral chain does not preponderantly go to the N atom
(\figsname\ref{fig:lengths_bonds_c4n_cation}d and
S8d), 
although this element is more electronegative than the C atoms
($\chi_{N}^{Pauling} = 3.04$ versus $\chi_{C}^{Pauling} = 2.55$),
Interestingly, it is the same group of atoms (\ce{C1}, \ce{C3} and \ce{N}) 
that shares more or less democratically
both the hole created by ionization and the excess electron attached to the neutral chain.

The fact that, for all even-members (\ce{C2N}, \ce{C4N}, \ce{C6N}) of astrophysical interest,
the present theoretical estimate for electron affinity $EA$ excellently
agrees with experiment \cite{Neumark:09} is another significant report. 
$EA$-data for molecular species of astrochemical interest are very scarce;
even values for small ``normal'' (i.e.,~nonradical) cyanopolyynes
\cite{Vuitton:09} continue to be missing.
Even for small(er) carbon chains, accurate $EA$-estimates are very challenging
for theory \cite{Sommerfeld:05}.

Last but not least,
our results suggest that astronomical detection should first focus on anion \ce{C4N-} chains rather
than on neutral \ce{C4N^0} chains. Letting alone their substantially lower enthalpy of formation
(cf.~Table~\ref{table:enthalpies} and \figname\ref{fig:H}),
at comparable abundances,  \ce{C4N-} anions should be much easier detectable than neutrals  
via rotational transition spectroscopy.
More quantitatively, in view of the different dipole momenta presently estimated (Table~\ref{table:D}),  
\ce{C4N-} anions transition intensities should be about hundred times stronger 
than for neutral chains. Distinguishing between neutral and anion \ce{C4N} chains
also appears to be feasible; the difference between the estimated rotational 
constants (Table~\ref{table:B}) exceeds by far the experimental resolution
currently achieved. 
More specific astrophysical details are presented separately.\cite{Baldea:2020c}
\section*{Acknowledgments}
Financial support for this research provided by
the Deu\-tsche For\-schungs\-ge\-mein\-schaft (DFG grant BA 1799/3-2),
and computational support from the State of Baden-W\"urttemberg through bwHPC/DFG
through grant INST 40/467-1 FUGG are gratefully acknowledged.
\providecommand{\latin}[1]{#1}
\makeatletter
\providecommand{\doi}
  {\begingroup\let\do\@makeother\dospecials
  \catcode`\{=1 \catcode`\}=2 \doi@aux}
\providecommand{\doi@aux}[1]{\endgroup\texttt{#1}}
\makeatother
\providecommand*\mcitethebibliography{\thebibliography}
\csname @ifundefined\endcsname{endmcitethebibliography}
  {\let\endmcitethebibliography\endthebibliography}{}

\renewcommand{\theequation}{S\arabic{equation}}
\setcounter{equation}{0}
\renewcommand{\thefigure}{S\arabic{figure}}
\setcounter{figure}{0}
\renewcommand{\thetable}{S\arabic{table}}
\setcounter{table}{0}
\renewcommand{\thesection}{S\arabic{section}}
\setcounter{section}{0}
\renewcommand{\thepage}{S\arabic{page}}
\setcounter{page}{1}
\renewcommand{\thefootnote}{\alph{footnote}}

\section{Appendix}
This appendix presents additional theoretical and computational details, additional tables and figures.

\begin{figure*}
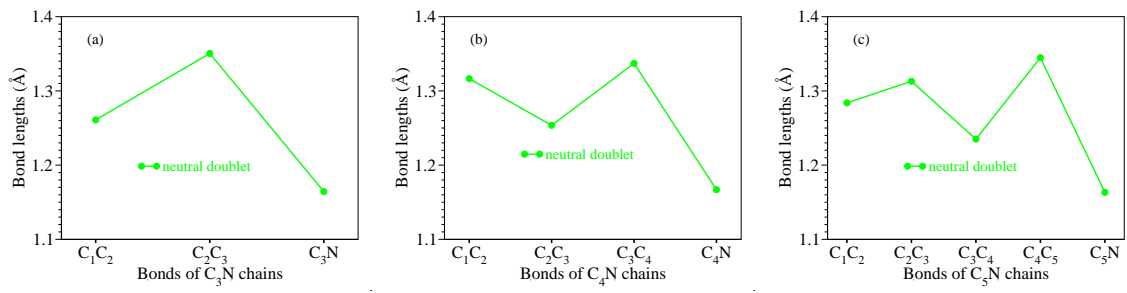

  \centerline{
    \includegraphics[width=0.3\textwidth]{fig_bond_lengths_c3n_neutral_b3lyp_MaxPople.eps}
    \includegraphics[width=0.3\textwidth]{fig_bond_lengths_c4n_neutral_b3lyp_MaxPople.eps}
    \includegraphics[width=0.3\textwidth]{fig_bond_lengths_c5n_neutral_b3lyp_MaxPople.eps}
    }
  \caption{Bond lengths (in angstrom) of neutral \ce{C4N}, \ce{C4N}, and \ce{C5N} chains in their electronic ground state.}
    \label{fig:lengths_CnN_only_neutral}
\end{figure*}

\begin{figure*}
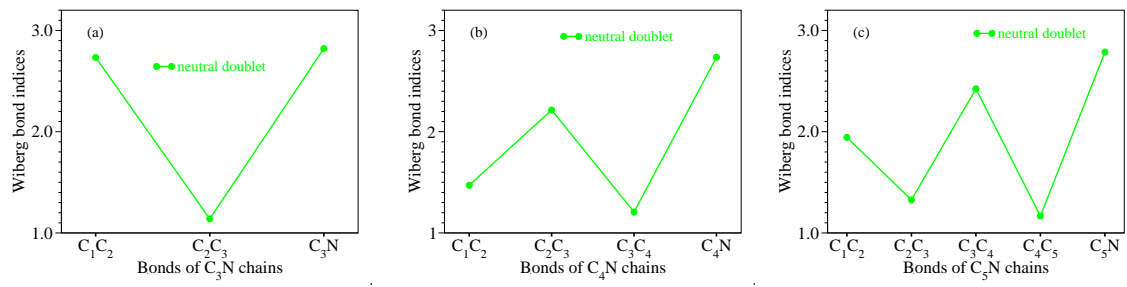

  \centerline{
    \includegraphics[width=0.3\textwidth]{fig_wiberg_indices_c3n_neutral_roccsd_t_augccpvtz.eps}
    \includegraphics[width=0.3\textwidth]{fig_wiberg_indices_c4n_neutral_roccsd_t_augccpvtz.eps}
    \includegraphics[width=0.3\textwidth]{fig_wiberg_indices_c5n_neutral_roccsd_t_augccpvtz.eps}
  }
\caption{Wiberg bond indices of neutral \ce{C4N}, \ce{C4N}, and \ce{C5N} chains in their electronic ground state.}
    \label{fig:wbi_CnN_only_neutral}
\end{figure*}

\begin{figure*}
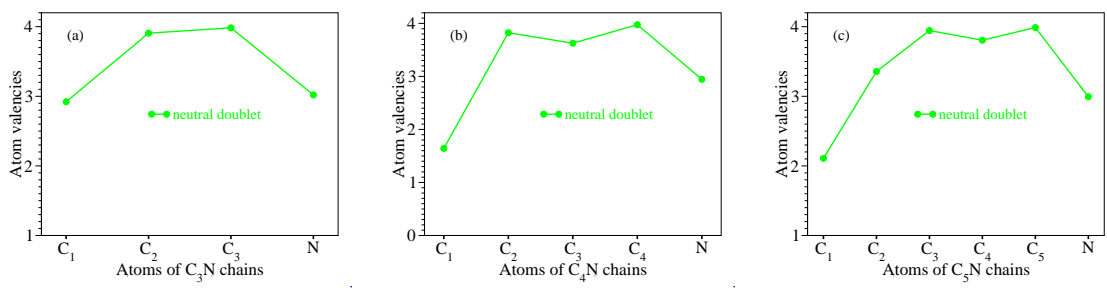

  \centerline{
    \includegraphics[width=0.3\textwidth]{fig_valencies_c3n_neutral_roccsd_t_augccpvtz.eps}
    \includegraphics[width=0.3\textwidth]{fig_valencies_c4n_neutral_roccsd_t_augccpvtz.eps}
    \includegraphics[width=0.3\textwidth]{fig_valencies_c5n_neutral_roccsd_t_augccpvtz.eps}
  }
 \caption{Wiberg valencies of neutral \ce{C4N}, \ce{C4N}, and \ce{C5N} chains in their electronic ground state.}
    \label{fig:valency_CnN_only_neutral}
\end{figure*}

\begin{figure*}
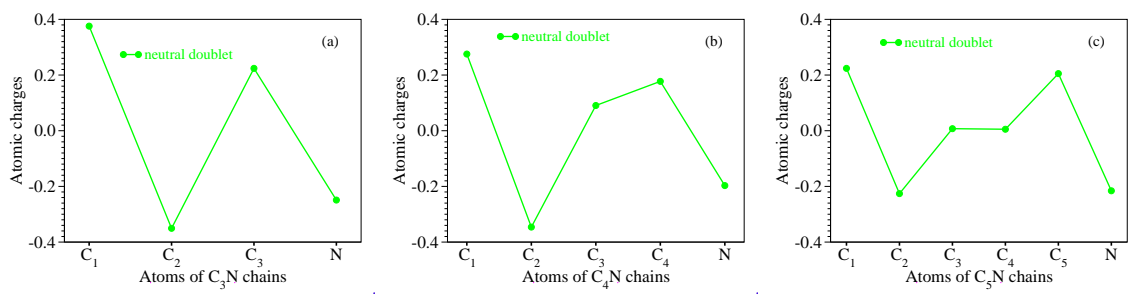

  \centerline{
    \includegraphics[width=0.3\textwidth]{fig_charges_c3n_neutral_roccsd_t_augccpvtz.eps}
    \includegraphics[width=0.3\textwidth]{fig_charges_c4n_neutral_roccsd_t_augccpvtz.eps}
    \includegraphics[width=0.3\textwidth]{fig_charges_c5n_neutral_roccsd_t_augccpvtz.eps}
}
    \caption{Atomic charges of neutral \ce{C4N}, \ce{C4N}, and \ce{C5N} chains in their electronic ground state.}
    \label{fig:charges_CnN_only_neutral}
\end{figure*}

\begin{figure*}
  \centerline{
    \includegraphics[width=0.3\textwidth]{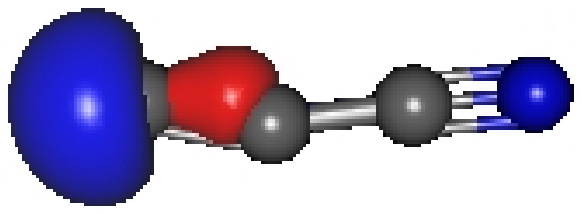}
    \includegraphics[width=0.3\textwidth]{c4n_doublet_rohf-homo-nondeg-alpha.eps}
    \includegraphics[width=0.3\textwidth]{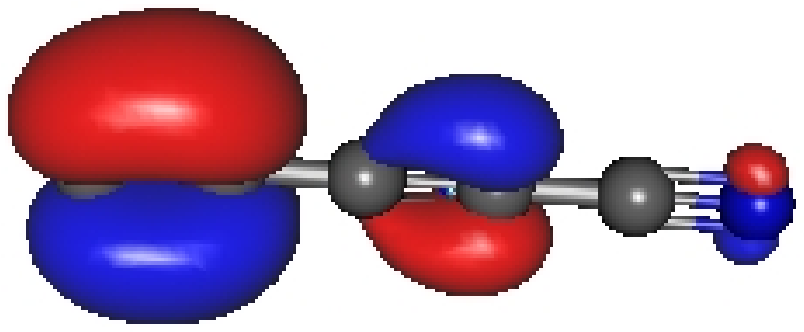}
   }
  \caption{HOMO of neutral doublet \ce{C3N^0}, \ce{C4N^0}, and \ce{C5N^0} chains.}
    \label{fig:homo-cxn-doublet}
\end{figure*}

\begin{figure*}
  \centerline{
    \includegraphics[width=0.3\textwidth]{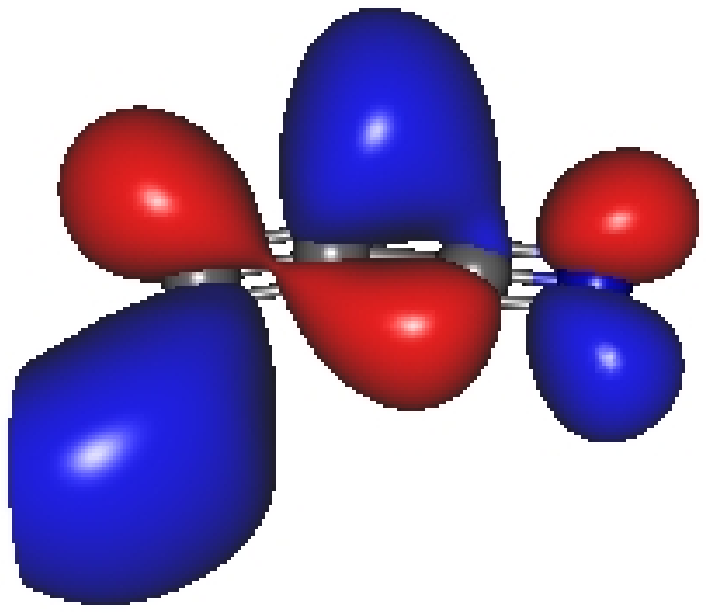}\
    \includegraphics[width=0.3\textwidth]{c4n_doublet_rohf-lumo-nondeg-alpha.eps}
    \includegraphics[width=0.3\textwidth]{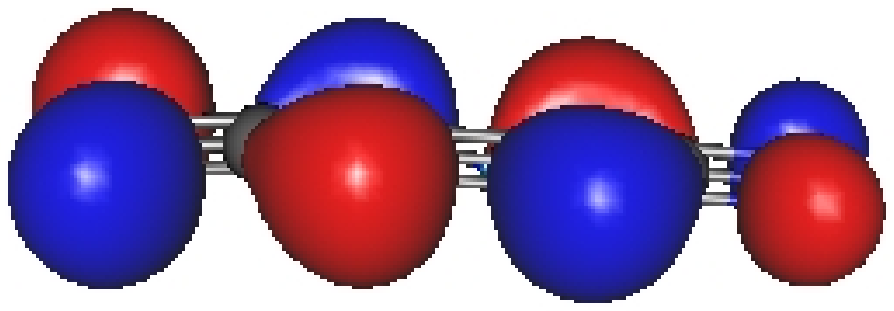}
  }
  \caption{LUMO of neutral doublet \ce{C3N^0}, \ce{C4N^0}, and \ce{C5N^0} chains.}
    \label{fig:lumo-cxn-doublet}
\end{figure*}

\begin{figure*}
  \includegraphics[width=0.4\textwidth]{fig_bond_length_differences_c4n_neutral_anion_b3lyp_MaxPople.eps}
  \includegraphics[width=0.4\textwidth]{fig_wiberg_index_differences_c4n_neutral_anion_roccsd_t_augccpvtz.eps}
  \includegraphics[width=0.4\textwidth]{fig_valency_differencies_c4n_neutral_anion_roccsd_t_augccpvtz.eps}
  \includegraphics[width=0.4\textwidth]{fig_charge_differences_c4n_neutral_anion_roccsd_t_augccpvtz.eps}
  \caption{Changes with respect to the neutral doublet \ce{C4N^0} of several molecular properties:
    (a) bond lengths (in angstrom),
    (b) Wiberg bond order indices, (c) Wiberg valencies and (d) atomic charges.}
    \label{fig:differences_lengths_bonds_c4n_anion}
\end{figure*}

\begin{figure*}
  \includegraphics[width=0.4\textwidth]{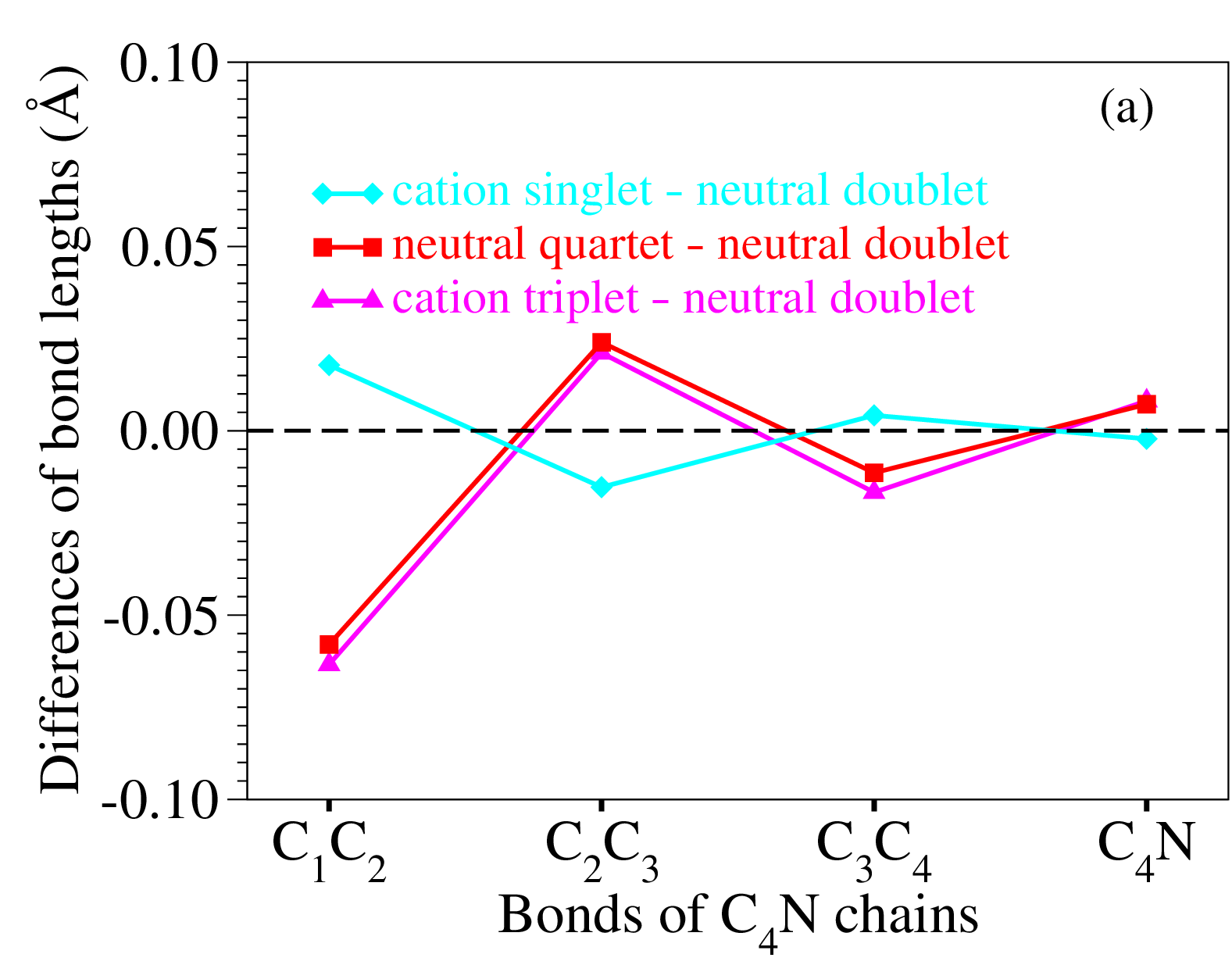}
  \includegraphics[width=0.4\textwidth]{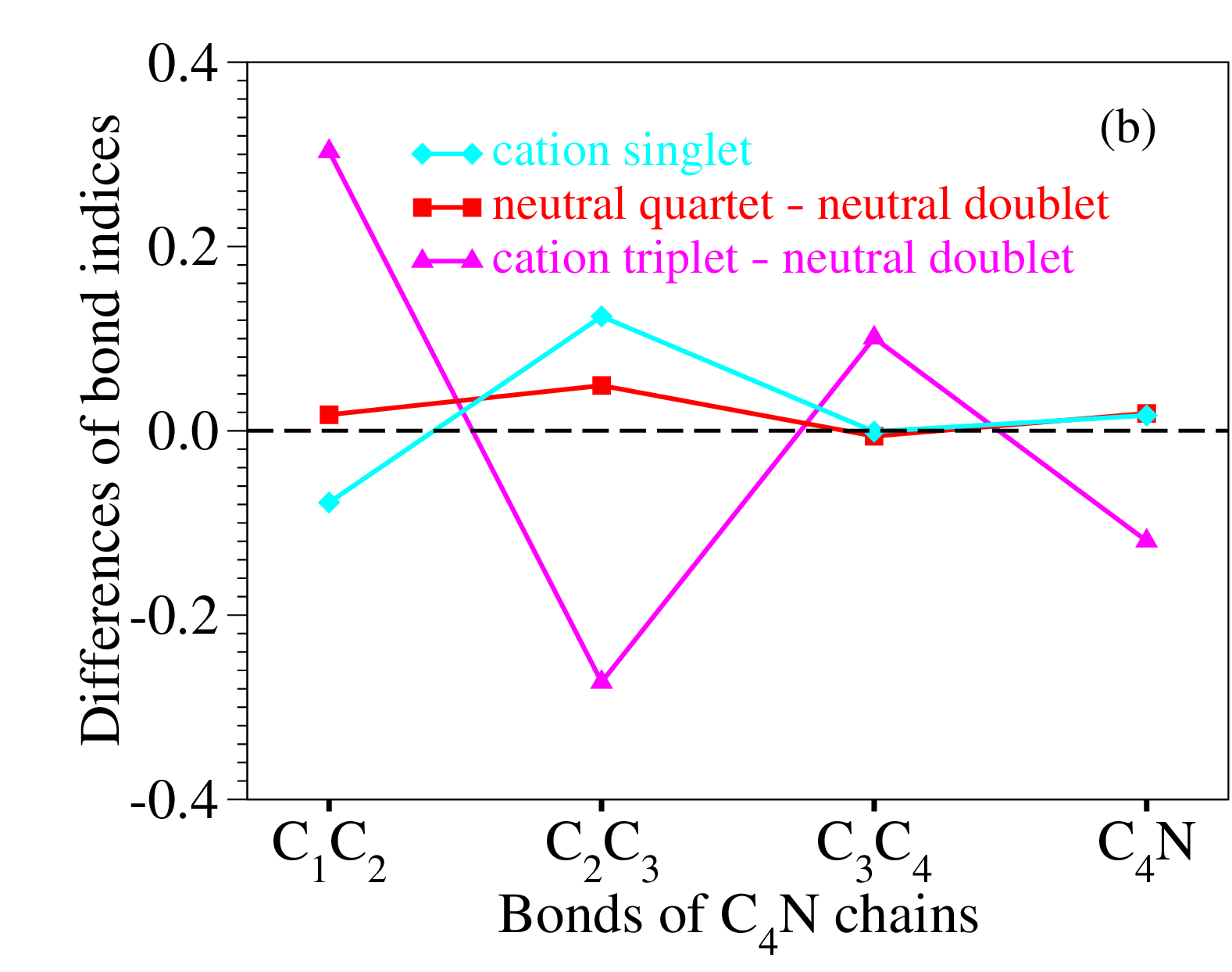}
  \includegraphics[width=0.4\textwidth]{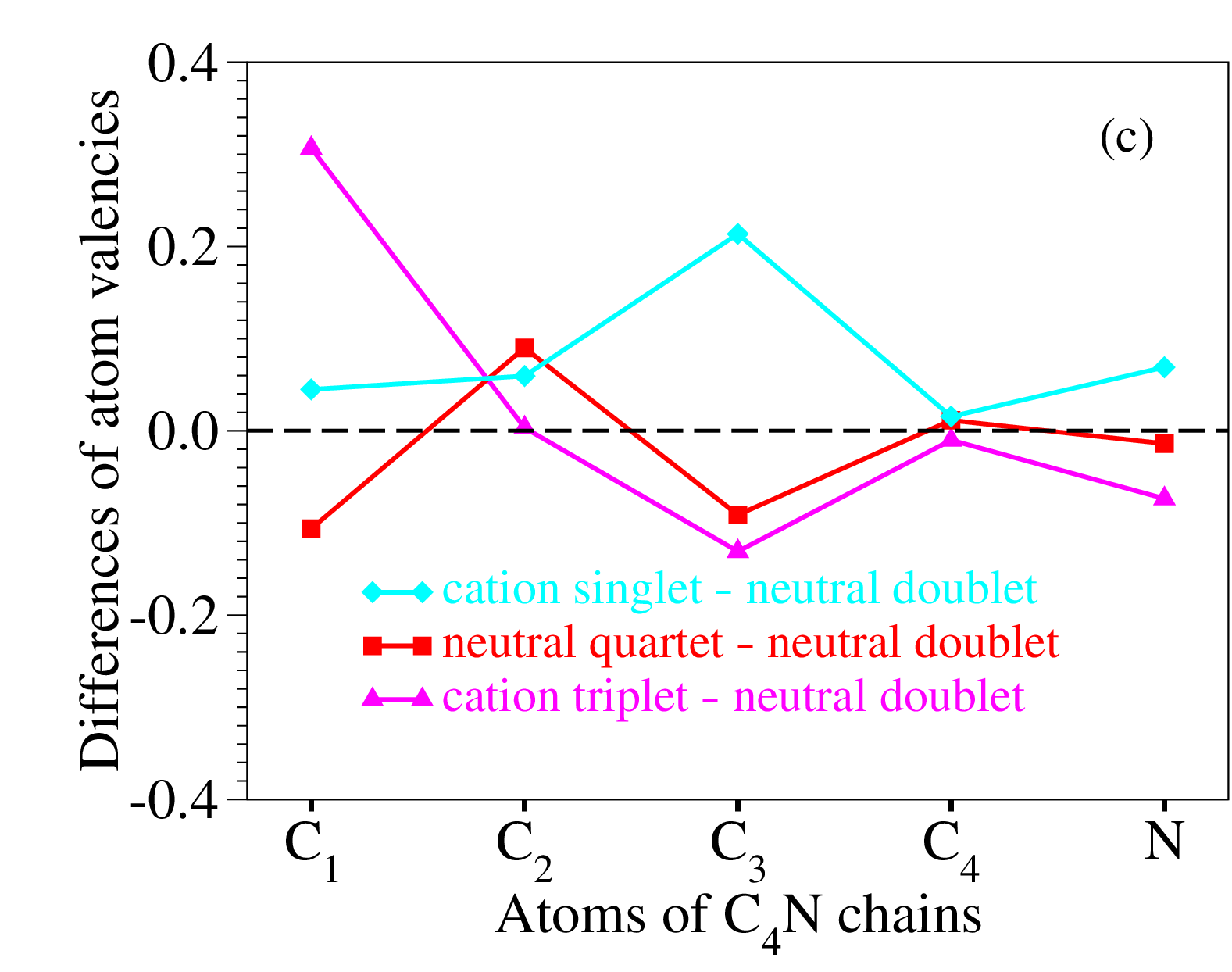}
  \includegraphics[width=0.4\textwidth]{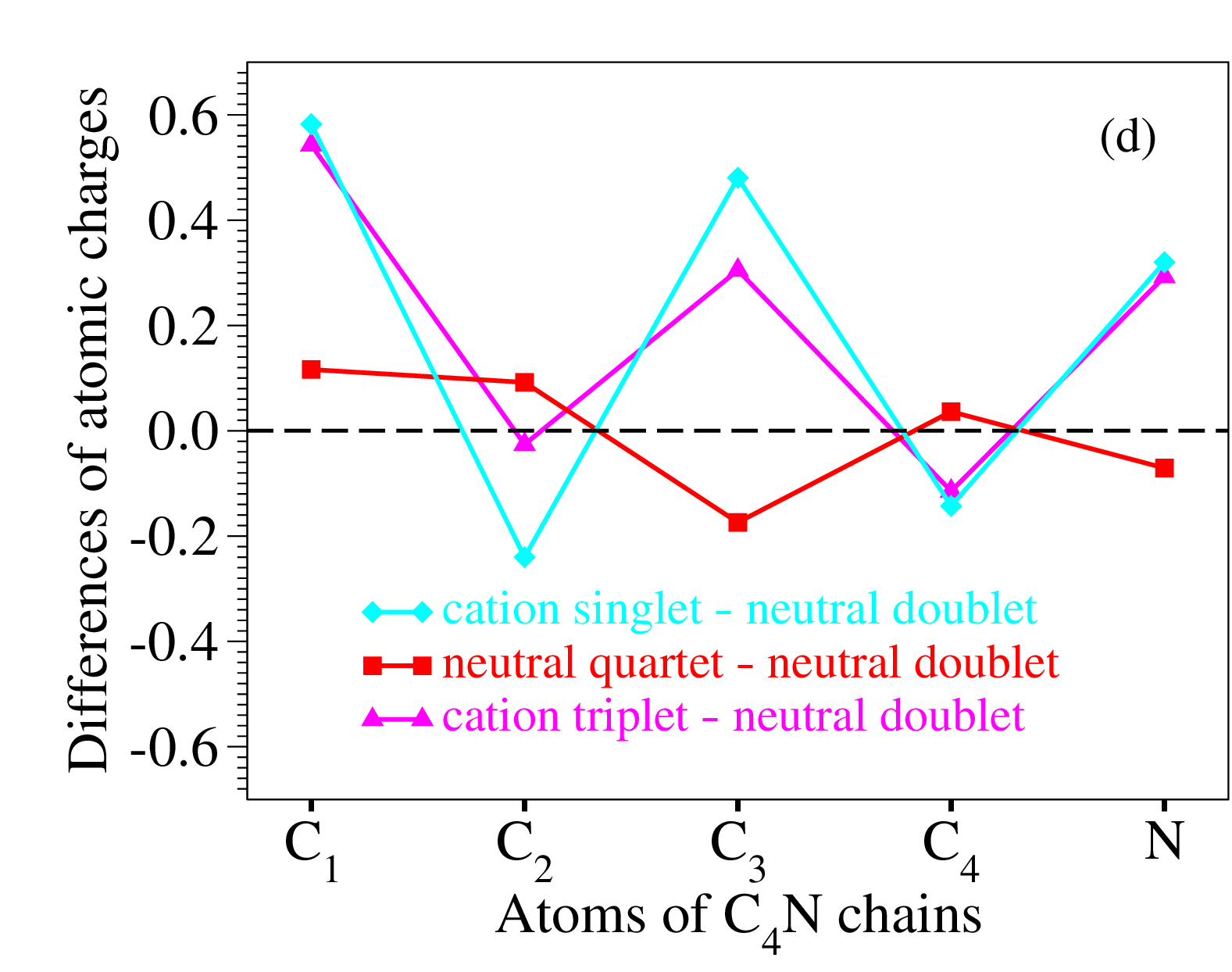}
  \caption{Changes with respect to the neutral doublet \ce{C4N^0} of several molecular properties:
    (a) bond lengths (in angstrom), (b) Wiberg bond order indices, (c) Wiberg valencies and (d) atomic charges.}
    \label{fig:differences_lengths_bonds_c4n_cation}
\end{figure*}

\begin{table*}
  \centering
  \caption{Cartesian coordinates (in angstrom), natural charges and Wiberg valencies of the atoms of the
    (most stable)
    \ce{C4N^0} neutral doublet ($\tilde{X} ^2 \Pi$).}
  \label{table:geom-doublet}
  \begin{tabular}{lccccc}
    \hline
 Atom  &        X     &      Y         &    Z        &   charge  & valence \\
\hline
 \ce{C1}    &   0.000000  &   0.000000   &    2.65410363   &   0.27537 &   1.6424 \\
 \ce{C2}    &   0.000000  &   0.000000   &    1.33769693   &  -0.34620 &   3.8239 \\
 \ce{C3}    &   0.000000  &   0.000000   &    0.08416284   &   0.09058 &   3.6263 \\
 \ce{C4}    &   0.000000  &   0.000000   &   -1.25284849   &   0.17730 &   3.9745 \\
      N     &   0.000000  &   0.000000   &   -2.41981278   &  -0.19705 &   2.9459 \\
    \hline
  \end{tabular}
\end{table*}
\begin{table*}
  \centering
  \caption{Cartesian coordinates (in angstrom), natural charges and Wiberg valencies of the atoms of the
    (metastable) \ce{C4N^0} neutral quartet ($\tilde{a} ^4 \Sigma^{-}$).}
  \label{table:geom-quadruplet}
  \begin{tabular}{lccccc}
    \hline
 Atom  &        X     &      Y         &    Z         &   charge  & valence \\
\hline
 \ce{C1}    &    0.000000  & 0.000000  &  -2.61882889 &  0.38999  & 1.5361 \\
 \ce{C2}    &    0.000000  & 0.000000  &  -1.36043777 & -0.25338  & 3.9155 \\
 \ce{C3}    &    0.000000  & 0.000000  &  -0.08294868 & -0.08296  & 3.5336 \\
 \ce{C4}    &    0.000000  & 0.000000  &   1.24266035 &  0.21504  & 3.9845 \\
      N     &    0.000000  & 0.000000  &   2.41676142 & -0.26869  & 2.9311 \\
    \hline
  \end{tabular}
\end{table*}
\begin{table*}
  \centering
  \caption{Cartesian coordinates (in angstrom), natural charges and Wiberg valencies of the atoms of the
    (most stable) bent \ce{C4N-} singlet ($^1 A^\prime$).}
  \label{table:geom-stable-singlet}
  \begin{tabular}{lccccc}
    \hline
     Atom  &        X     &      Y         &    Z        &   charge  & valence \\
\hline
 \ce{C1}   &   2.452143   &  -0.396162     &   0.000000  & -0.12258  &  2.4933 \\
 \ce{C2}   &   1.261801   &   0.069056     &   0.000000  & -0.46722  &  3.9608 \\
 \ce{C3}   &   0.078147   &   0.674522     &   0.000000  & -0.20040  &  3.3211 \\
 \ce{C4}   &  -1.153197   &   0.041285     &   0.000000  &  0.26342  &  3.9784 \\
     N     &  -2.261910   &  -0.333173     &   0.000000  & -0.47322  &  2.9340 \\
    \hline
  \end{tabular}
\end{table*}
\begin{table*}
  \centering
  \caption{Cartesian coordinates (in angstrom), natural charges and Wiberg valencies of the atoms of the
    (metastable, nearly) linear \ce{C4N-} singlet ($^1 \Sigma^{-}$).}
  \label{table:geom-metastable-singlet}
  \begin{tabular}{lccccc} 
    \hline
 Atom  &       X       &     Y         &    Z        &   charge    & valence \\
\hline
 \ce{C1}    &     5.057640  &    0.060822   &   0.000000  & -0.16047    &  2.4257 \\
 \ce{C2}    &     3.768464  &    0.050142   &   0.000000  & -0.39695    &  3.9445 \\
 \ce{C3}    &     2.475903  &    0.039685   &   0.000000  & -0.15945    &  3.6447 \\
 \ce{C4}    &     1.158166  &    0.029423   &   0.000000  &  0.20914    &  3.9811 \\
      N     &    -0.030173  &    0.019929   &   0.000000  & -0.49227    &  2.8156 \\
    \hline
  \end{tabular}
\end{table*}
\begin{table*}
  \centering
  \caption{Cartesian coordinates (in angstrom), natural charges and Wiberg valencies of the atoms of the
    most stable linear \ce{C4N-} triplet ($^3 \Sigma^{-}$).}
  \label{table:geom-triplet}
  \begin{tabular}{lccccc}
    \hline
      Atom  &        X     &      Y      &      Z        &   charge  & valence \\
\hline
 \ce{C1}    &    0.000000  &   0.000000  &   -2.65415168 & -0.17449  &   1.9735 \\
 \ce{C2}    &    0.000000  &   0.000000  &   -1.36306952 & -0.32969  &   3.8275 \\
 \ce{C3}    &    0.000000  &   0.000000  &   -0.07142727 & -0.24430  &   3.2586 \\
 \ce{C4}    &    0.000000  &   0.000000  &    1.24775715 &  0.27001  &   3.9430 \\
      N     &    0.000000  &   0.000000  &    2.43504970 & -0.52154  &   2.7338 \\
    \hline
  \end{tabular}
\end{table*}
\begin{table*}
  \centering
  \caption{Cartesian coordinates (in angstrom), natural charges and Wiberg valencies of the atoms of the
    linear singlet \ce{C4N+} cation ($^1 \Sigma^{+}$).}
  \label{table:geom-cation-singlet}
  \begin{tabular}{lccccc}
    \hline
      Atom  &        X     &      Y       &      Z        &   charge  & valence \\
\hline
 \ce{C1}    &    0.000000  &    0.000000  &    2.66019710 &   0.85764 &  1.6873 \\
 \ce{C2}    &    0.000000  &    0.000000  &    1.32621289 &  -0.58628 &  3.8834 \\
 \ce{C3}    &    0.000000  &    0.000000  &    0.08794830 &   0.57113 &  3.8399 \\
 \ce{C4}    &    0.000000  &    0.000000  &   -1.25330642 &   0.03453 &  3.9900 \\
      N     &    0.000000  &    0.000000  &   -2.41804446 &   0.12297 &  3.0149 \\
    \hline
  \end{tabular}
\end{table*}
\begin{table*}
  \centering
  \caption{Cartesian coordinates (in angstrom), natural charges and Wiberg valencies of the atoms of the
    (metastable) linear triplet \ce{C4N+} cation ($^3 \Sigma^{+}$).}
  \label{table:geom-cation-triplet}
  \begin{tabular}{lccccc}
    \hline
 Atom  &        X     &      Y      &    Z        &   charge  & valence \\
\hline
 C1    &    0.000000  &   0.000000  &  2.61065996  &  0.81831  &  1.9491 \\
 C2    &    0.000000  &   0.000000  &  1.35771105  & -0.37213  &  3.8271 \\
 C3    &    0.000000  &   0.000000  &  0.08308407  &  0.39549  &  3.4954 \\
 C4    &    0.000000  &   0.000000  & -1.23722188  &  0.06271  &  3.9643 \\
 N     &    0.000000  &   0.000000  & -2.41219988  &  0.09561  &  2.8723 \\
 \hline
  \end{tabular}
\end{table*}
\begin{table*}
  \tiny 
  \centering
  \caption{Bond metric data for \ce{C4N} chains at geometries optimized using
    several exchange-correlation functionals and basis sets.
    Bond lengths $l$ between atoms XY (in angstrom),
    angles $\alpha$ between atoms $\widehat{\mbox{\ce{XYZ}}}$ (in degrees).
    Whenever angles between adjacent bonds are indicated, the geometries were optimized without imposing symmetry constraints.
}
  \label{table:lengths-bonds-pbe0-neumark}
  \begin{tabular}{ccclllllllllll}
    \hline
    Species & Method & Property & \ce{C1C2} & $\widehat{\mbox{\ce{C1C2C3}}}$ & \ce{C2C3} & $\widehat{\mbox{\ce{C2C3C4}}}$ & \ce{C3C4} & $\widehat{\mbox{\ce{C3C4N}}}$ & \ce{C4N} \\
    \hline
bent \ce{C4N-} singlet& RB3LYP/6-311++G(3df, 3pd) & $l$, $\alpha$ & 1.2780 & 174.3 & 1.3295 & 125.7 & 1.3846 & 171.4 & 1.1702 \\
                      & RPBE0/6-311++G(3df, 3pd)   &               & 1.2792 & 174.3 & 1.3287 & 125.0 & 1.3847 & 171.6 & 1.1688 \\
\ce{C4N-} triplet     & UB3LYP/6-311++G(3df, 3pd) & $l$, $\alpha$ & 1.2912 & 179.8 & 1.2917 & 178.7 & 1.3193 & 180.0 & 1.1874 \\
                      & UB3LYP/aug-cc-pVTZ        &               & 1.2913 &       & 1.2920 &       & 1.3197 &       & 1.1874 \\
                      & UPBE0/6-311++G(3df, 3pd)  &               & 1.2938 & 179.8 & 1.2903 & 178.7 & 1.3209 & 180.0 & 1.1849 \\
                      & UM06-2X/6-311++G(3df, 3pd) &               & 1.2935 &       & 1.2877 &       & 1.3300 &       & 1.1774 \\
                      & UB2GP-PLYP/6-311++G(3df, 3pd)    &       & 1.2924 &       & 1.2862 &       & 1.3293 &       & 1.1781 \\
                      & ROCCSD(T)/aug-cc-PVTZ    &               & 1.3040 &       & 1.3027 &       & 1.3343 &       & 1.1944 \\
                      &                                  &               &        &       &        &       &        &       &        \\
\ce{C4N^0} doublet    & UB3LYP/6-311++G(3df, 3pd) & $l$, $\alpha$ & 1.3165 & 179.8 & 1.2536 & 178.8 & 1.3371 & 180.0 & 1.1670 \\
                      & UB3LYP/aug-cc-pVTZ        &               & 1.3169 &       & 1.2537 &       & 1.3377 &       & 1.1671 \\
                      & UPBE0/6-311++G(3df, 3pd)  &               & 1.3193 & 179.8 & 1.2522 & 178.8 & 1.3386 & 180.0 & 1.1653 \\
                      & UB2GP-PLYP/6-311++G(3df, 3pd)    &       & 1.3258 &       & 1.2330 &       & 1.3594 &       & 1.1539 \\
                      & ROCCSD(T)/aug-cc-PVTZ    &               & 1.3357 &       & 1.2594 &       & 1.3581 &       & 1.1755 \\
                      &                          &               &        &       &        &       &        &       &        \\ 
\ce{C4N^0} quartet    & UB3LYP/6-311++G(3df, 3pd) &  $l$, $\alpha$ & 1.2585 & 179.9 & 1.2776 & 178.8 & 1.3257 & 179.9 & 1.1742 \\
                      & UB3LYP/aug-cc-pVTZ        &                & 1.2585 &       & 1.2777 &       & 1.3262 &       & 1.1742 \\
                      & UPBE0/6-311++G(3df, 3pd)  &               & 1.2626 &       & 1.2735 &       & 1.3289 &       & 1.1717 \\
                      &                          &               &        &       &        &       &        &       &        \\ 
\ce{C4N+} singlet     & RB3LYP/6-311++G(3df, 3pd) & $l$, $\alpha$ & 1.3343 & 178.8 & 1.2383 & 179.7 & 1.3413 & 179.6 & 1.1648 \\
                      & RPBE0/6-311++G(3df, 3pd)  &               & 1.3361 & 178.7 & 1.2374 & 179.6 & 1.3421 & 179.7 & 1.1636 \\
                      & B2GP-PLYP/6-311++G(3df, 3pd)  &               & 1.3326 &       & 1.2459 &       & 1.3407 &       & 1.1736 \\
                      & RCCSD(T)/aug-cc-PVTZ     &               & 1.3343 &       & 1.2383 &       & 1.3413 &       & 1.1648 \\
                      &                          &               &        &       &        &       &        &       &        \\ 
\ce{C4N+} triplet     & UB3LYP/6-311++G(3df, 3pd) &               & 1.2531 & 179.8 & 1.2747 & 178.9 & 1.3204 & 179.7 & 1.1751 \\
                      & UPBE0/6-311++G(3df, 3pd)  &               & 1.2564 &       & 1.2722 &       & 1.3216 &       & 1.1737 \\
\hline
  \end{tabular}
\end{table*}

\begin{figure*}
  \centerline{
    \includegraphics[width=0.4\textwidth]{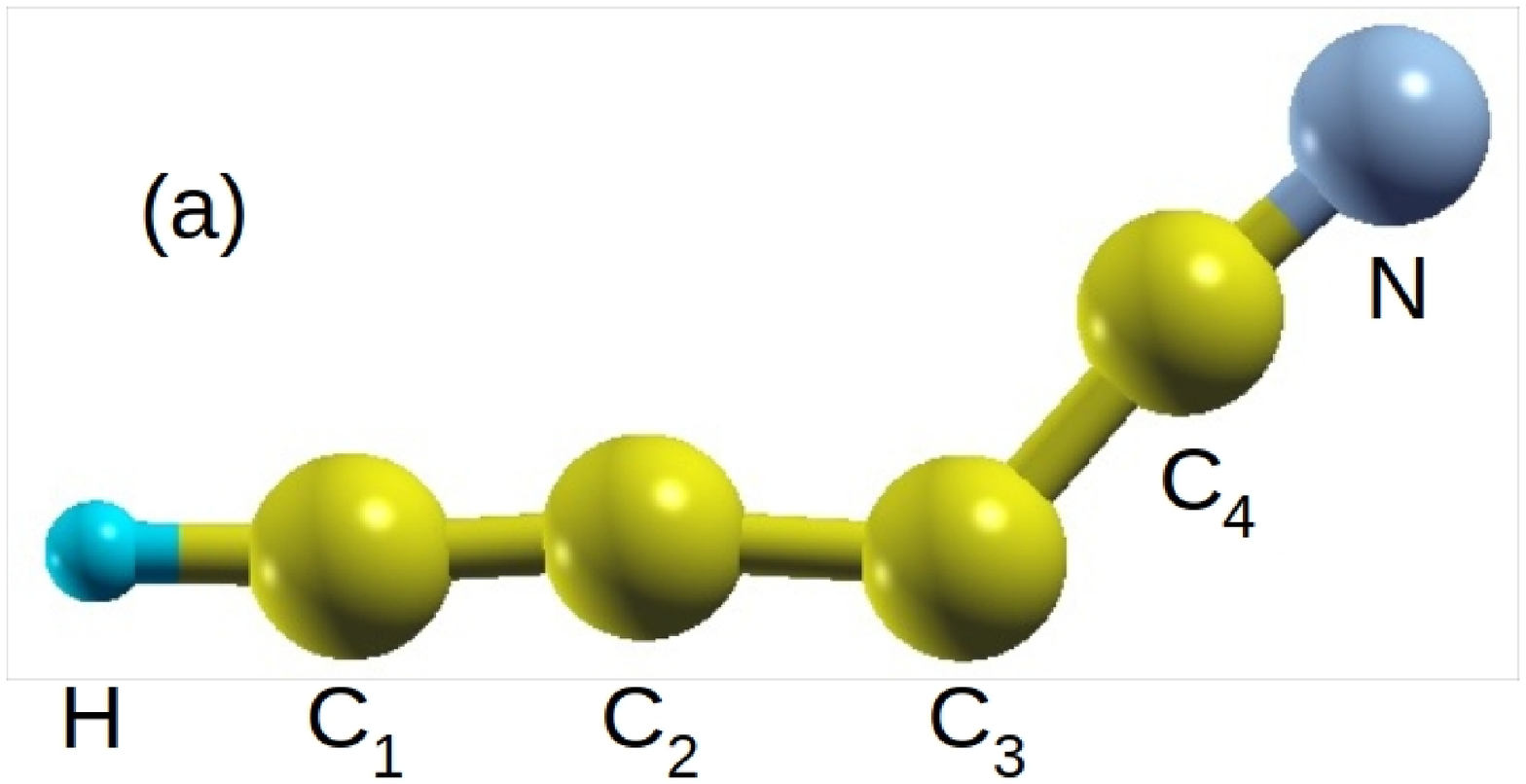}\hspace*{10ex}
    \includegraphics[width=0.4\textwidth]{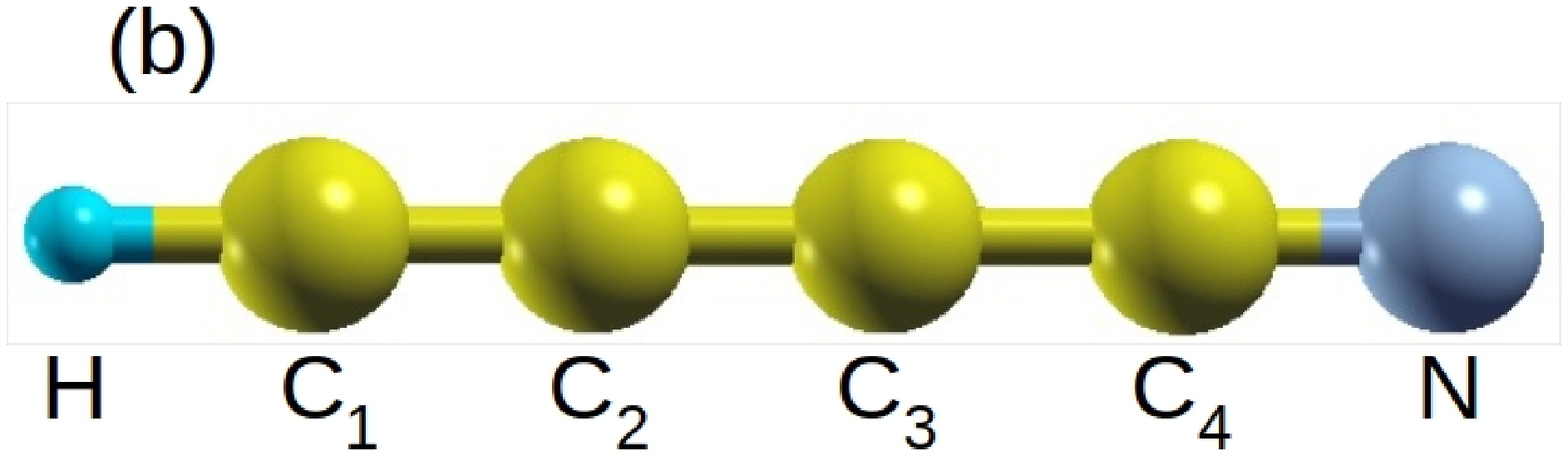}
  }
  \caption{Geometries of singlet and triplet \ce{HC4N} chains (left and right panels, respectively)
    investigated in the present paper.}
    \label{fig:geometries-hc4n}
\end{figure*}

\begin{table*}
  \centering
  \caption{Natural charges and Wiberg valencies of the atoms of the bent stable \ce{HC4N} (second and third columns)
    and \ce{C4N-} (fourth and fifth columns) singlets.}
  \label{table:bent-singlet-hc4n-c4n-}
  \begin{tabular}{lcccc}
    \hline
      Atom  &     charge     &  valence &   charge  & valence \\
\hline
      H     &     0.23992    &   0.9448 &   ---     &   ---   \\
 \ce{C1}    &    -0.02018    &   3.6818 & -0.12258  &  2.4933 \\
 \ce{C2}    &    -0.19260    &   3.9373 & -0.46722  &  3.9608 \\
 \ce{C3}    &     0.06473    &   3.0972 & -0.20040  &  3.3211 \\
 \ce{C4}    &     0.15035    &   3.9731 &  0.26342  &  3.9784 \\ 
      N     &    -0.24222    &   2.9193 & -0.47322  &  2.9340 \\
    \hline
  \end{tabular}
\end{table*}

\begin{table*}
  \centering
  \caption{Natural charges and Wiberg valencies of the atoms of the stable \ce{HC4N} (second and third columns)
    and \ce{C4N-} (fourth and fifth columns) triplets.}
  \label{table:triplet-hc4n-c4n-}
  \begin{tabular}{lcccc}
    \hline
      Atom  &     charge       &  valence  &   charge    &  valence \\
\hline
      H     &    0.24010       &  0.9448   &      ---    &     ---  \\
 \ce{C1}    &   -0.06097       &  3.4399   &   -0.17449  &   1.9735 \\
 \ce{C2}    &   -0.21113       &  3.9731   &   -0.32969  &   3.8275 \\
 \ce{C3}    &    0.16184       &  2.8127   &   -0.24430  &   3.2586 \\
 \ce{C4}    &    0.14022       &  3.9897   &    0.27001  &   3.9430 \\
      N     &   -0.27006       &  2.7173   &   -0.52154  &   2.7338 \\
    \hline
  \end{tabular}
\end{table*}

\begin{figure*}
  \includegraphics[width=0.4\textwidth]{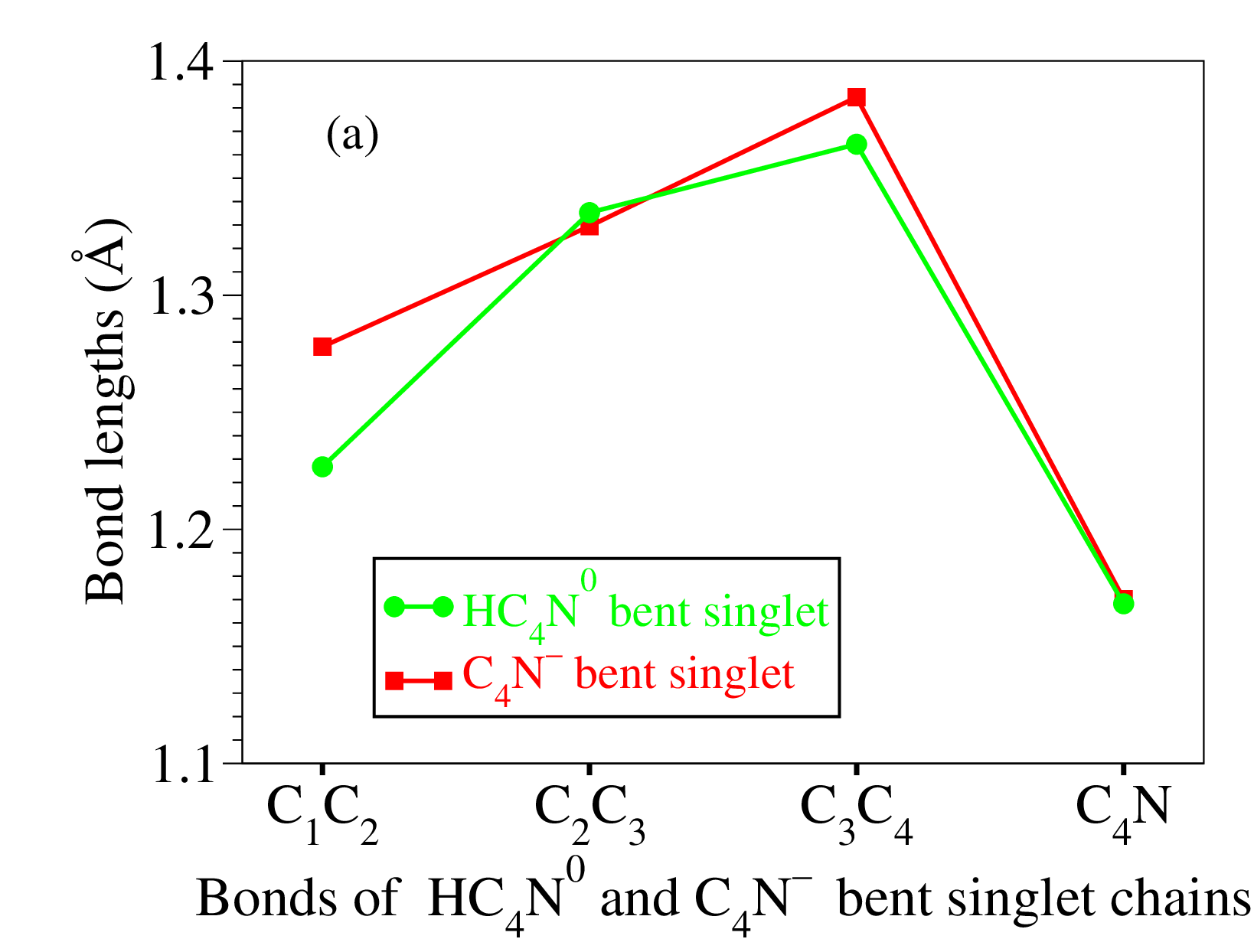}
  \includegraphics[width=0.4\textwidth]{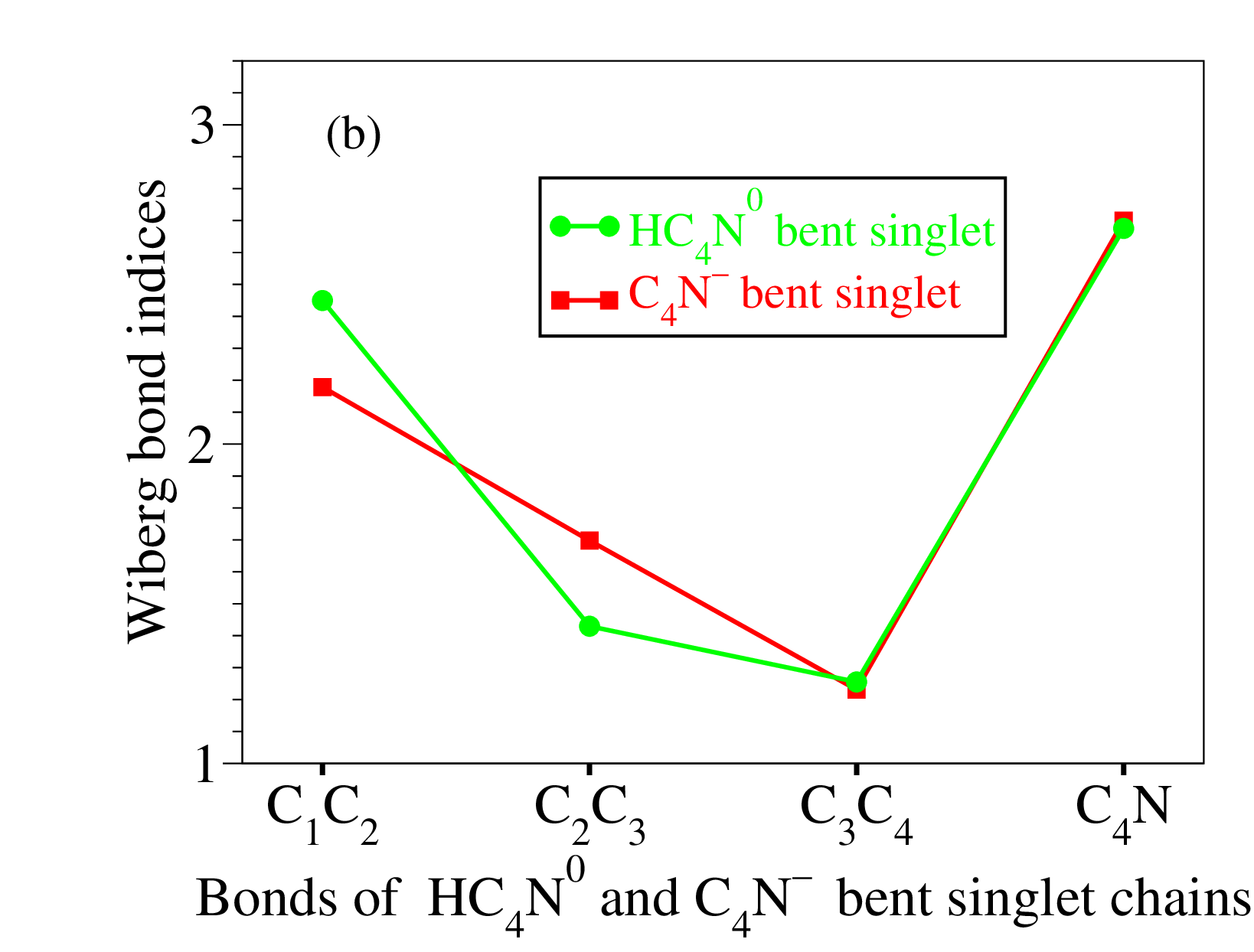}
  \includegraphics[width=0.4\textwidth]{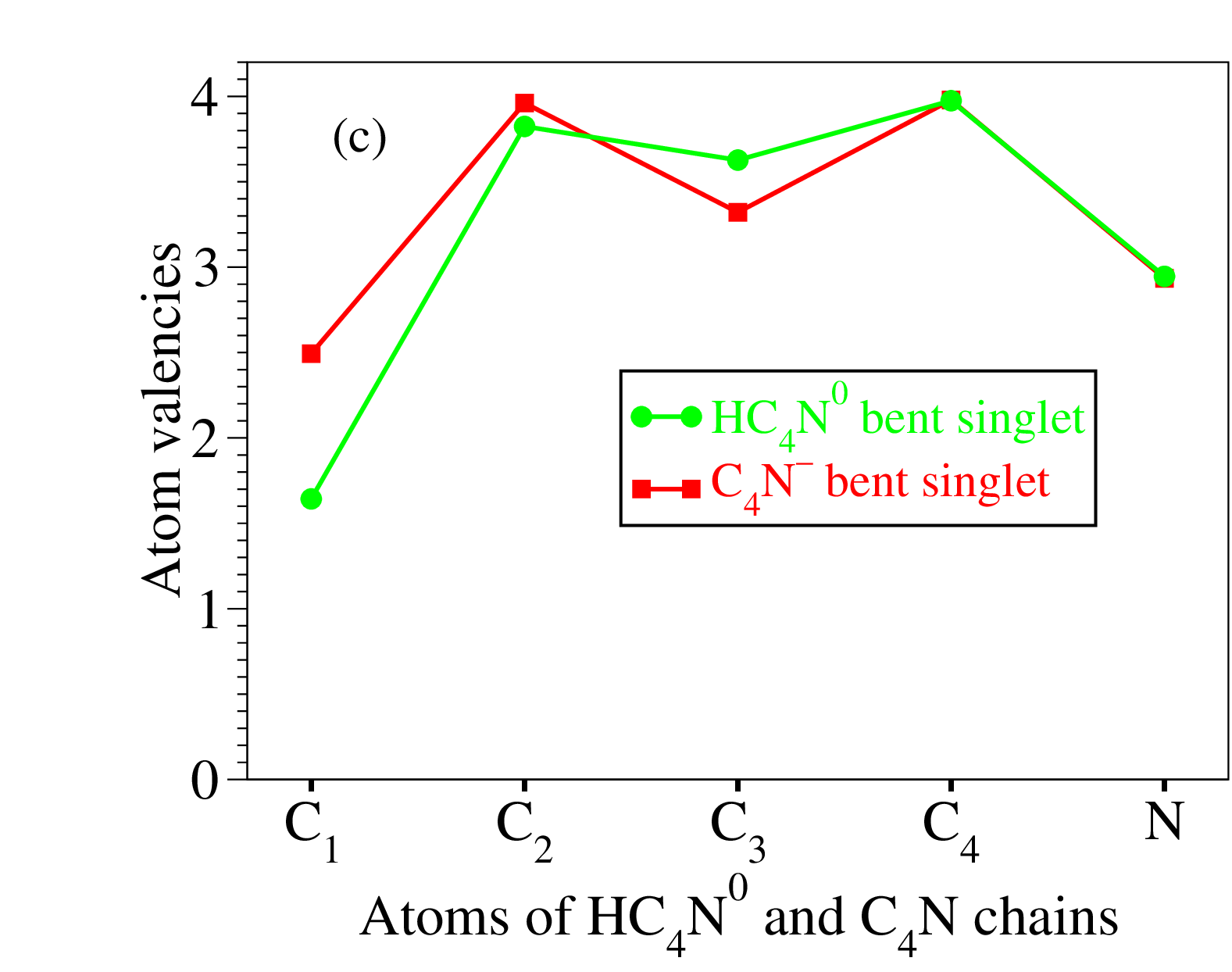}
  \includegraphics[width=0.4\textwidth]{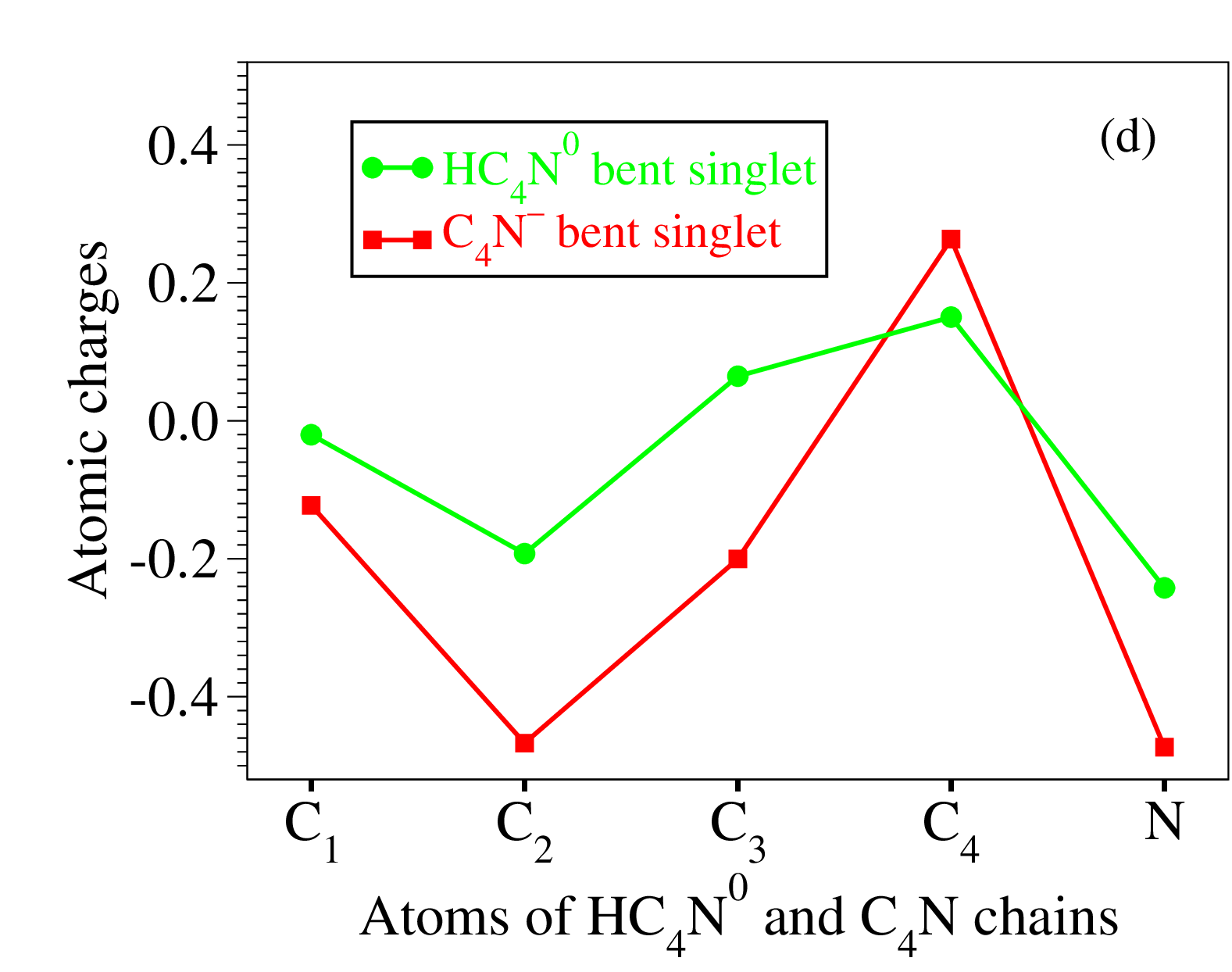}
  \caption{(a) Bond lengths (in angstrom), (b) Wiberg bond indices, (c) Wiberg valencies
    and (d) atomic charges of the isoelectronic \ce{HC4N} and \ce{C4N-} singlet bent chains considered in this paper.}
    \label{fig:lengths_bonds_hc4n-c4n_anion-bent-singlet}
\end{figure*}

\begin{figure*}
  \includegraphics[width=0.4\textwidth]{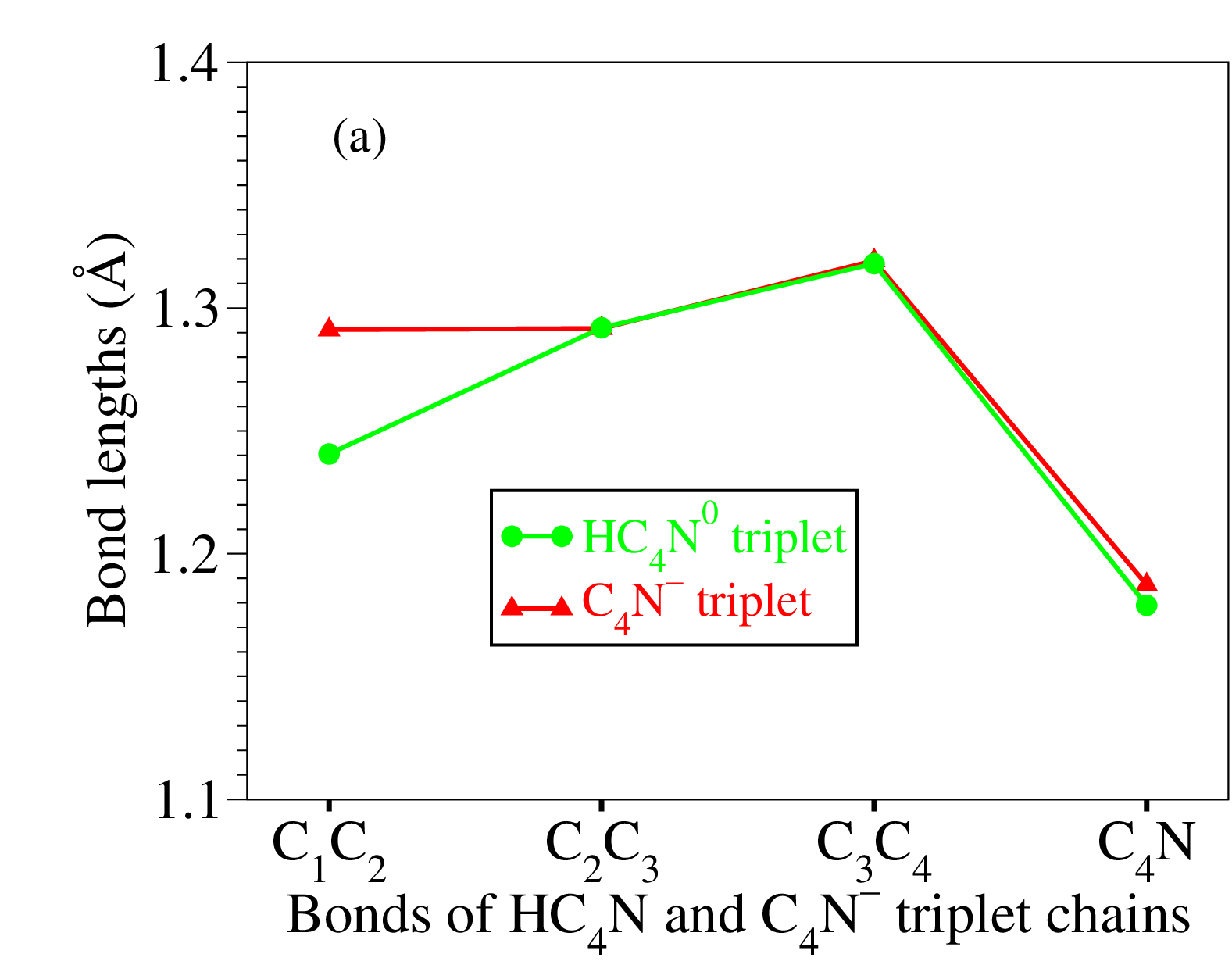}
  \includegraphics[width=0.4\textwidth]{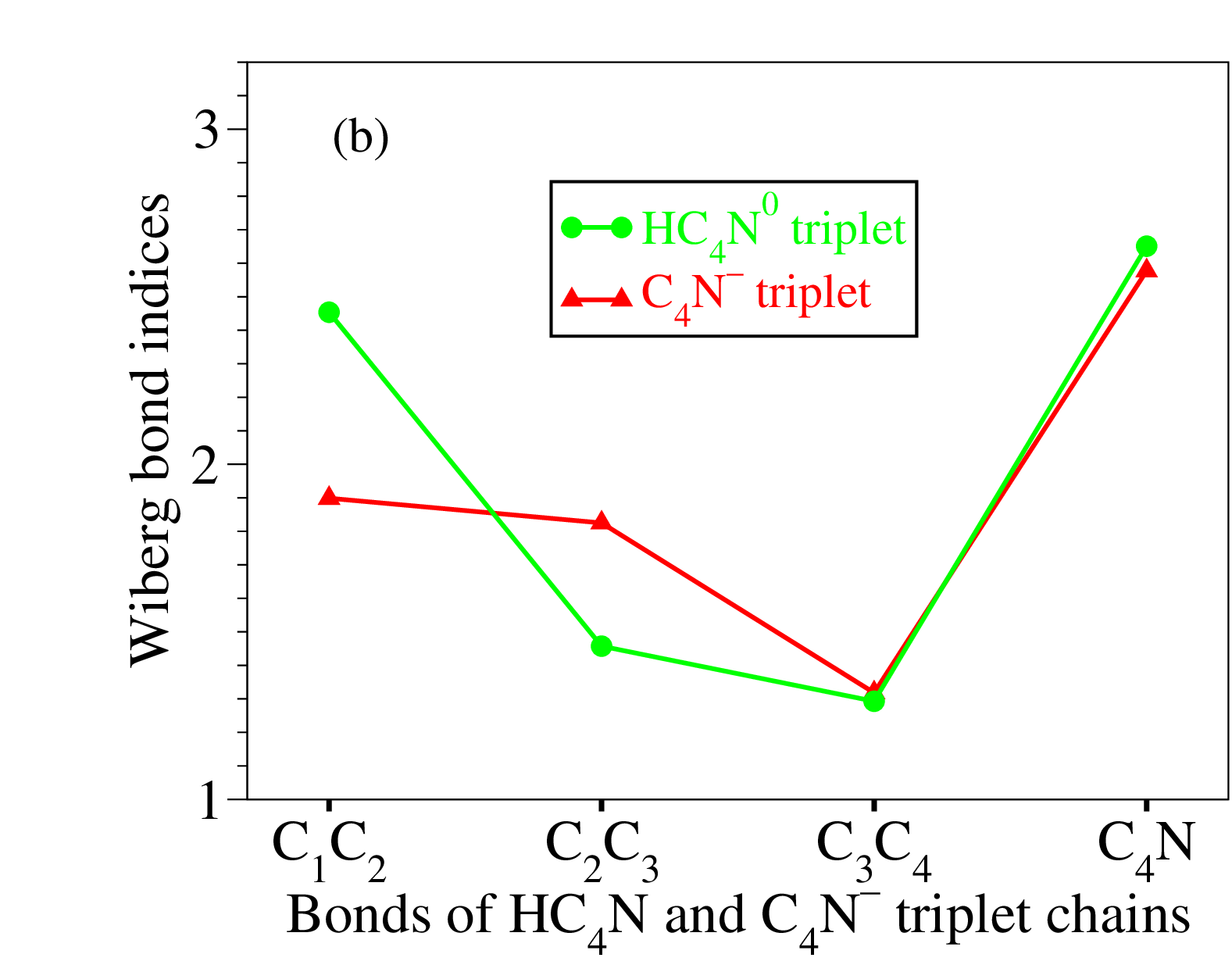}
  \includegraphics[width=0.4\textwidth]{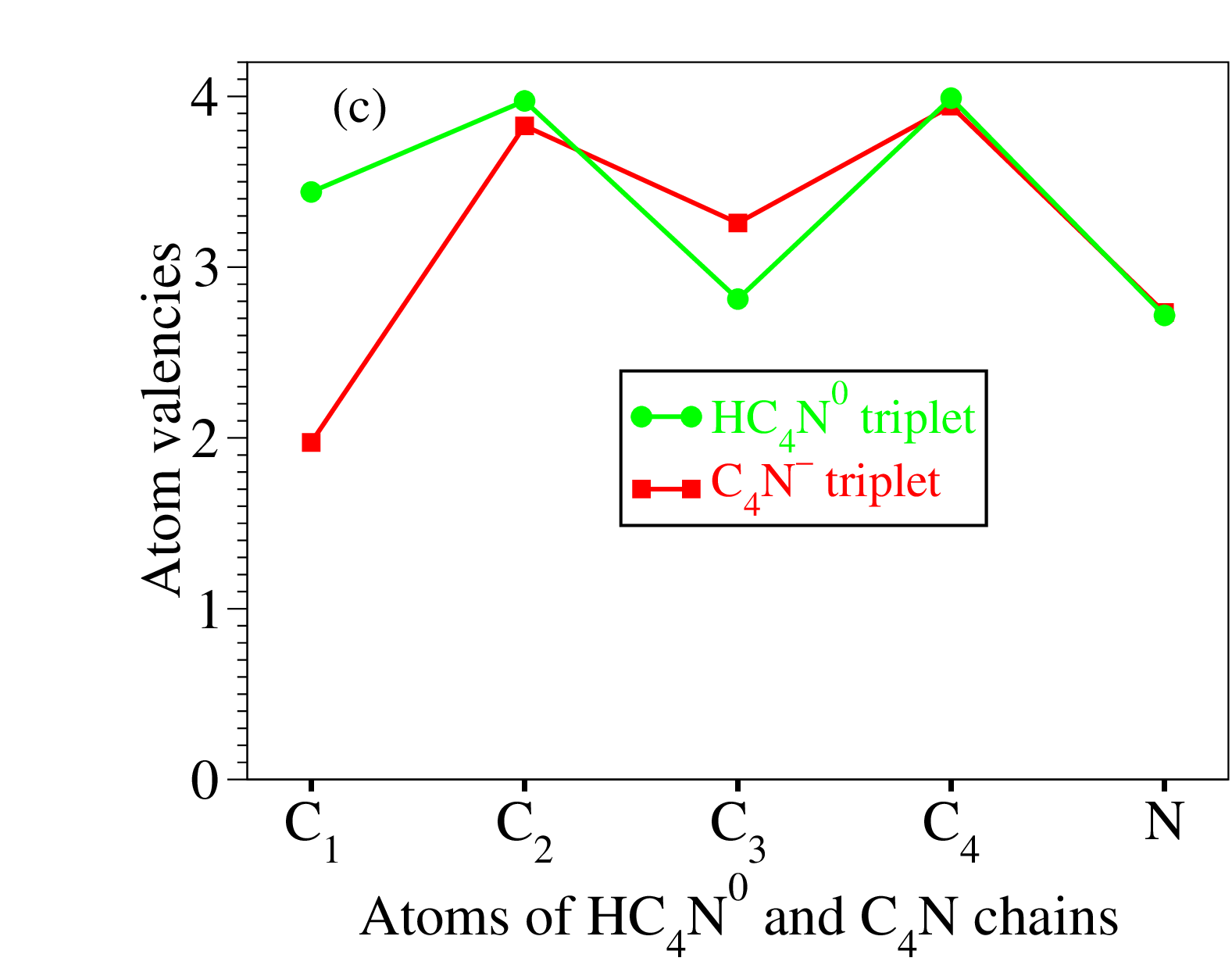}
  \includegraphics[width=0.4\textwidth]{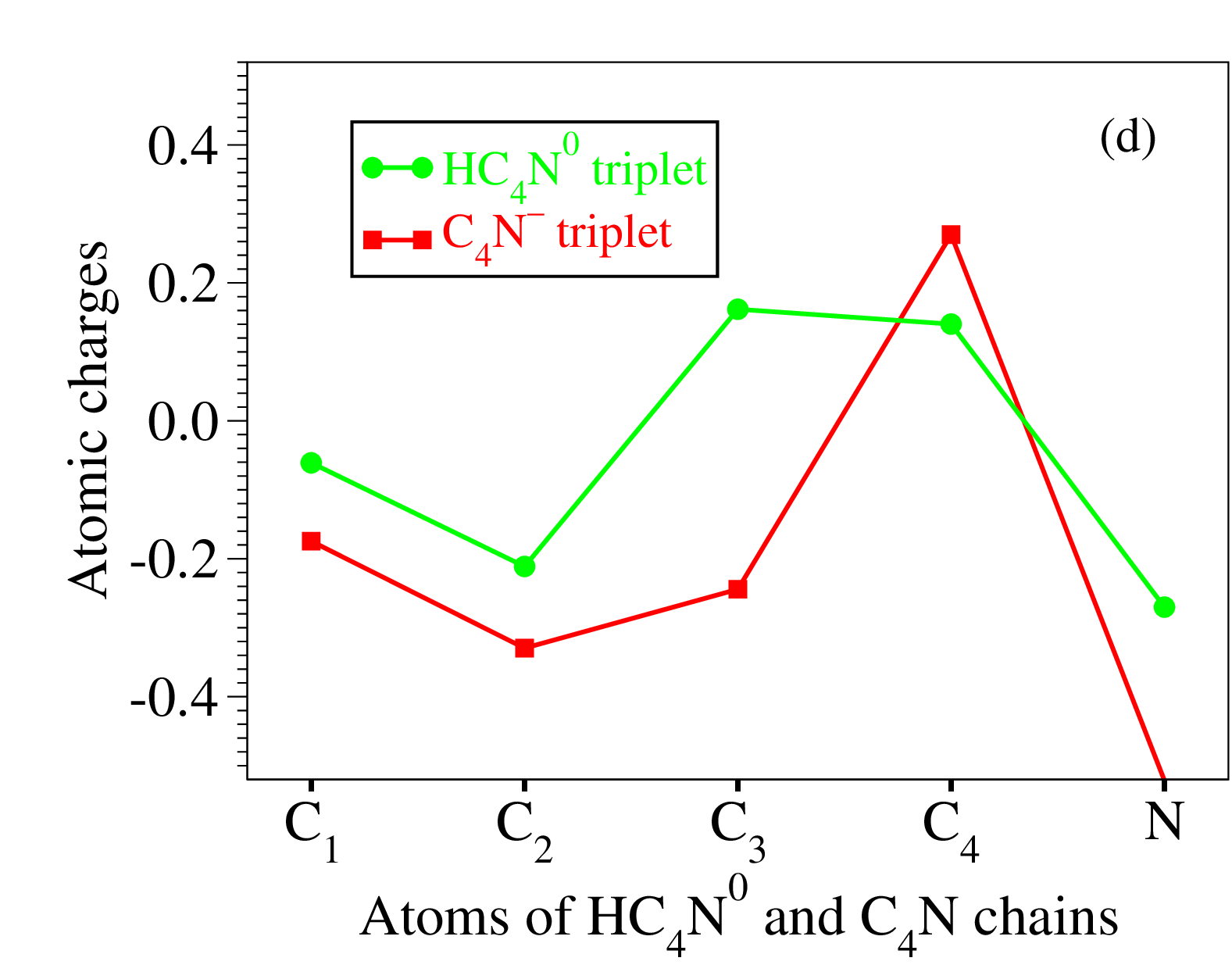}
    \caption{(a) Bond lengths (in angstrom), (b) Wiberg bond indices, (c) Wiberg valencies
    and (d) atomic charges of \ce{HC4N} and \ce{C4N-} triplet chains considered in this paper.}
    \label{fig:lengths_bonds_hc4n-c4n_anion-triplet}
\end{figure*}

\begin{figure*}
  \includegraphics[width=0.4\textwidth]{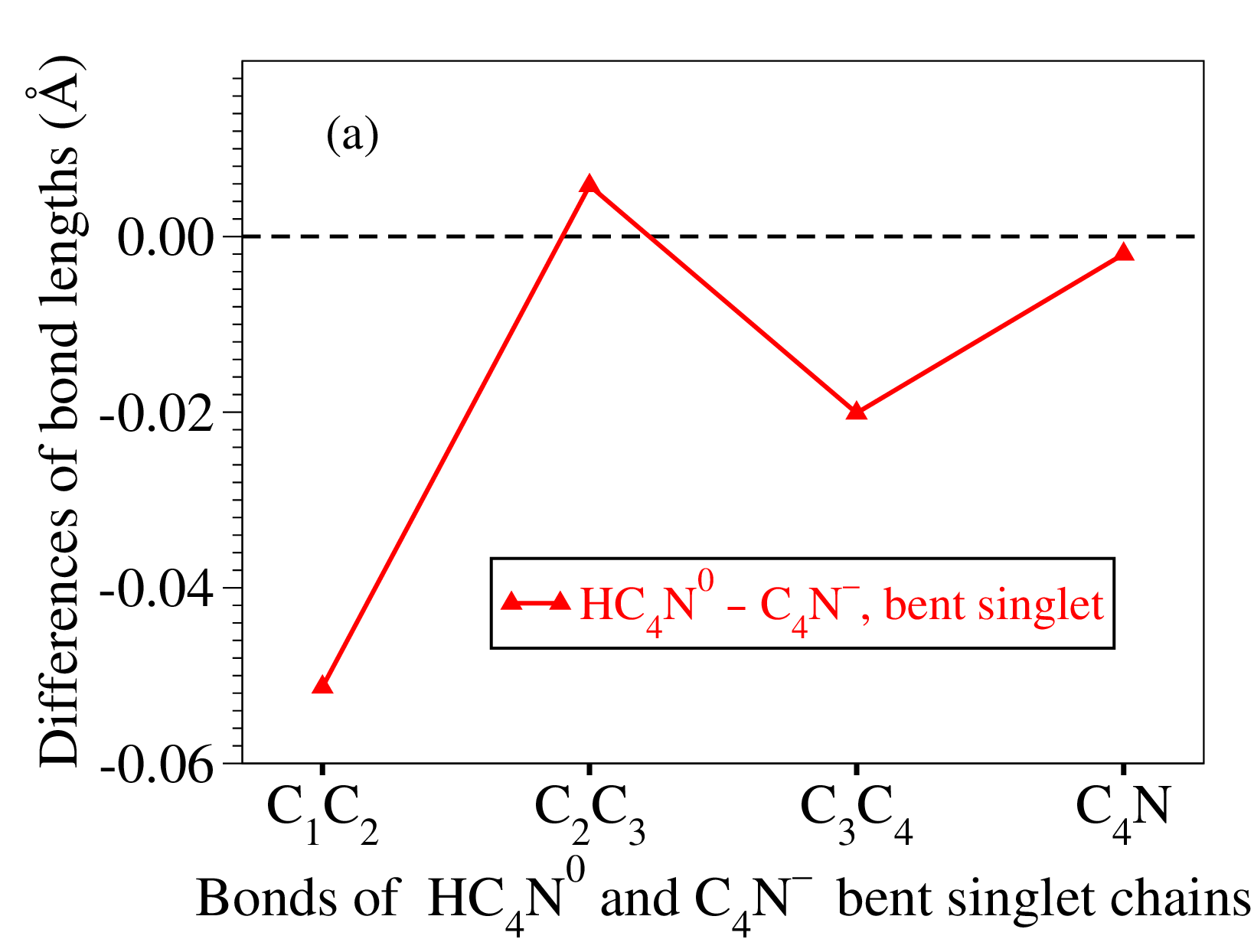}
  \includegraphics[width=0.4\textwidth]{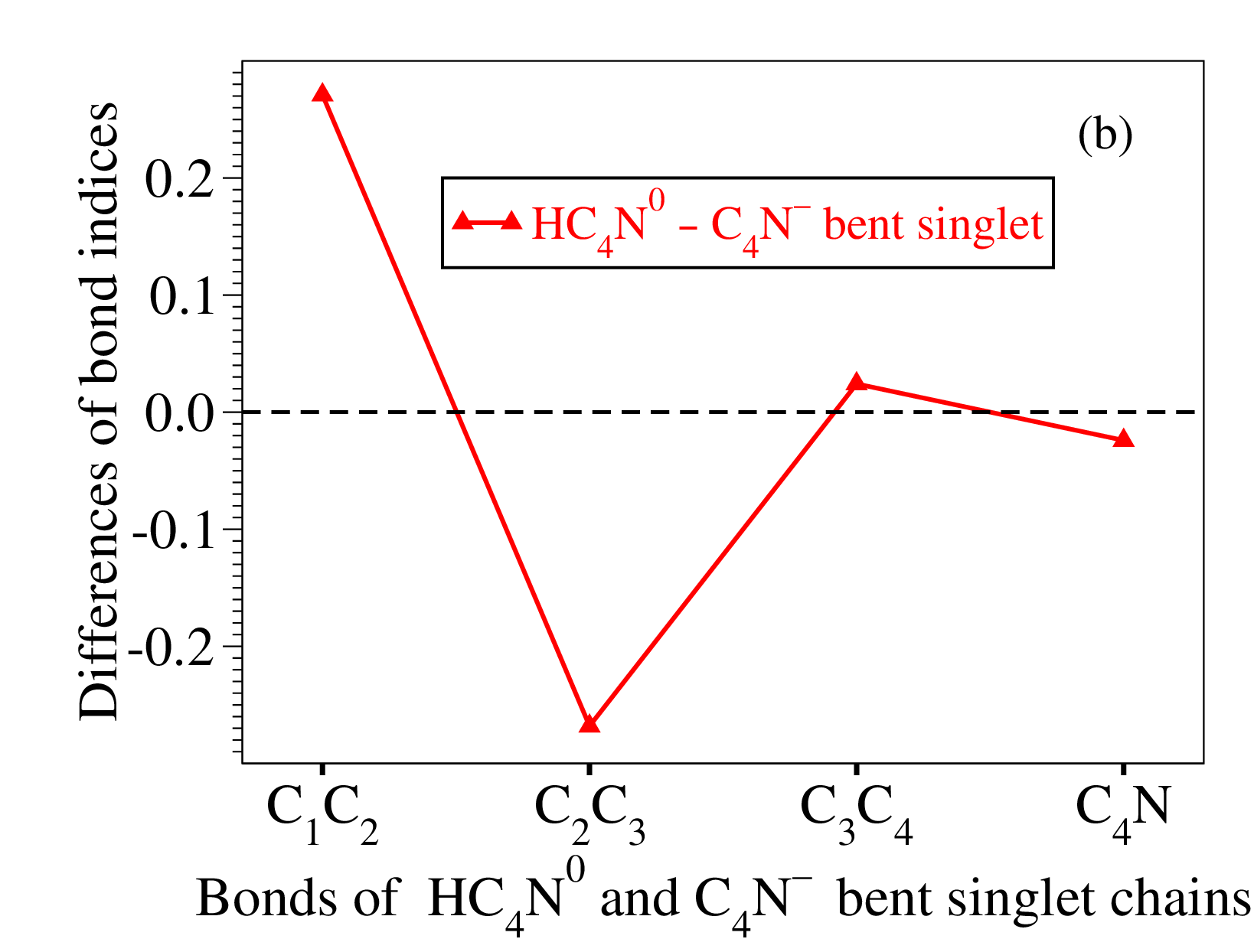}
  \includegraphics[width=0.4\textwidth]{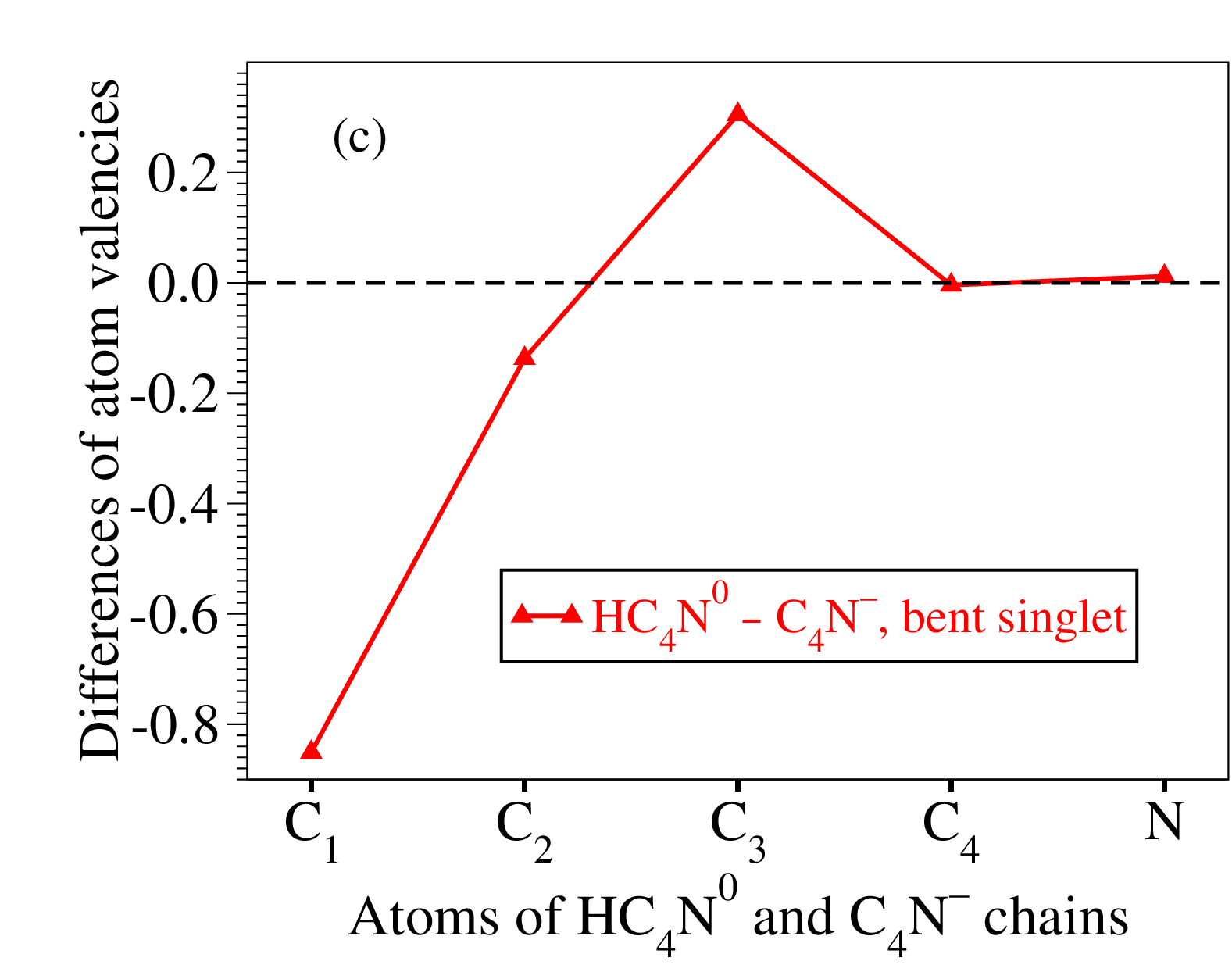}
  \includegraphics[width=0.4\textwidth]{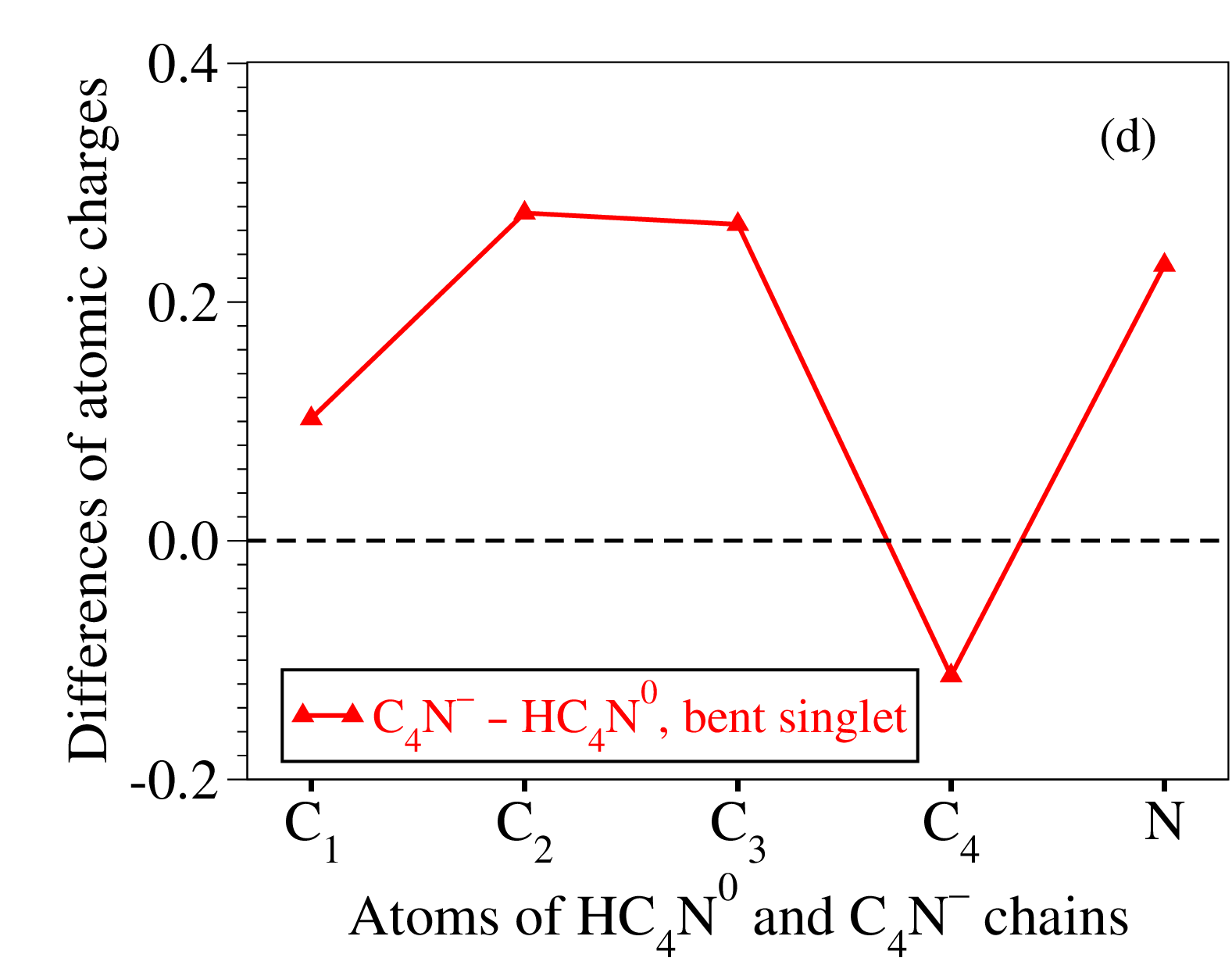}
  \caption{Differences between several molecular properties of the isoelectronic
    \ce{HC4N} and \ce{C4N-} singlet bent chains considered in this paper: (a) bond lengths (in angstrom), (b) Wiberg bond indices,
  (c) Wiberg valencies and (d) atomic charges.}
    \label{fig:differences_lengths_bonds_hc4n-c4n_anion-bent-singlet}
\end{figure*}

\begin{figure*}
  \includegraphics[width=0.4\textwidth]{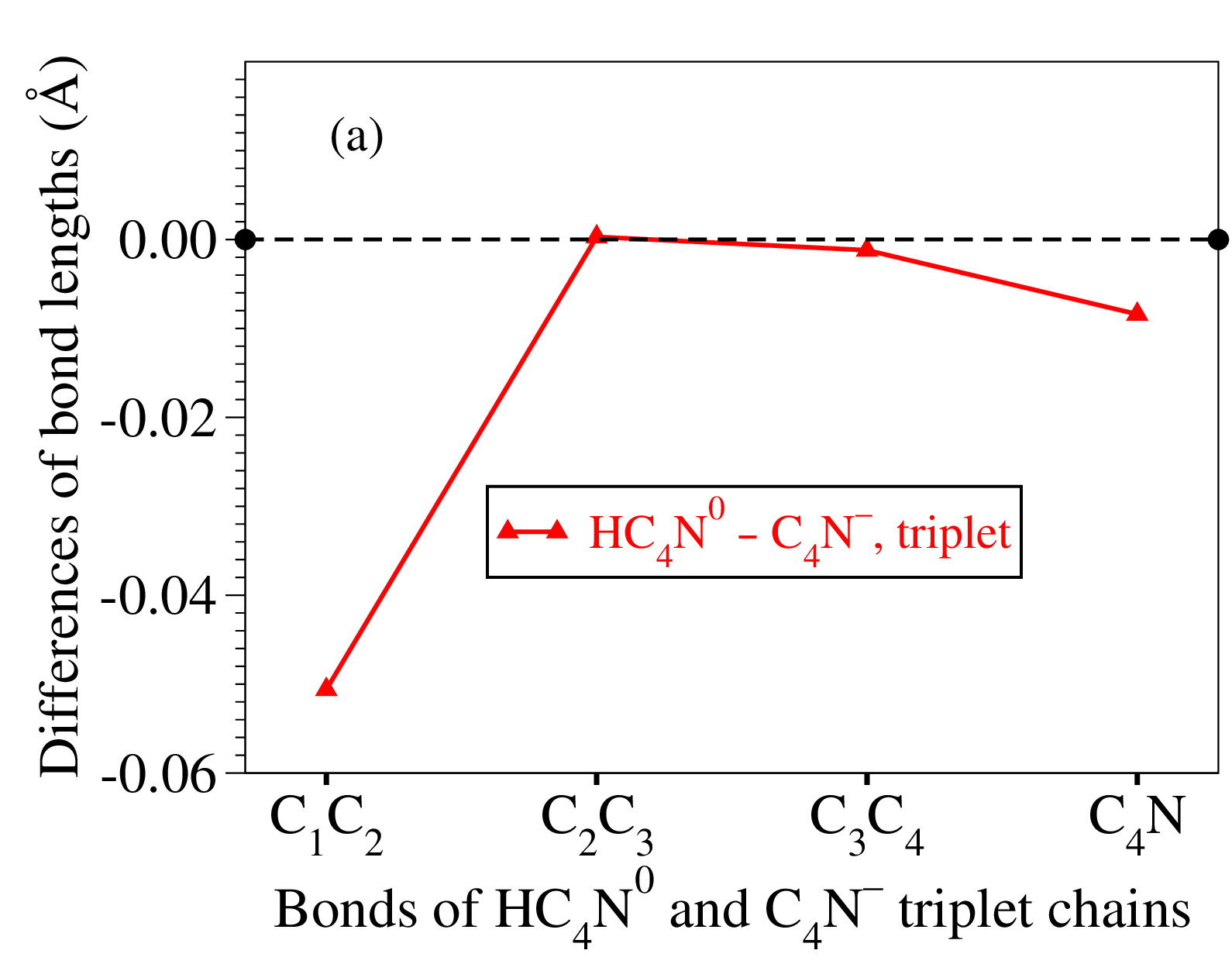}
  \includegraphics[width=0.4\textwidth]{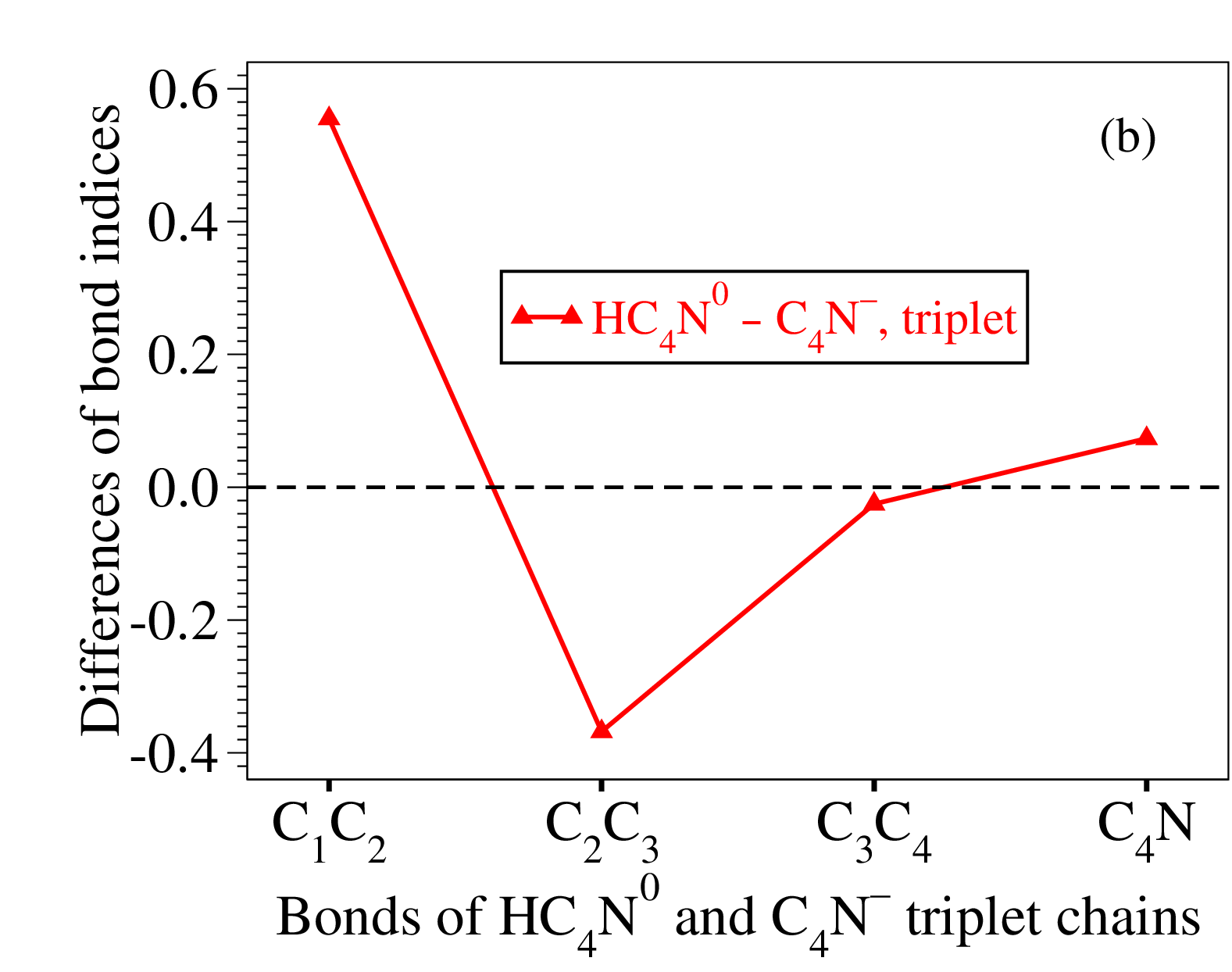}
  \includegraphics[width=0.4\textwidth]{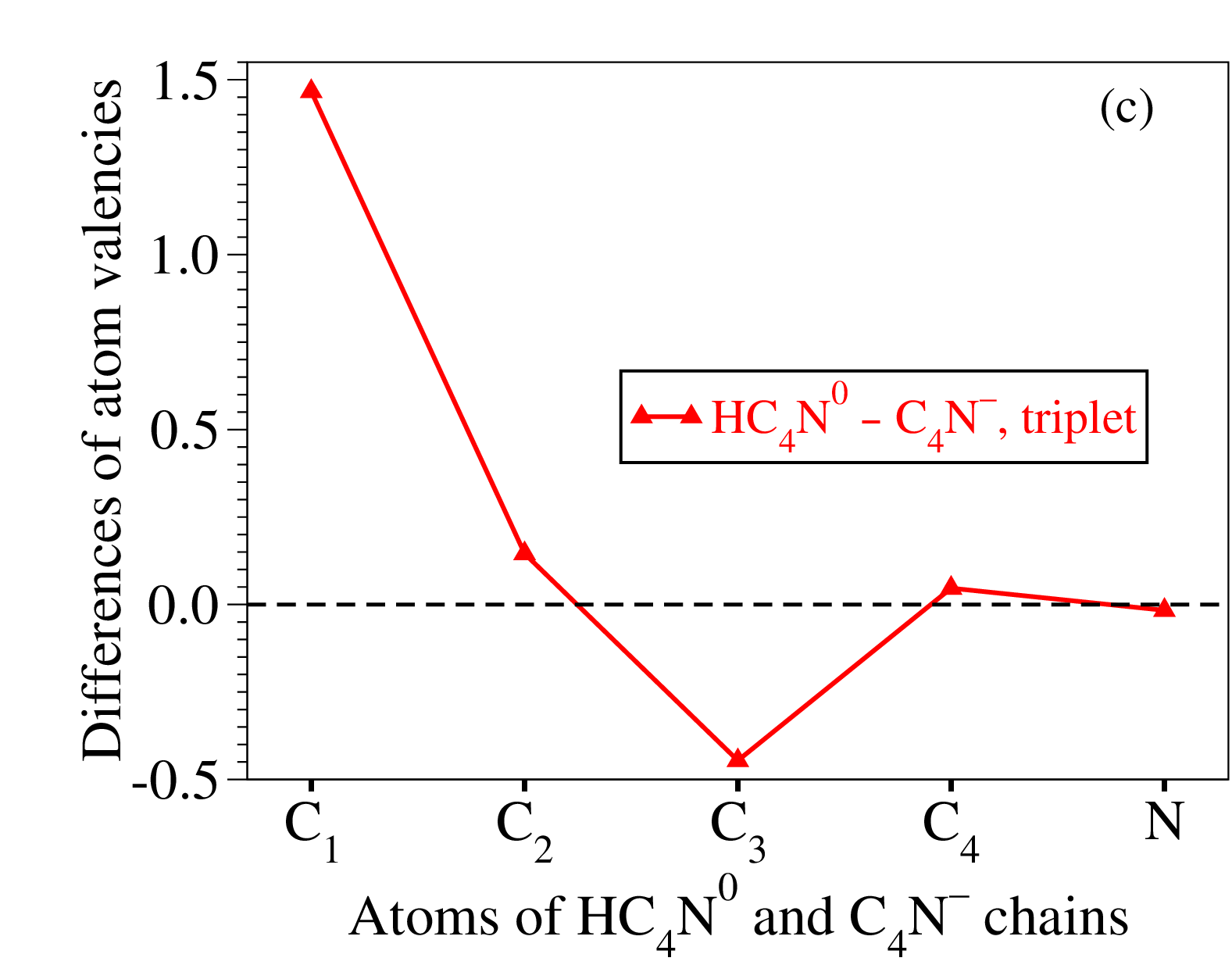}
  \includegraphics[width=0.4\textwidth]{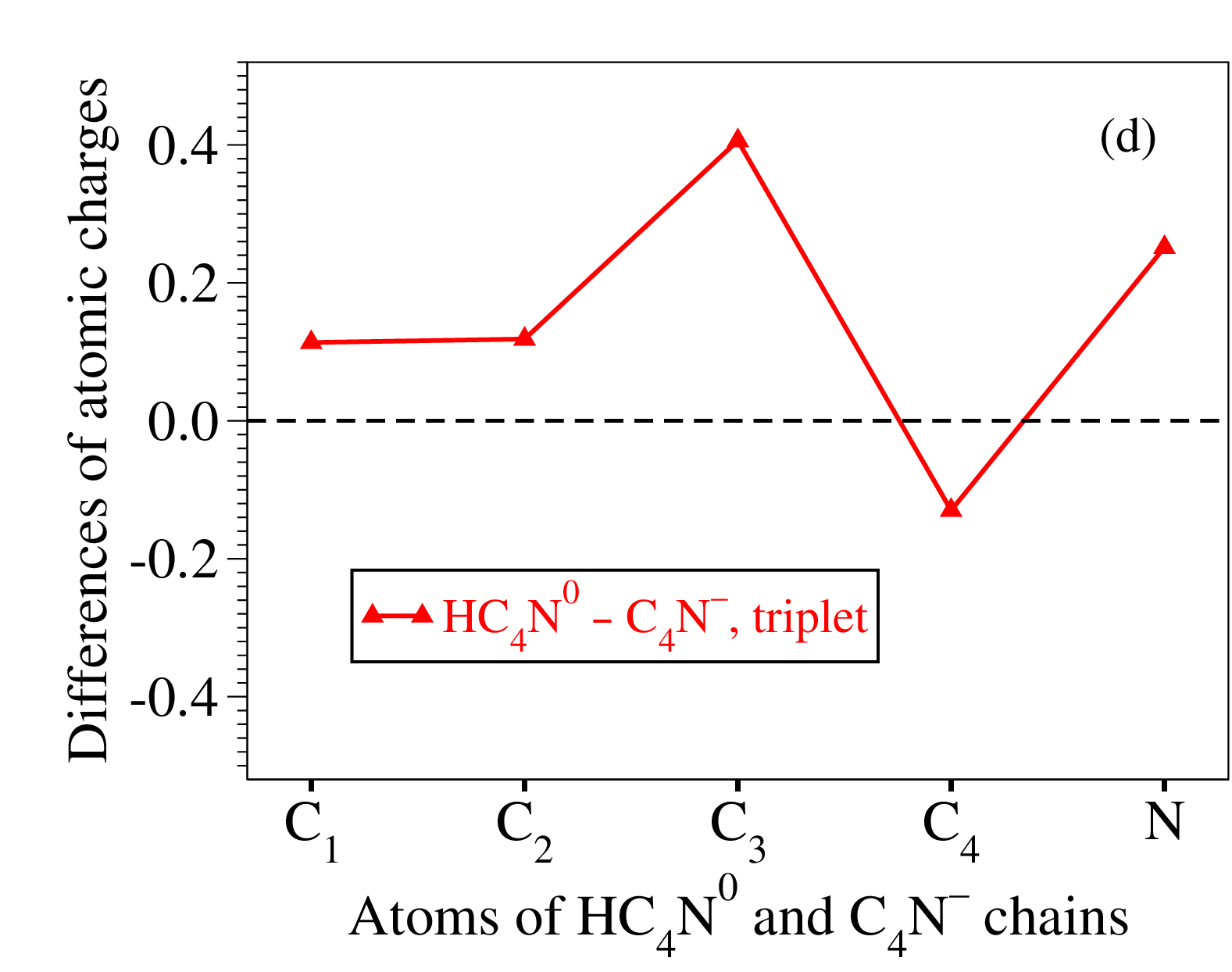}
  \caption{Differences between several molecular properties of the isoelectronic
    \ce{HC4N} and \ce{C4N-} linear triplet chains considered in this paper:
    (a) bond lengths (in angstrom), (b) Wiberg bond indices,
    (c) Wiberg valencies and (d) atomic charges.}
    \label{fig:differences_lengths_bonds_hc4n-c4n_anion-triplet}
\end{figure*}

\begin{figure*}
  \includegraphics[width=0.4\textwidth]{fig_ir_c4n-_hc4n_singlet.eps}
  \includegraphics[width=0.4\textwidth]{fig_raman_c4n-_hc4n_singlet.eps}
  \caption{(a) Infrared and (b) Raman spectra of \ce{HC4N} and \ce{C4N-} bent singlet chains considered in this paper.}
    \label{fig:vibr-hc4n-c4n-_singlet}
\end{figure*}

\begin{figure*}
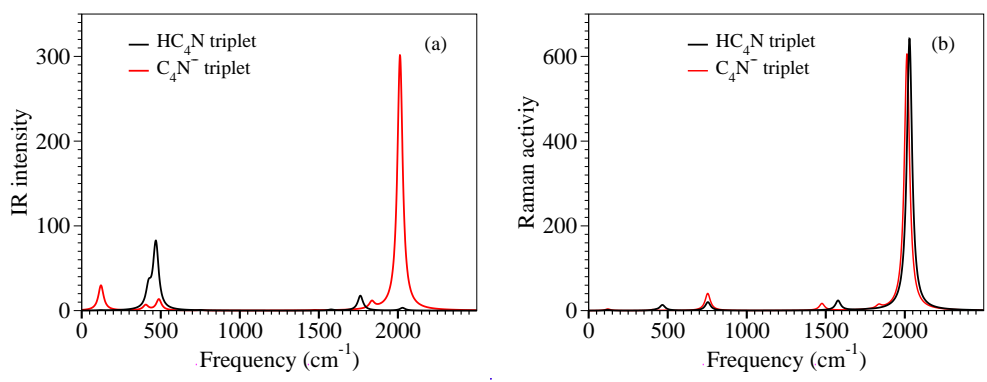

  \includegraphics[width=0.4\textwidth]{fig_ir_c4n-_hc4n_triplet.eps}
  \includegraphics[width=0.4\textwidth]{fig_raman_c4n-_hc4n_triplet.eps}
  \caption{(a) Infrared and (b) Raman spectra of \ce{HC4N} and \ce{C4N-} triplet chains considered in this paper.}
    \label{fig:vibr-hc4n-c4n-_triplet}
\end{figure*}

\begin{figure*}
  \centerline{
    \includegraphics[width=0.4\textwidth]{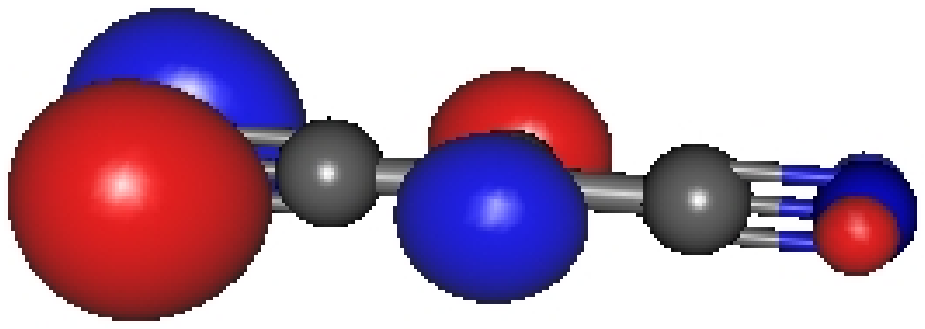}\hspace*{10ex}
    \includegraphics[width=0.4\textwidth]{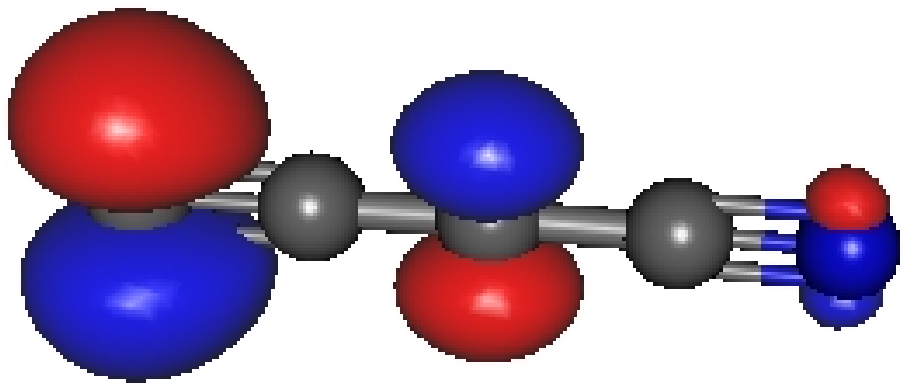}\hspace*{10ex}
  }
  \centerline{
    \includegraphics[width=0.4\textwidth]{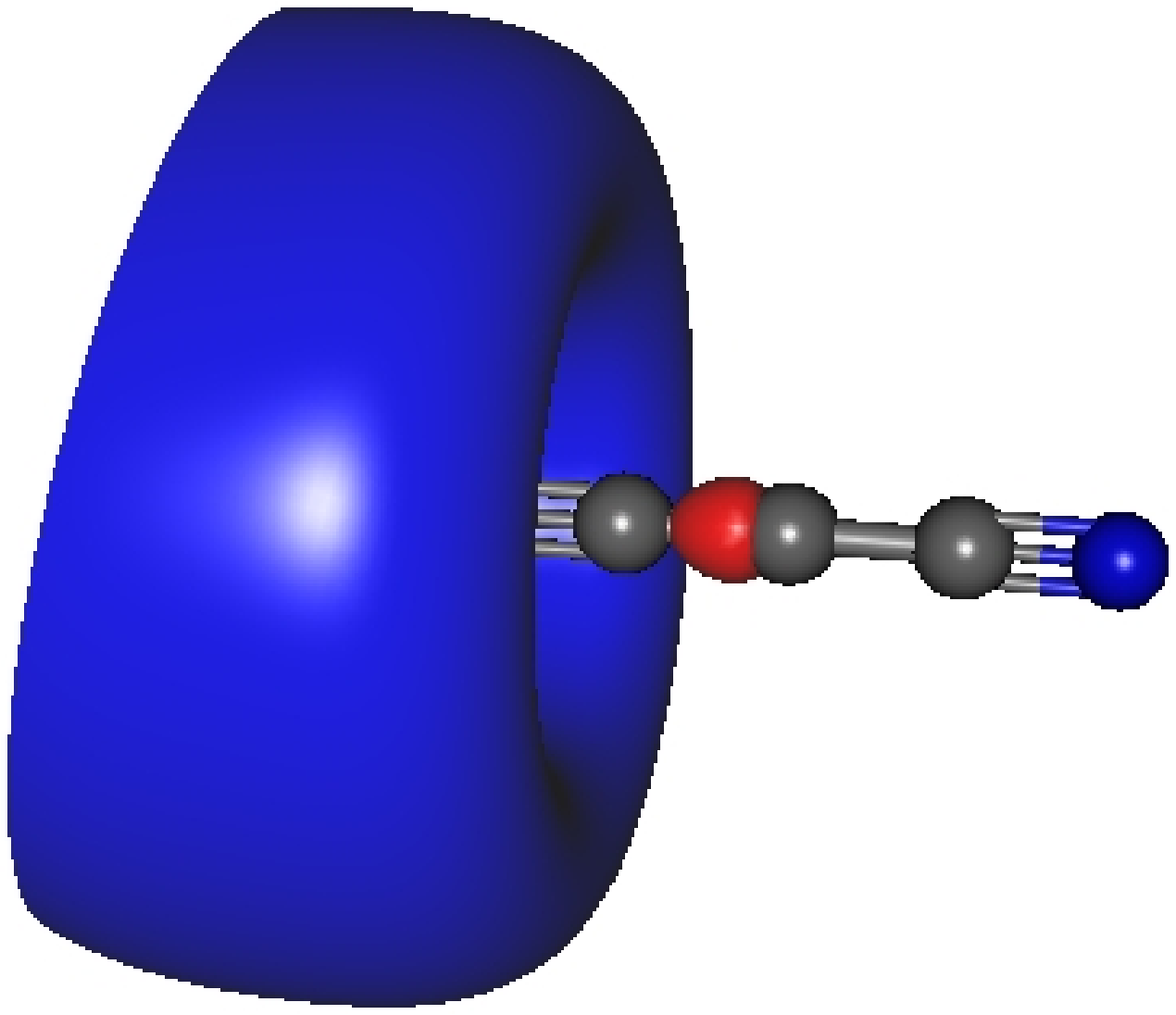}\hspace*{10ex}
  }
  \caption{Degenerate HOMO and HOMO-1 (upper left and right panel, respectively) and LUMO (lower panel)
    of the neutral \ce{C4N^0} quartet ($\tilde{a} ^4 \Sigma^{-}$).}
    \label{fig:homo-lumo-c4n-quadruplet}
\end{figure*}

\begin{figure*}
  \centerline{
    \includegraphics[width=0.4\textwidth]{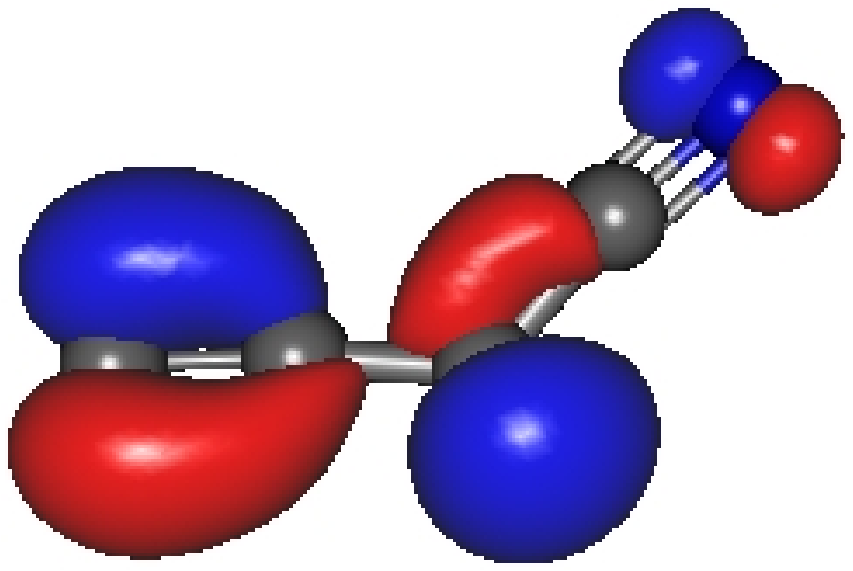}
  }
  \centerline{
    \includegraphics[width=0.4\textwidth]{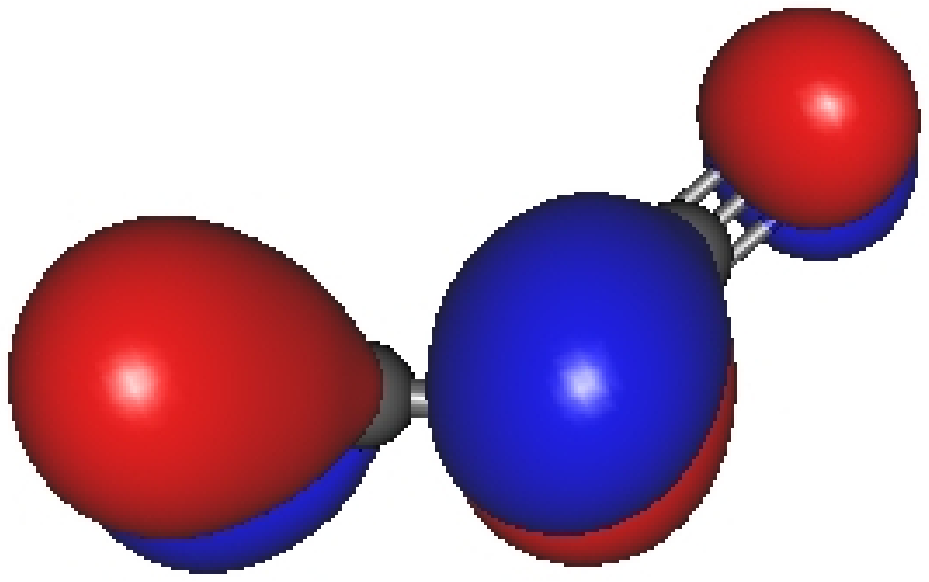}
  }
  \caption{HOMO and LUMO (upper and lower panel, respectively)
    of the bent \ce{C4N-} singlet ($^1 A^\prime$).}
    \label{fig:homo-lumo-c4n-_singlet_bent}
\end{figure*}

\begin{figure*}
    \centerline{
    \includegraphics[width=0.4\textwidth]{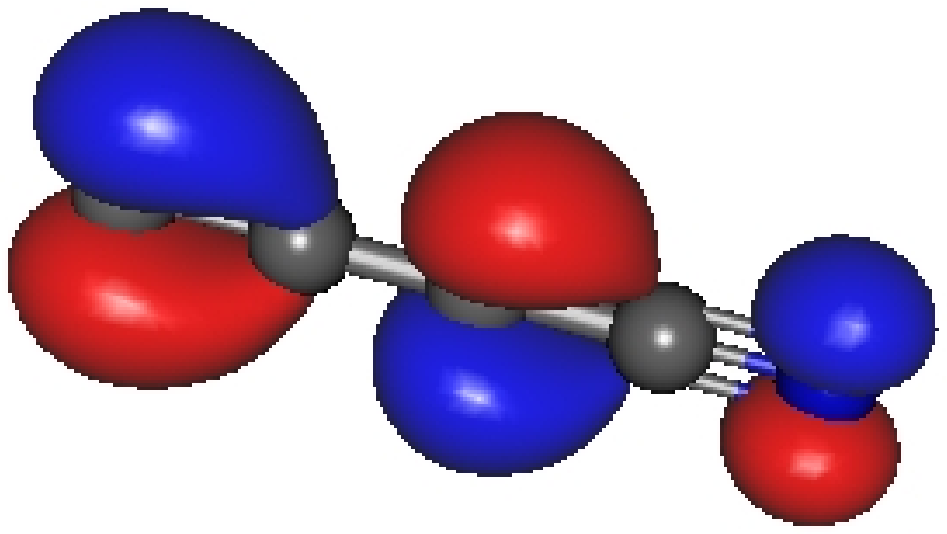}
  }
  \centerline{
    \includegraphics[width=0.4\textwidth]{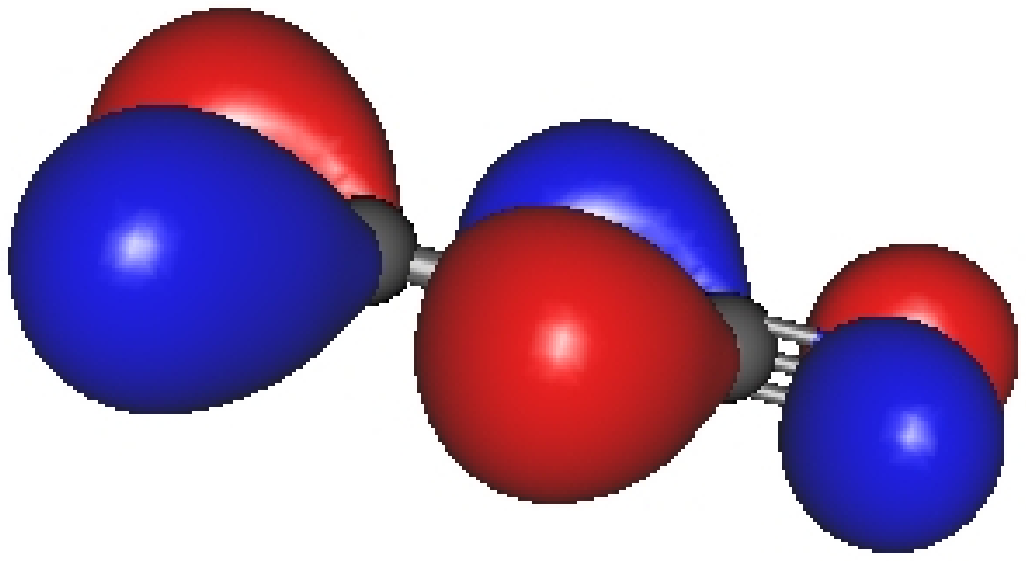}\hspace*{10ex}
    \includegraphics[width=0.4\textwidth]{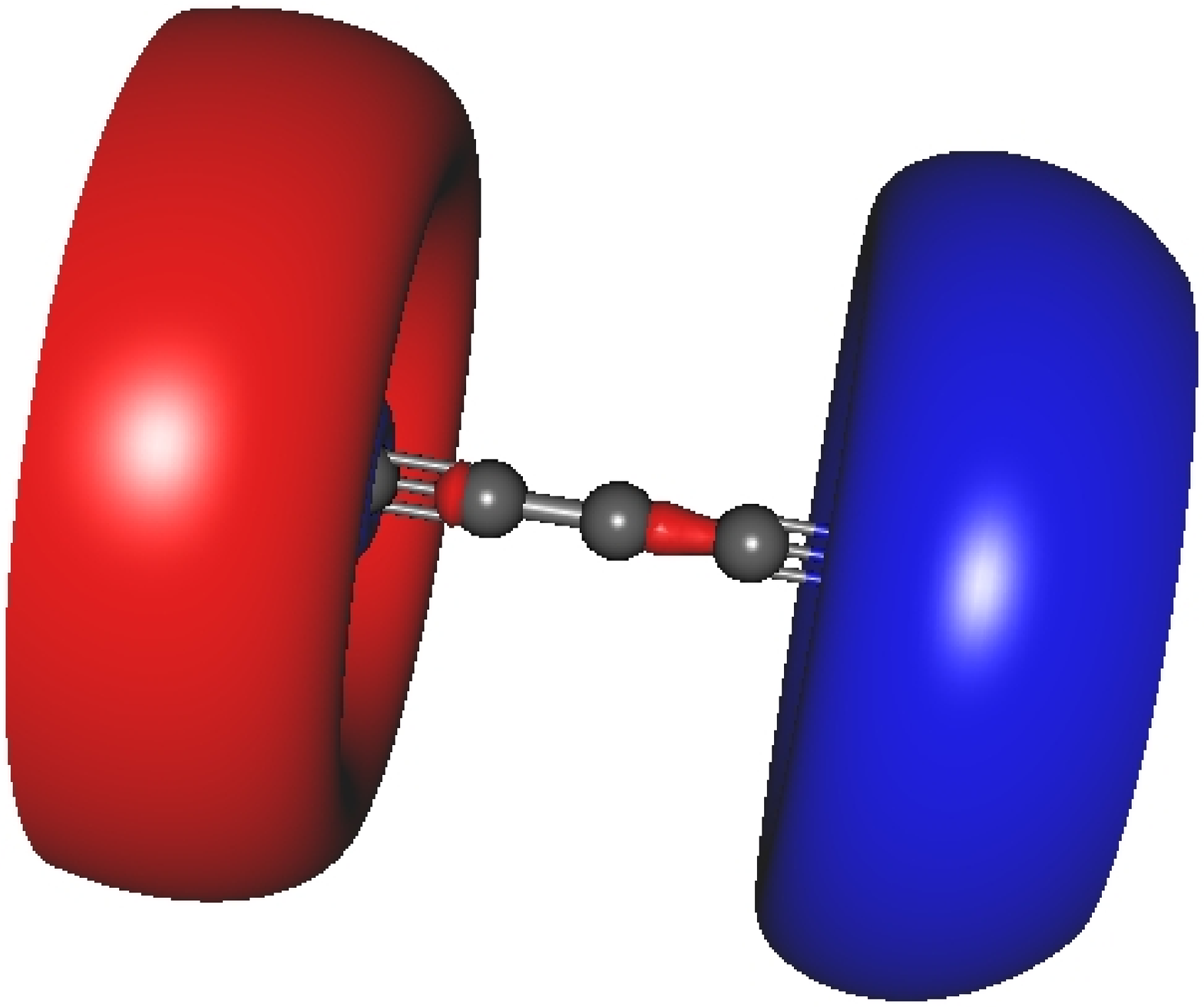}
  }
  \caption{HOMO (upper panel) and nearly degenerate LUMO and LUMO+1
    (lower left and right panel, respectively) of the linear \ce{C4N-} singlet ($^1 \Sigma^{-}$).}
    \label{fig:lumo-lumo+1-deg-c4n-_singlet_linear}
\end{figure*}

\begin{figure*}
  \centerline{
    \includegraphics[width=0.4\textwidth]{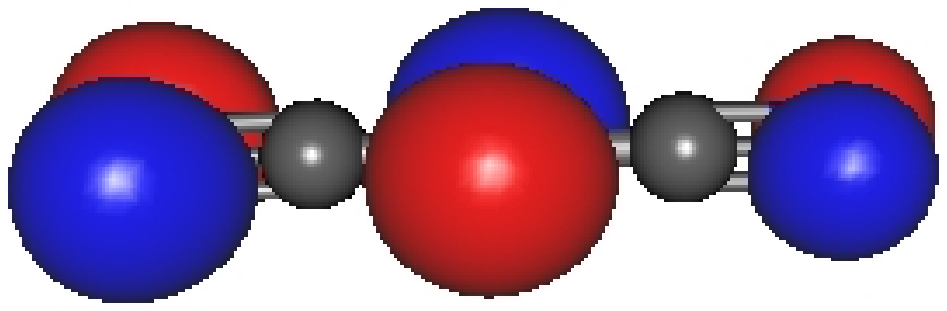}\hspace*{10ex}
    \includegraphics[width=0.4\textwidth]{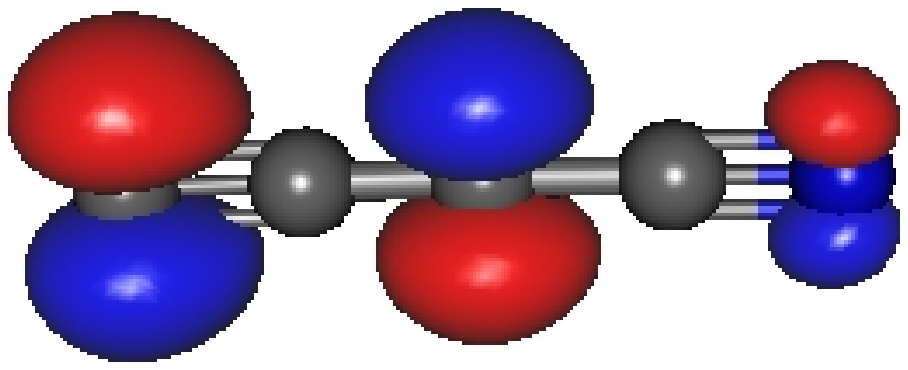}
  }
  \caption{HOMO and LUMO (left and right panel, respectively) of the \ce{C4N+} triplet ($^3 \Sigma^{+}$).}
    \label{fig:homo-lumo-c4n+_triplet}
\end{figure*}

\begin{table*}
  \centering
  \caption{Values of the vertical and adiabatic doublet-quartet splitting
    ($\Delta_{DQ}^{0}\left(\mathbf{R}_{D,Q}^{0}\right) \equiv \mathcal{E}^{0}_{Q}\left(\mathbf{R}_{D,Q}^{0}\right) - \mathcal{E}^{0}_{DE}\left(\mathbf{R}_{D,Q}^{0}\right)$ and 
    $\Delta_{DQ}^{ad} \equiv \mathcal{E}^{0}_{Q}\left(\mathbf{R}_{Q}^{0}\right) - \mathcal{E}^{0}_{D}\left(\mathbf{R}_{D}^{0}\right)$,
    respectively) computed without and with corrections due to zero point motion at geometries
    ($\mathbf{R}^{0}_x, x=D, Q$) optimized using the largest Pople basis sets
    6-311++G(3df, 3pd) and several exchange-correlation functionals.}
  \label{table:Delta-neutral-pbe0-m062x}
  \begin{tabular}{ccccc}
    \hline
                                                      &             & B3LYP & PBE0  & M06-2X \\
    \hline
    $\Delta_{DQ}^{0}\left(\mathbf{R}_{D}^{0}\right)$     & uncorrected & 1.167 & 0.921 & 0.979 \\
                                                      &  corrected  & 1.182 & 0.946 & 0.971 \\
                                                      &             &       &       &       \\
   $\Delta_{DQ}^{0}\left(\mathbf{R}_{Q}^{0}\right)$      & uncorrected & 1.062 & 0.742 & 0.777 \\
                                                      &  corrected  & 1.076 & 0.766 & 0.770 \\
                                                      &             &       &       &       \\
                               $\Delta_{DQ}^{0, ad}$    & uncorrected & 1.167 & 0.839 & 0.889 \\  
                                                      &  corrected  & 1.182 & 0.864 & 0.881 \\
    \hline
  \end{tabular}
\end{table*}
\begin{table*}
  \centering
  \caption{Values of the vertical 
    ($\Delta_{bS,T}^{-}\left(\mathbf{R}_{T}^{-}\right) \equiv 
    \mathcal{E}^{-}_{T}\left(\mathbf{R}_{T}^{-}\right) - \mathcal{E}^{-}_{bS}\left(\mathbf{R}_{T}^{-}\right)$
    $\Delta_{bS,T}^{-}\left(\mathbf{R}_{bS}^{-}\right) \equiv
    \mathcal{E}^{-}_{T}\left(\mathbf{R}_{bS}^{-}\right) - \mathcal{E}^{-}_{bS}\left(\mathbf{R}_{bS}^{-}\right)$
    $\Delta_{lS,T}^{-}\left(\mathbf{R}_{lS}^{-}\right) \equiv
    \mathcal{E}^{-}_{T}\left(\mathbf{R}_{lS}^{-}\right) - \mathcal{E}^{-}_{lS}\left(\mathbf{R}_{lS}^{-}\right)$)
    and adiabatic singlet-triplet splitting
    computed without and with corrections due to zero point motion using geometries
    ($\mathbf{R}^{-}_x, x=T, bS, lS$) optimized using the largest Pople basis sets
    6-311++G(3df, 3pd) and several exchange-correlation functionals.}
  \label{table:Delta-anion-pbe0-m062x}
  \begin{tabular}{ccccc}
    \hline
                                                         &             & B3LYP & PBE0  & M06-2X    \\
    \hline
       $ -\Delta_{bS,T}^{-}\left(\mathbf{R}_{T}^{-}\right)$ & uncorrected & 0.785  & 0.917 &  0.615  \\    
                                                         & corrected   & 0.791  & 0.923 &  0.615  \\    
                                                         &             &        &       &         \\
      $ -\Delta_{bS,T}^{-}\left(\mathbf{R}_{bS}^{-}\right)$ & uncorrected & 0.103  & 0.224 & -0.013  \\
                                                         &   corrected & 0.109  & 0.230 & -0.013  \\
                                                         &             &        &       &         \\
                                  $ -\Delta_{bS,T}^{-, ad}$ & uncorrected & 0.527  & 0.661 &  0.503  \\    
                                                         &  corrected  & 0.533  & 0.667 &  0.503  \\    
    \hline
  \end{tabular}
\end{table*}
\begin{table*}
  \centering
  \label{table:Delta-anion-pascoli}
  \caption{Values of adiabatic anion singlet-triplet splittings obtained within unrestricted ab initio methods
    with zero-point motion corrections. Values in italics are deduced from \citet{Pascoli:99}.}
         \begin{tabular}{cccc}
    \hline
    Method   & Basis set             & $-\Delta_{T,bS}^{-, ad}$ &$-\Delta_{T,lS}^{-, ad}$ \\
    \hline
     B3LYP   & 6-311G$^\ast$          & \emph{0.57}           & \emph{0.81}          \\
     B3LYP   & aug-cc-pVTZ           & \emph{0.53}           & \emph{0.78}          \\
     B3LYP   & 6-311++G(3df, 3pd)    & 0.533                 & 0.791                \\
             &                       &                       &                      \\
    QCISD    & 6-311G$^\ast$          & \emph{0.40}           & \emph{0.87}          \\
    QCISD    & 6-311++G(3df, 3pd)    & 0.374                 & 0.824                \\
             &                       &                       &                      \\
    QCISD(T) & 6-311G$^\ast$          & \emph{0.27}           & \emph{0.72}          \\
    QCISD(T) & 6-311++G(3df, 3pd)    & 0.243                 & 0.671                \\
             &                       &                       &                      \\
    CCSD    & 6-311G$^\ast$          & \emph{0.39}           & \emph{0.87}          \\
    CCSD    & 6-311++G(3df, 3pd)    & 0.367                 & 0.822                \\
             &                       &                       &                      \\
    CCSD(T) & 6-311G$^\ast$          & \emph{0.25}           & \emph{0.71}          \\
    CCSD(T) & 6-311++G(3df, 3pd)    & 0.234                 & 0.653                \\    
    \hline
  \end{tabular}   
\end{table*}

\begin{table*}
  \centering
  \caption{Values of vertical and adiabatic cation singlet-triplet splitting
    ($\Delta_{ST}^{+}\left(\mathbf{R}_{S,T}^{+}\right) \equiv \mathcal{E}^{+}_{T}\left(\mathbf{R}_{S,T}^{+}\right) - \mathcal{E}^{+}_{S}\left(\mathbf{R}_{S,T}^{+}\right)$)
    and $\Delta_{ST}^{+, ad} \equiv \mathcal{E}^{+}_{T}\left(\mathbf{R}_{T}^{+}\right) - \mathcal{E}^{+}_{S}\left(\mathbf{R}_{S}^{+}\right)$, respectively)
    computed without and with corrections due to zero point motion with geometries $\mathbf{R}_{S,T}^{+}$ optimized 
     using several exchange-correlation functionals and 6-311++G(3df, 3pd) basis sets.}
  \label{table:Delta-cation-pbe0-m062x}
  \begin{tabular}{ccccc}
    \hline
                                                      &             & B3LYP & PBE0  & M06-2X \\
    \hline
    $\Delta_{ST}^{+}\left(\mathbf{R}_{S}^{+}\right)$     & uncorrected & 1.517 & 1.250 & 1.451 \\ 
                                                      &  corrected  & 1.489 & 1.251 & 1.441 \\
                                                      &             &       &       &       \\
   $\Delta_{ST}^{+}\left(\mathbf{R}_{T}^{+}\right)$      & uncorrected & 1.046 & 0.796 & 0.965 \\
                                                      &  corrected  & 1.018 & 0.797 & 0.955 \\ 
                                                      &             &       &       &       \\
                               $\Delta_{ST}^{+, ad}$     & uncorrected & 1.311 & 1.052 & 1.247 \\
                                                       &  corrected  & 1.283 & 1.054 & 1.236 \\
    \hline
  \end{tabular}
\end{table*}

\clearpage

\begin{table*}
  \centering
  \caption{Longitudinal (nonvanishing $A$ only for bent anion singlet) 
    and perpendicular ($B=C$ except for the bent anion singlet) 
    rotational constants of the \ce{C4N} chains investigated in this paper computed by using methods indicated in the second column. }
  \label{table:B-pbe0-m062x}
  \begin{tabular}{rrrll}
    \hline
    Species         &  Method                                         & $A$\,(GHz) &       $B$\,(GHz) &      $C$\,(GHz) \\
    \hline
    neutral doublet &  UB3LYP/6-311++G(3df, 3pd)  &                        &         2.44239  & \\
                    &  UPBE0/6-311++G(3df, 3pd)   &                        &         2.44128  & \\
                    &  UM06-2X/6-311++G(3df, 3pd) &   &         2.43646  & \\
                    &  UB2GP-PLYP/6-311++G(3df, 3pd)         &                        &         2.44310  & \\
                    &  UHF/3-21G \citep{Pauzat:91}&                        &         2.4075   & \\
                    &  UHF/svp \citep{Pauzat:91}                      &                        &         2.3963   & \\
                    &                                                 &                        &                  & \\
    neutral quartet &  UB3LYP/6-311++G(3df, 3pd)  &                        &         2.46635  & \\
                    &  UPBE0/6-311++G(3df, 3pd)   &                        &         2.46586  & \\
                    &  UM06-2X/6-311++G(3df, 3pd) &   &         2.46171  & \\
                    &                            &                        &                    &                    \\
    anion triplet   &  UB3LYP/6-311++G(3df, 3pd)  &                        &         2.42267  & \\
                    &  UPBE0/6-311++G(3df, 3pd)   &                        &         2.42220  & \\
                    &  UM06-2X/6-311++G(3df, 3pd) &   &         2.42084  & \\
                    &  UB2GP-PLYP/6-311++G(3df, 3pd)          &                        &         2.42361  & \\
                    &                            &                        &                    &                    \\
bent anion singlet  &  RB3LYP/6-311++G(3df, 3pd)  &         56.30860       &         2.82435  & 2.68945 \\
                    &  RPBE0/6-311++G(3df, 3pd )  &         54.50451       &         2.84356  & 2.70256 \\
                    &  RM06-2X/6-311++G(3df, 3pd)& 46.19743 &   2.92536  & 2.75115 \\
                    &                            &                        &                    &                    \\
cation singlet      &  RB3LYP/6-311++G(3df, 3pd)  &                        &         2.44330  & \\
                    &  RPBE0/6-311++G(3df, 3pd)   &                        &         2.44262  & \\
                    &  RM06-2X/6-311++G(3df, 3pd)  &  &         2.44031  & \\
                    &  RB2GP-PLYP/6-311++G(3df, 3pd)         &                        &         2.42933  & \\
                    &                            &                        &                    &                    \\
cation triplet      &  UB3LYP/6-311++G            &                        &         2.47931  & \\
                    &  UPBE0/6-311++G             &                        &         2.47907  &  \\
                    &  UM06-2X/6-311++G(3df, 3pd) &   &         2.47802  &  \\
    \hline
  \end{tabular}
\end{table*}

\clearpage

\begin{table*}
  \centering
  \caption{Values of the dipole momentum $\mathbf{D}$ (field independent basis, debye) at various levels of theory
    indicated in the second column. Notice that the value in italics obtained by \citet{Pauzat:91}
    within the UHF/svp approach is somewhat different 
    from that of our calculations at the same level of theory.}
  \label{table:D-full}
  \begin{tabular}{ccrrrr}
    \hline
    Species              &          Method              & $D_{X}$ & $D_{Y}$ & $D_{Z}$ & $D_{total}$ \\
    \hline                            
    neutral doublet      &     B3LYP/6-311++G(3df, 3pd) &  0.0000 &  0.0000 &  0.3347 & 0.3347 \\
                         &     B3LYP/aug-cc-pVTZ        &  0.0000 &  0.0000 &  0.3393 & 0.3393 \\
                         &  UCCSD(T)/6-311++G(3df, 3pd) &  0.0000 &  0.0000 &  0.0907 & 0.0907 \\
                         &  UCCSD(T)/aug-cc-pvtz        &  0.0000 &  0.0000 &  0.0990 & 0.0990 \\
                         & ROCCSD(T)/6-311++G(3df, 3pd) &  0.0000 &  0.0000 &  0.4512 & 0.4512 \\ 
                         & ROCCSD(T)/aug-cc-pVTZ        &  0.0000 &  0.0000 &  0.4436 & 0.4436 \\
                         & UHF/3-21g                    &  0.0000 &  0.0000 &  0.0544 & 0.0544 \\
                         & UHF/svp                      &  0.0000 &  0.0000 &  0.1119 & 0.1119 \\
                         & UHF/svp \citep{Pauzat:91}    &  0.0000 &  0.0000 & \emph{0.14} & \emph{0.14} \\
                         & UHF/6-311++G(3df, 3pd)       &  0.0000 &  0.0000 &  0.0587 & 0.0587 \\
                         & UHF/aug-cc-pvtz              &  0.0000 &  0.0000 &  0.0654 & 0.0654 \\
                         & ROHF/3-21g                   &  0.0000 &  0.0000 &  0.5486 & 0.5486 \\
                         & ROHF/svp                     &  0.0000 &  0.0000 &  0.6216 & 0.6216 \\
                         & ROHF/6-311++G(3df, 3pd)      &  0.0000 &  0.0000 &  0.7821 & 0.7821 \\
                         & ROHF/aug-cc-pVTZ             &  0.0000 &  0.0000 &  0.7781 & 0.7781 \\
                         &                              &         &         &   \\
    neutral quartet      &     B3LYP/6-311++G(3df, 3pd) &  0.0000 &  0.0000 & 3.4628 & 3.4628 \\    
                         &     B3LYP/aug-cc-pVTZ        &  0.0000 &  0.0000 & 3.4586 & 3.4586 \\
                         & UCCSD(T)/6-311++G(3df, 3pd)  &  0.0000 &  0.0000 & 3.2558 & 3.2558 \\
                         & ROCCSD(T)/6-311++G(3df, 3pd) &  0.0000 &  0.0000 & 4.5003 & 4.5003 \\
                         & ROCCSD(T)/aug-cc-pVTZ        &  0.0000 &  0.0000 & 4.4940 & 4.4940 \\
                         &    UHF/3-21G                 &  0.0000 &  0.0000 & 2.9749 & 2.9749 \\
                         &    UHF/svp                   &  0.0000 &  0.0000 & 3.1581 & 3.1581 \\
                         &  UHF/6-311++G(3df, 3pd)      &  0.0000 &  0.0000 & 3.2558 & 3.2558 \\
                         &  UHF/aug-ccpVTZ              &  0.0000 &  0.0000 & 3.2479 & 3.2479 \\
                         & ROHF/3-21G                   &  0.0000 &  0.0000 & 3.8729 & 3.8729 \\
                         & ROHF/svp                     &  0.0000 &  0.0000 & 4.2865 & 4.2865 \\
                         & ROHF/6-31++G(3df, 3pd)       &  0.0000 &  0.0000 & 4.5003 & 4.5003 \\
                         & ROHF/aug-cc-pvtz             &  0.0000 &  0.0000 & 4.4940 & 4.4940 \\
                         &                              &         &         &   \\
    anion triplet        &     B3LYP/6-311++G(3df, 3pd) &  0.0000 &  0.0000 & 2.9398 & 2.9398  \\
                         &     B3LYP/aug-cc-pVTZ        &  0.0000 &  0.0000 & 2.9340 & 2.9400  \\  
                         &  UCCSD(T)/6-311++G(3df, 3pd) &  0.0000 &  0.0000 & 4.4930 & 4.4930  \\
                         & ROCCSD(T)/6-311++G(3df, 3pd) &  0.0000 &  0.0000 & 2.2379 & 2.2379  \\
                         & ROCCSD(T)/aug-cc-pVTZ        &  0.0000 &  0.0000 & 2.2447 & 2.2447  \\
                         &     UHF/3-21G                &  0.0000 &  0.0000 & 4.5640 & 4.5640  \\
                         &     UHF/svp                  &  0.0000 &  0.0000 & 4.4277 & 4.4277  \\
                         &     UHF/6-311++G(3df, 3pd)   &  0.0000 &  0.0000 & 4.4930 & 4.4930  \\
                         &     UHF/aug-cc-pVTZ          &  0.0000 &  0.0000 & 4.5002 & 4.5002  \\
                         &    ROHF/3-21G               &  0.0000 &  0.0000 & 2.4519 & 2.4519  \\
                         &    ROHF/svp                  &  0.0000 &  0.0000 & 2.3422 & 2.3422  \\
                         &    ROHF/6-311++G(3df, 3pd)   &  0.0000 &  0.0000 & 2.2379 & 2.2379  \\
                         &    ROHF/aug-cc-pVTZ          &  0.0000 &  0.0000 & 2.5956 & 2.5956  \\
                         &                              &         &         &         &         \\ 
    cation triplet       &     B3LYP/6-311++G(3df, 3pd) &  0.0000 &  0.0000  &  4.8254 & 4.8254 \\
                         &     B3LYP/aug-cc-pVTZ        &  0.0000 &  0.0000  &  4.8236 & 4.8236 \\
                         &  UCCSD(T)/6-311++G(3df, 3pd) &  0.0000 &  0.0000  &  4.1447 & 4.1447 \\
                         & ROCCSD(T)/6-311++G(3df, 3pd) &  0.0000 &  0.0000  &  6.0065 & 6.0065 \\
                         & ROCCSD(T)/aug-cc-pVTZ        &  0.0000 &  0.0000  &  6.0002 & 6.0002 \\
                         &     UHF/3-21G                &  0.0000 &  0.0000  &  4.1310 & 4.1310 \\  
                         &     UHF/svp                  &  0.0000 &  0.0000  &  4.8990 & 4.8990 \\  
                         &     UHF/6-311++G(3df, 3pd)   &  0.0000 &  0.0000  &  4.8682 & 4.8682 \\  
                         &     UHF/aug-cc-pVTZ          &  0.0000 &  0.0000  &  4.8457 & 4.8457 \\  
                         &    ROHF/3-21G                &  0.0000 &  0.0000  &  9.5571 & 9.5571 \\  
                         &    ROHF/svp                  &  0.0000 &  0.0000  & 11.0789 &11.0789 \\  
                         &    ROHF/6-311++G(3df, 3pd)   &  0.0000 &  0.0000  & 10.8968 &10.8968 \\   
                         &    ROHF/aug-cc-pVTZ          &  0.0000 &  0.0000  &  6.0002 & 6.0002 \\   
    \hline
  \end{tabular}
\end{table*}

\begin{table*}
  \footnotesize
  \centering
  \caption{Values of the quadrupole momentum $\mathbf{Q}$ (field independent basis, debye-angstrom)
    of the \ce{C4N} chains investigated in this paper obtained using geometries optimized as
    indicated in the second column.}
  \label{table:Q}
  \begin{tabular}{cccccccc}
    \hline
    Species              &          Method              & $Q_{xx}$    & $Q_{yy}$   &  $Q_{zz}$    & $Q_{xy}$    &     $Q_{xz}$    &  $Q_{yz}$  \\
    \hline                            
    neutral doublet      &     B3LYP/6-311++G(3df, 3pd) & -26.3541   & -27.9983   &  -42.2421  &  0.0000   &  0.0000   &  0.0000  \\
                         &     B3LYP/aug-cc-pVTZ        & -26.3443   & -27.9421   &  -42.2635  &  0.0000   &  0.0000   &  0.0000  \\
                         &  UCCSD(T)/6-311++G(3df, 3pd) & -26.4146   & -28.2057   &  -42.2981  &  0.0000   &  0.0000   &  0.0000  \\
                         &  UCCSD(T)/aug-cc-pvtz        & -26.3955   & -28.1379   &  -42.3180  &  0.0000   &  0.0000   &  0.0000  \\
                         & ROCCSD(T)/6-311++G(3df, 3pd) & -28.4353   & -26.8169   &  -41.6984  &  0.0000   &  0.0000   &  0.0000  \\
                         & ROCCSD(T)/aug-cc-pVTZ        & -26.8005   & -28.3699   &  -41.7230  &  0.0000   &  0.0000   &  0.0000  \\
                         & UHF/3-21g                    & -28.2685   & -26.4478   &  -41.7598  &  0.0000   &  0.0000   &  0.0000  \\
                         & UHF/svp                      & -26.5232   & -28.3367   &  -42.6564  &  0.0000   &  0.0000   &  0.0000  \\
                         & UHF/6-311++G(3df, 3pd)       & -26.4489   & -28.2494   &  -42.2687  &  0.0000   &  0.0000   &  0.0000  \\ 
                         & UHF/aug-cc-pvtz              & -28.1845   & -26.4327   &  -42.2852  &  0.0000   &  0.0000   &  0.0000  \\ 
                         &                              &            &            &            &           &           &          \\
    neutral quartet      &     B3LYP/6-311++G(3df, 3pd) & -27.5520   & -27.5520   &  -30.7287  &  0.0000   &  0.0000   &  0.0000  \\
                         &     B3LYP/aug-cc-pVTZ        & -27.5250   & -27.5250   &  -30.7352  &  0.0000   &  0.0000   &  0.0000  \\
                         &  UCCSD(T)/6-311++G(3df, 3pd) & -27.6065   & -27.6065   &  -30.5244  &  0.0000   &  0.0000   &  0.0000  \\
                         & ROCCSD(T)/6-311++G(3df, 3pd) & -28.0387   & -28.0387   &  -29.5375  &  0.0000   &  0.0000   &  0.0000  \\
                         & ROCCSD(T)/aug-cc-pVTZ        & -27.9995   & -27.9995   &  -29.5470  &  0.0000   &  0.0000   &  0.0000  \\
                         &                              &            &            &            &           &           &          \\
    anion triplet        &     B3LYP/6-311++G(3df, 3pd) &  -31.7023  & -31.7023   &  -71.4631  &  0.0000   &  0.0000   &  0.0000  \\
                         &     B3LYP/aug-cc-pVTZ        &  -31.6879  & -31.6879   &  -71.5144  &  0.0000   &  0.0000   &  0.0000  \\
                         &  UCCSD(T)/6-311++G(3df, 3pd) &  -31.8018  & -31.8018   &  -71.7508  &  0.0000   &  0.0000   &  0.0000  \\
                         & ROCCSD(T)/6-311++G(3df, 3pd) &  -32.1142  & -32.1142   &  -70.2281  &  0.0000   &  0.0000   &  0.0000  \\
                         & ROCCSD(T)/aug-cc-pVTZ        &  -32.0834  & -32.0834   &  -70.2791  &  0.0000   &  0.0000   &  0.0000  \\
                         &                              &            &            &            &           &           &          \\
    bent anion singlet   &     B3LYP/6-311++G(3df, 3pd) &  -63.8640  & -36.8339   &  -30.3058  &   1.4754   &   0.0000  &  0.0000  \\
                         &     B3LYP/aug-cc-pVTZ        &  -63.8640  & -36.8339   &  -30.3058  &   1.4754   &   0.0000  &  0.0000  \\ 
                         &  RCCSD(T)/6-311++G(3df, 3pd) &  -62.9911  & -37.1863   &  -30.6481  &   1.1594   &   0.0000  &  0.0000  \\
                         &  RCCSD(T)/aug-cc-pVTZ        &  -63.0000  & -37.1600   &  -30.6336  &   1.1690   &   0.0000  &  0.0000  \\
                         &                              &            &            &            &            &           &          \\   
    linear anion singlet &     B3LYP/6-311++G(3df, 3pd) &  -29.8650  & -33.8601   &  -71.3765  &   0.0000   &   0.0000  &  0.0000  \\
                         &     B3LYP/aug-cc-pVTZ        &  -29.9174  & -33.8018   &  -71.4570  &   0.0000   &   0.0000  &  0.0000  \\
                         &  RCCSD(T)/6-311++G(3df, 3pd) &  -30.2092  & -34.2094   &  -70.3492  &   0.0000   &   0.0000  &  0.0000  \\
                         &  RCCSD(T)/aug-cc-pVTZ        &  -34.1124  & -30.2533   &  -70.4022  &   0.0000   &   0.0000  &  0.0000  \\
                         &                              &            &            &            &            &           &          \\   
    cation singlet       &     B3LYP/6-311++G(3df, 3pd) &  -23.7770  & -23.7770   &  -15.6956  &  0.0000    &   0.0000  &  0.0000  \\
                         &     B3LYP/aug-cc-pVTZ        &  -23.7770  & -23.7770   &  -15.6957  &  0.0000    &   0.0000  &  0.0000  \\
                         &  RCCSD(T)/6-311++G(3df, 3pd) &  -23.7366  & -23.7366   &  -15.7048  &  0.0000    &   0.0000  &  0.0000  \\
                         &  RCCSD(T)/aug-cc-pVTZ        &  -24.3542  & -24.3542   &  -14.1423  &  0.0000    &   0.0000  &  0.0000  \\
                         &                              &            &            &            &            &           &          \\
    cation triplet       &     B3LYP/6-311++G(3df, 3pd) &  -23.4092  & -24.8186   &  -6.0974   &  0.0000    &   0.0000  &  0.0000  \\
                         &     B3LYP/aug-cc-pVTZ        &  -24.7575  & -23.3930   &  -6.1030   &  0.0000    &   0.0000  &  0.0000  \\
                         &  UCCSD(T)/6-311++G(3df, 3pd) &  -25.0731  & -23.7379   &  -7.4195   &  0.0000    &   0.0000  &  0.0000  \\
                         & ROCCSD(T)/6-311++G(3df, 3pd) &  -24.0058  & -25.3116   &  -5.2179   &  0.0000    &   0.0000  &  0.0000  \\
                         & ROCCSD(T)/aug-cc-pVTZ        &  -25.2346  & -23.9786   &  -5.2274   &  0.0000    &   0.0000  &  0.0000  \\
    \hline                                                                                                                         
  \end{tabular}
\end{table*}

\begin{table*}
  \tiny 
  \centering
  \caption{Values of the higher vibrational frequencies (in cm$^{-1}$) of the presently investigated molecular species obtained via B3LYP/6-311++G(3df, 3pd) calculations.}
  \label{table:vibrations}
        { 
  \begin{tabular}{@{}r@{\:}r@{\:}r@{\:}r@{\:}r@{\:}r@{\:}r@{\:}r@{\:}r@{}}        
    \hline
Description & \ce{C4N^0} doublet &  \ce{C4N^0} quartet & bent \ce{C4N-} singlet & \ce{C4N-} triplet & \ce{C4N+} singlet & \ce{C4N+} triplet & \ce{HC4N} singlet  & \ce{HC4N} triplet \\
\hline
symmetric stretch (breath.) &          752.62  &              765.66  &              827.08  &         753.17  &         756.09  &         775.07  &          847.97  &       754.26    \\
out-of-phase \ce{C1C2}---\ce{C4N} stretch &         1421.85  &             1559.49  &             1320.77  &        1475.77  &        1418.37  &        1579.14  &         1376.99  &      1577.41    \\
in-phase \ce{C1C2}---\ce{C4N} stretch &         1989.43  &             1753.86  &             1898.25  &        1835.45  &        2198.39  &        1956.48  &         1997.43  &      1762.66    \\
CN stretch &         2181.87  &             2071.92  &             2149.62  &        2013.49  &        2325.12  &        2138.33  &         2156.55  &      2029.44    \\
CH stretch &  ---               &   ---                  &   ---                  &    ---            &    ---            &      ---          &         3449.65  &      3446.77    \\
    \hline
  \end{tabular}
  }
\end{table*}

\begin{table*}
  \centering
  \caption{Values of the vertical and adiabatic doublet-triplet electron attachment energies
    ($EA_{TD}^{vert}\left(\mathbf{R}\right) \equiv \mathcal{E}^{0}_{D}\left(\mathbf{R}\right) - \mathcal{E}^{-}_{T}\left(\mathbf{R}\right)$
    and
    $EA_{TD}^{ad} \equiv \mathcal{E}^{0}_{D}\left(\mathbf{R}_{D}^{0}\right) - \mathcal{E}^{-}_{T}\left(\mathbf{R}_{T}^{-}\right)$,
    respectively) computed without and with corrections due to zero point motion
    using the neutral doublet ($\mathbf{R} = \mathbf{R}_{D}^{0}$) and
    anion triplet ($\mathbf{R} = \mathbf{R}_{T}^{-}$)
    B3LYP/6-311++G(3df, 3pd) optimum geometries.}
  \label{table:EA-triplet@doublet-b3lyp}
  \begin{tabular}{cccccc}
    \hline
                                                 &             & EOM-ROCCSD &  B3LYP &  LC-BLYP& LC-$\omega$PBE \\
    \hline                                                                                
 $EA_{TD}^{vert}\left(\mathbf{R}_{D}^{0}\right)$ & uncorrected & 3.027      & 3.217  &   3.479 & 3.514   \\    
                                                 &  corrected  & 3.017      & 3.207  &   3.469 & 3.504   \\    
                                                 &             &            &        &         &         \\
 $EA_{TD}^{vert}\left(\mathbf{R}_{T}^{-}\right)$ & uncorrected & 3.199      & 3.360  &   3.670 & 3.690   \\    
                                                 &  corrected  & 3.189      & 3.350  &   3.659 & 3.679   \\
                                                 &             &            &        &         &         \\
                                  $EA_{TD}^{ad}$ & uncorrected & 3.109      & 3.285  &   3.497 & 3.545   \\    
                                                 &  corrected  & 3.099      & 3.274  &   3.486 & 3.534   \\    
    \hline
  \end{tabular}
\end{table*}

\begin{table*}
  \centering
  \caption{Values of the vertical and adiabatic doublet-triplet electron attachment energies
    ($EA_{TD}^{vert}\left(\mathbf{R}\right) \equiv \mathcal{E}^{0}_{D}\left(\mathbf{R}\right) - \mathcal{E}^{-}_{T}\left(\mathbf{R}\right)$
    and
    $EA_{TD}^{ad} \equiv \mathcal{E}^{0}_{D}\left(\mathbf{R}_{D}^{0}\right) - \mathcal{E}^{-}_{T}\left(\mathbf{R}_{T}^{-}\right)$,
    respectively) computed without and with corrections due to zero point motion using
    the neutral doublet $\mathbf{R}_{D}^{0}$ and anion triplet $\mathbf{R}_{T}^{-}$
    geometries optimized within B3LYP/6-311++G(3df, 3pd) and PBE0/6-311++G(3df, 3pd).}
  \label{table:EA-triplet@doublet-pbe0}
  \begin{tabular}{cccccccccc}
    \hline
                                                 &             &  B3LYP & PBE0 & EOM-ROCCSD@B3LYP & EOM-ROCCSD@PBE0  \\
    \hline
 $EA_{TD}^{vert}\left(\mathbf{R}_{D}^{0}\right)$ & uncorrected & 3.217  & 3.288 &        3.027    & 3.006      \\    
                                                 &  corrected  & 3.207  & 3.275 &        3.017    & 2.993      \\    
                                                 &             &        &       &                 &            \\
 $EA_{TD}^{vert}\left(\mathbf{R}_{T}^{-}\right)$ & uncorrected & 3.360  & 3.431 &        3.199    & 3.175      \\    
                                                 &  corrected  & 3.350  & 3.418 &        3.189    & 3.162      \\
                                                 &             &        &       &                 &            \\
                                  $EA_{TD}^{ad}$ & uncorrected & 3.285  & 3.355 &        3.109    & 3.086      \\    
                                                 &  corrected  & 3.274  & 3.342 &        3.099    & 3.073      \\    
    \hline
  \end{tabular}
\end{table*}

\begin{table*}
  \centering
  \caption{Values of the vertical and adiabatic doublet-triplet electron attachment $EA$
    computed  without and with corrections due to zero point motion
    using the neutral doublet $\mathbf{R}_{D}^{0}$ and anion triplet $\mathbf{R}_{T}^{-}$ geometries
    optimized by means of several functionals and 6-311++G(3df, 3pd) basis sets.}
  \label{table:EA-optim-various-functionals}
  \begin{tabular}{cccccccccc}
    \hline
                                                 &             &  B3LYP & PBE0  & M06-2X \\
    \hline
 $EA_{TD}^{vert}\left(\mathbf{R}_{D}^{0}\right)$ & uncorrected & 3.217  & 3.288 & 3.304 \\
                                                 &  corrected  & 3.207  & 3.275 & 3.317 \\
                                                 &             &        &       &       \\
 $EA_{TD}^{vert}\left(\mathbf{R}_{T}^{-}\right)$ & uncorrected & 3.360  & 3.431 & 3.273 \\
                                                 &  corrected  & 3.350  & 3.418 & 3.285 \\
                                                 &             &        &       &       \\
                                  $EA_{TD}^{ad}$ & uncorrected & 3.285  & 3.355 & 3.386 \\
                                                 &  corrected  & 3.274  & 3.342 & 3.398 \\
    \hline
  \end{tabular}
\end{table*}

\begin{table*}
  \centering
   \caption{Values of the vertical and adiabatic doublet-singlet ionization energy
    ($IP_{SD}^{vert}\left(\mathbf{R}\right) \equiv \mathcal{E}^{+}_{S}\left(\mathbf{R}\right) - \mathcal{E}^{0}_{D}\left(\mathbf{R}\right)$
     and
     $IP_{SD}^{ad} \equiv \mathcal{E}^{+}_{S}\left(\mathbf{R}_{S}^{+}\right) - \mathcal{E}^{0}_{D}\left(\mathbf{R}_{D}^{0}\right)$,
     respectively) computed without and with with corrections due to zero point motion 
     using the neutral doublet ($\mathbf{R} = \mathbf{R}_{D}^{0}$) and cation singlet
     ($\mathbf{R} =\mathbf{R}_{S}^{+}$) B3LYP/6-311++G(3df, 3pd) optimum geometries.}
  \label{table:IP-singlet@doublet-b3lyp} 
  \begin{tabular}{cccccccccc}
    \hline
                                                    &             & EOM-ROCCSD &  B3LYP&  LC-BLYP &  LC-$\omega$PBE\\
    \hline                                                                                          
    $IP_{SD}^{vert}\left(\mathbf{R}_{D}^{0}\right)$ & uncorrected & 9.802      & 9.812 &  10.258  &  10.226 \\    
                                                    &  corrected  & 9.842      & 9.852 &  10.297  &  10.265 \\    
                                                    &             &            &       &          &         \\
   $IP_{SD}^{vert}\left(\mathbf{R}_{S}^{+}\right)$  & uncorrected & 9.797      & 9.780 &  10.225  &  10.194 \\    
                                                    & corrected   & 9.836      & 9.819 &  10.265  &  10.233 \\    
                                                    &             &            &       &          &         \\
                                     $IP_{SD}^{ad}$  & uncorrected & 9.783      & 9.794 &  10.215  &  10.187 \\    
                                                    &  corrected  & 9.823      & 9.833 &  10.254  &  10.227 \\    
    \hline
  \end{tabular}
\end{table*}

\clearpage

\begin{table*}
  \centering
   \caption{Values of the vertical and adiabatic doublet-singlet ionization energy
    ($IP_{SD}^{vert}\left(\mathbf{R}\right) \equiv \mathcal{E}^{+}_{S}\left(\mathbf{R}\right) - \mathcal{E}^{0}_{D}\left(\mathbf{R}\right)$ and
     $IP_{SD}^{ad} \equiv \mathcal{E}^{+}_{S}\left(\mathbf{R}_{S}^{+}\right) - \mathcal{E}^{0}_{D}\left(\mathbf{R}_{D}^{0}\right)$,
     respectively) computed without and with corrections due to zero point motion using 6-311++G(3df, 3pd) basis sets 
     and the neutral doublet ($\mathbf{R} = \mathbf{R}_{D}^{0}$) and cation singlet ($\mathbf{R} = \mathbf{R}_{S}^{+}$)
     geometries optimized within B3LYP/6-311++G(3df, 3pd) and PBE0/6-311++G(3df, 3pd).}
  \label{table:IP-singlet@doublet-pbe0} 
  \begin{tabular}{cccccccccc}
    \hline
                                                    &             &  B3LYP & PBE0   & EOM-ROCCSD@B3LYP & EOM-ROCCSD@PBE0 \\
    \hline                                                                                                   
    $IP_{SD}^{vert}\left(\mathbf{R}_{D}^{0}\right)$ & uncorrected & 9.812  & 9.874  &   9.802          &  9.805 \\    
                                                    &  corrected  & 9.852  & 9.915  &   9.842          &  9.845 \\    
                                                    &             &        &        &                  &        \\
   $IP_{SD}^{vert}\left(\mathbf{R}_{S}^{+}\right)$  & uncorrected & 9.780  & 9.844  &   9.797          &  9.801 \\    
                                                    & corrected   & 9.819  & 9.884  &   9.836          &  9.841 \\    
                                                    &             &        &        &                  &        \\
                                     $IP_{SD}^{ad}$  & uncorrected & 9.794  & 9.857  &   9.783          &  9.800 \\    
                                                    &  corrected  & 9.833  & 9.897  &   9.823          &  9.840 \\    
    \hline
  \end{tabular}
\end{table*}

\begin{table*}
  \centering
   \caption{Values of the vertical and adiabatic doublet-singlet ionization energy $IP$
     computed without and with corrections due to zero point motion using
     the neutral doublet $\mathbf{R}_{D}^{0}$ and cation singlet $\mathbf{R}_{S}^{+}$ geometries
     optimized by means of several functionals and 6-311++G(3df, 3pd) basis sets.}
  \label{table:IP-singlet@doublet-various-functionals} 
  \begin{tabular}{cccccccccc}
    \hline
                                                    &             &  B3LYP & PBE0   & M06-2X   \\
    \hline                                                                                                   
    $IP_{SD}^{vert}\left(\mathbf{R}_{D}^{0}\right)$     & uncorrected & 9.812  & 9.874  & 9.835   \\    
                                                    &  corrected  & 9.852  & 9.915  & 9.946   \\    
                                                    &             &        &        &         \\
   $IP_{SD}^{vert}\left(\mathbf{R}_{S}^{+}\right)$      & uncorrected & 9.780  & 9.844  & 9.812   \\    
                                                    & corrected   & 9.819  & 9.884  & 9.822   \\    
                                                    &             &        &        &         \\
                                      $IP_{SD}^{ad}$  & uncorrected & 9.794   & 9.857  & 9.822  \\    
                                                   &  corrected  & 9.833   & 9.897  & 9.832  \\    
    \hline
  \end{tabular}
\end{table*}

\begin{table*}
  \footnotesize
  \centering
  \caption{Quadrupole moment $\mathbf{Q}$ (field independent basis, debye-angstrom) of
    the isoelectronic \ce{C4N-} and \ce{HC4N} chains computed as indicated in the second column.}
  \label{table:Q-c4n-_vs_hc4n}
  \begin{tabular}{cccccccc}
    \hline
    Species              &          Method              & $Q_{xx}$    & $Q_{yy}$   &  $Q_{zz}$    & $Q_{xy}$    &     $Q_{xz}$    &  $Q_{yz}$  \\
    \hline                            
    \ce{C4N-} triplet    &     B3LYP/6-311++G(3df, 3pd) &  -34.2070  & -68.9571   &  -31.7034  & -9.6536    &  0.0000   &  0.0000 \\
                         &     B3LYP/aug-cc-pVTZ        &  -34.1971  & -69.0040   &  -31.6890  & -9.6702    &  0.0000   &  0.0000 \\
                         &  UCCSD(T)/6-311++G(3df, 3pd) &  -34.3255  & -69.2273   &  -31.8029  & -9.7128    &  0.0000   &  0.0000 \\
                         & ROCCSD(T)/6-311++G(3df, 3pd) &  -34.5067  & -67.8347   &  -32.1155  & -9.2375    &  0.0000   &  0.0000 \\
                         & ROCCSD(T)/aug-cc-pVTZ        &  -34.4814  & -67.8801   &  -32.0848  & -9.2581    &  0.0000   &  0.0000 \\
                         &                              &            &            &            &            &           &         \\
   \ce{HC4N} triplet     &     B3LYP/6-311++G(3df, 3pd) &  -28.4399  & -27.7588   &   -28.4371 & -0.1693    &  0.0000   &  0.0000 \\
                         &     B3LYP/aug-cc-pVTZ        &  -28.4196  & -27.7593   &   -28.4116 & -0.1574    &  0.0000   &  0.0000 \\
                         &  UCCSD(T)/6-311++G(3df, 3pd) &  -28.4547  & -27.4563   &   -28.5638 & -0.3883    &  0.0000   &  0.0000 \\                          & ROCCSD(T)/6-311++G(3df, 3pd) &  -29.0226  & -28.2790   &   -29.0194 & -0.1844    &  0.0000   &  0.0000 \\
                         & ROCCSD(T)/aug-cc-pVTZ        &  -28.9932  & -28.2743   &   -28.9830 & -0.1694    &  0.0000   &  0.0000 \\ 
\hline
 bent \ce{C4N-} singlet  &     B3LYP/6-311++G(3df, 3pd) &  -63.8640  & -36.8339   &  -30.3058  &   1.4754   &   0.0001  &  0.0001 \\
                         &     B3LYP/aug-cc-pVTZ        &  -63.8640  & -36.8339   &  -30.3058  &   1.4754   &   0.0001  &  0.0001 \\ 
                         &  RCCSD(T)/6-311++G(3df, 3pd) &  -62.9911  & -37.1863   &  -30.6481  &   1.1594   &   0.0000  &  0.0001  \\
                         &  RCCSD(T)/aug-cc-pVTZ        &  -63.0000  & -37.1600   &  -30.6336  &   1.1690   &   0.0001  &  0.0001  \\
                        &                              &            &            &            &            &           &          \\
    \ce{HC4N} singlet    &     B3LYP/6-311++G(3df, 3pd) &  -27.3591  & -30.3776   &  -27.1872  & -4.3257    &   0.0000  &  0.0000 \\
                         &     B3LYP/aug-cc-pVTZ        &  -27.3407  & -30.3437   &  -27.1738  & -4.3184    &   0.0000  &  0.0000 \\                          &  RCCSD(T)/6-311++G(3df, 3pd) &  -27.1081  & -30.8650   &  -27.7089  & -4.6705    &   0.0000  &  0.0000 \\
                         &  RCCSD(T)/aug-cc-pVTZ        &  -27.0845  & -30.8217   &   -27.6860 &  -4.6599   &  0.0000   &  0.0000 \\
 \hline                                                                                                                         
  \end{tabular}
\end{table*}

\clearpage

\begin{figure*}
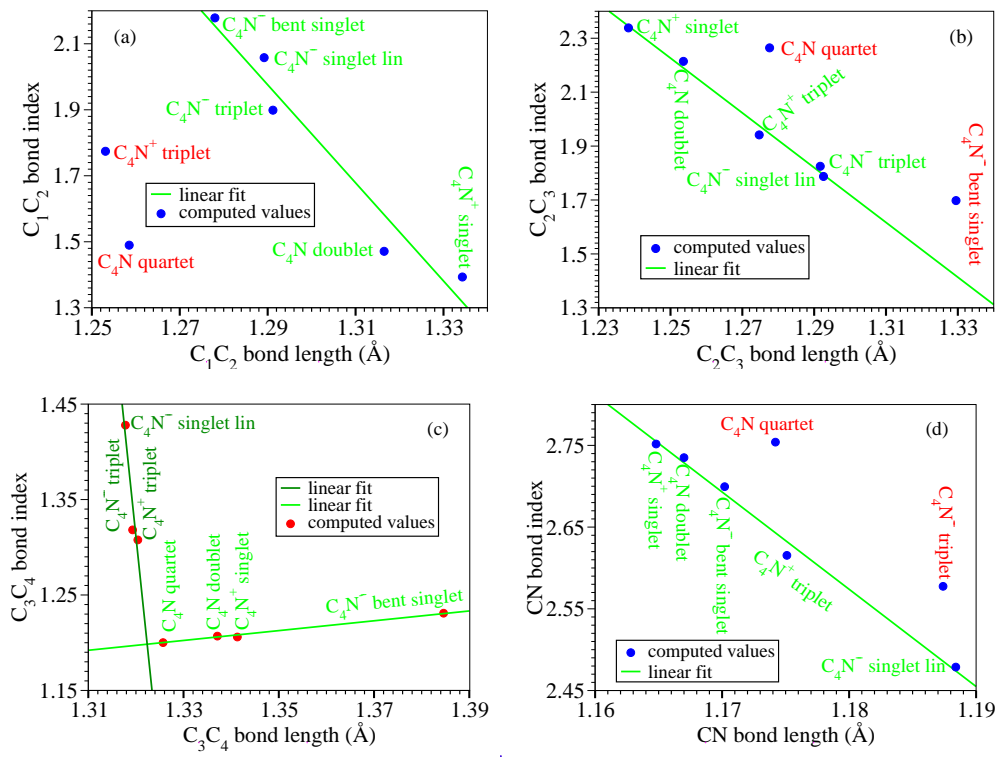

  \includegraphics[width=0.4\textwidth]{fig_C1C2_wbi_vs_lengths.eps}
  \includegraphics[width=0.4\textwidth]{fig_C2C3_wbi_vs_lengths.eps}
  \includegraphics[width=0.4\textwidth]{fig_C3C4_wbi_vs_lengths.eps}
  \includegraphics[width=0.4\textwidth]{fig_CN_wbi_vs_lengths.eps}
  \caption{Bond order indices versus bond lengths of the \ce{C4N} chains investigated in the present paper.
    The linear fitting line suggests possible correlations.}
    \label{fig:bond-index-vs-bond-length}
\end{figure*}

\begin{table*}
  \centering
  \caption{Reorganization energies
    $\lambda_{a}^{b} \equiv \mathcal{E}_{a}\left(\mathbf{R}_{b}\right) - \mathcal{E}_{a}\left(\mathbf{R}_{a}\right)$
    of the \ce{C4N} anions --- triplet ($T^{-}$), bent singlet ($bS^{-}$) and
    (metastable) linear singlet ($lS^{-}$) --- with respect to the neutral doublet ($D$).}
  \label{table:lambda-anions}
  \begin{tabular}{crrrrrrrrr} 
    \hline
Functional &     $\lambda_{T^{-}}^{D}$ & $\lambda_{D}^{T^{-}}$ & $\lambda_{bS^{-}}^{D}$ & $\lambda_{D}^{bS^{-}}$ & $\lambda_{lS^{-}}^{D}$ & $\lambda_{D}^{lS^{-}}$ \\
    \hline
     B3LYP & 0.067       &          0.076    &       0.338       &      0.488       &     0.082         & 0.081            \\
     PBE0  & 0.067       &          0.076    &       0.342       &      0.509       &     0.087         & 0.087            \\
     M06-2X & 0.081       &         -0.113    &       0.446       &      0.600       &     0.111         & 0.110            \\
     \hline
  \end{tabular}
\end{table*}

\begin{table*}
  \centering
  \caption{Reorganization energies
    $\lambda_{a}^{b} \equiv \mathcal{E}_{a}\left(\mathbf{R}_{b}\right) - \mathcal{E}_{a}\left(\mathbf{R}_{a}\right)$
    of the \ce{C4N} singlet ($S^{+}$) and triplet ($T^{+}$) cations with respect to the neutral doublet ($D$).}
  \label{table:lambda-cations}
  \begin{tabular}{crrrrrrr} 
    \hline
 Functional & $\lambda_{S^{+}}^{D}$ & $\lambda_{D}^{S^{+}}$ & $\lambda_{T^{+}}^{D}$ & $\lambda_{D}^{T^{+}}$ \\
    \hline
     B3LYP &          0.019       &     0.014       &   0.124          & 0.124   \\ 
     PBE0  &          0.018       &     0.013       &   0.121          & 0.121   \\
     M06-2X &          0.014       &     0.010       &   0.138          & 0.138   \\
     \hline
  \end{tabular}
\end{table*}

\begin{table*}
  \centering
  \caption{Dissociation of neutral and anion \ce{C4N} chains. 
    Enthalpies of reaction at zero (subscript 0)
    and room temperature (subscript RT) computed by several CBS protocols.\cite{g16} All values (in kcal/mol)
    refer to the electronic ground states.}
  \label{table:dissociation-c4n_si}
  \begin{tabular}{lcrlclclrr}
    \hline
        No. & Species  & Method              &  & Reaction &           &       &           &  $\Delta_{r} H_{0}^{0}$ & $\Delta_{r} H_{RT}^{0}$ \\
    \hline
         1  &   \ce{C4N} & $ \ce{C4N} $ & $ \to $ & $ \ce{C} $ & $ + $ & $ \ce{C3N}     $  & CBS-QB3  & 139.4 & 140.1 \\
            &            & $ \ce{C4N} $ & $ \to $ & $ \ce{C} $ & $ + $ & $ \ce{C3N}     $  & CBS-APNO & 137.3 & 138.0 \\
            &            & $ \ce{C4N} $ & $ \to $ & $ \ce{C} $ & $ + $ & $ \ce{C3N}     $  & CBS-4M   & 138.6 & 139.6 \\[1ex]
         2  &            & $ \ce{C4N} $ & $ \to $ & $ \ce{C2}$ & $ + $ & $ \ce{C2N}    $   & CBS-QB3  & 152.0 & 152.8 \\
            &            & $ \ce{C4N} $ & $ \to $ & $ \ce{C2}$ & $ + $ & $ \ce{C2N}    $   & CBS-APNO & 152.0 & 155.9 \\
            &            & $ \ce{C4N} $ & $ \to $ & $ \ce{C2}$ & $ + $ & $ \ce{C2N}    $   & CBS-4M   & 155.2 & 156.3 \\[1ex]
         3  &            & $ \ce{C4N} $ & $ \to $ & $ \ce{C3}$ & $ + $ & $ \ce{CN}     $   & CBS-QB3  &  95.3 &  96.4 \\
            &            & $ \ce{C4N} $ & $ \to $ & $ \ce{C3}$ & $ + $ & $ \ce{CN}     $   & CBS-APNO &  94.2 &  95.3 \\
            &            & $ \ce{C4N} $ & $ \to $ & $ \ce{C3}$ & $ + $ & $ \ce{CN}     $   & CBS-4M   & 103.0 & 104.3 \\[1ex]
         4  &            & $ \ce{C4N} $ & $ \to $ & $ \ce{C4}$ & $ + $ & $ \ce{N}      $   & CBS-QB3  & 159.6 & 160.5 \\
            &            & $ \ce{C4N} $ & $ \to $ & $ \ce{C4}$ & $ + $ & $ \ce{N}      $   & CBS-APNO & 157.8 & 158.5  \\
            &            & $ \ce{C4N} $ & $ \to $ & $ \ce{C4}$ & $ + $ & $ \ce{N}      $   & CBS-4M   & 156.6 & 157.7  \\
    \hline
         5a &  \ce{C4N-} & $ \ce{C4N-}$ & $ \to $ & $ \ce{C} $ & $ + $ & $ \ce{C3N-}   $ & CBS-QB3    & 109.1 & 109.4 \\
            &            & $ \ce{C4N-}$ & $ \to $ & $ \ce{C} $ & $ + $ & $ \ce{C3N-}   $ & CBS-APNO   & 110.8 & 111.6 \\
            &            & $ \ce{C4N-}$ & $ \to $ & $ \ce{C} $ & $ + $ & $ \ce{C3N-}   $ & CBS-4M     & 116.6 & 117.5 \\[1ex]
         5b &            & $  \ce{C4N-} $ & $ \to $ & $\ce{C-}$ & $ + $ & $ \ce{C3N}   $ & CBS-QB3    & 184.4 & 185.4 \\
            &            & $  \ce{C4N-} $ & $ \to $ & $\ce{C-}$ & $ + $ & $ \ce{C3N}   $ & CBS-APNO   & 183.9 & 184.8 \\
            &            & $  \ce{C4N-} $ & $ \to $ & $\ce{C-}$ & $ + $ & $ \ce{C3N}   $ & CBS-4M     & 190.9 & 191.8 \\[2ex]
         6a &            & $ \ce{C4N-} $ & $ \to $ & $  \ce{C2} $ & $ + $ & $ \ce{C2N-}$ & CBS-QB3    & 160.6 & 161.6 \\
            &            & $ \ce{C4N-} $ & $ \to $ & $  \ce{C2} $ & $ + $ & $ \ce{C2N-}$ & CBS-APNO   & 166.2 & 170.0 \\
            &            & $ \ce{C4N-} $ & $ \to $ & $  \ce{C2} $ & $ + $ & $ \ce{C2N-}$ & CBS-4M     & 165.5 & 166.6 \\[1ex]
         6b &            & $ \ce{C4N-} $ & $ \to $ & $  \ce{C2-}$ & $ + $ & $ \ce{C2N} $ & CBS-QB3    & 151.0 & 152.1 \\
            &            & $ \ce{C4N-} $ & $ \to $ & $  \ce{C2-}$ & $ + $ & $ \ce{C2N} $ & CBS-APNO   & 152.6 & 153.5 \\
           &            & $ \ce{C4N-} $ & $ \to $ & $  \ce{C2-}$ & $ + $ & $ \ce{C2N} $ & CBS-4M      & 156.9 & 158.0 \\[2ex]
         7a &            & $ \ce{C4N-} $ & $ \to $ & $ \ce{C3} $ & $ + $ & $ \ce{CN-}  $ & CBS-QB3    &  77.6 &  79.0 \\
            &            & $ \ce{C4N-} $ & $ \to $ & $ \ce{C3} $ & $ + $ & $ \ce{CN-}  $ & CBS-APNO   &  79.7 & 81.0 \\
            &            & $ \ce{C4N-} $ & $ \to $ & $ \ce{C3} $ & $ + $ & $ \ce{CN-}  $ & CBS-4M     &  88.4 & 89.7 \\[1ex]
         7b &            & $ \ce{C4N-} $ & $ \to $ & $ \ce{C3-} $ & $ + $ & $ \ce{CN}  $ & CBS-QB3    & 122.0 & 123.3 \\
            &            & $ \ce{C4N-} $ & $ \to $ & $ \ce{C3-} $ & $ + $ & $ \ce{CN}  $ & CBS-APNO   & 122.4 & 123.4 \\
            &            & $ \ce{C4N-} $ & $ \to $ & $ \ce{C3-} $ & $ + $ & $ \ce{CN}  $ & CBS-4M     & 130.2 & 131.5 \\[2ex]
         8a &            & $ \ce{C4N-} $ & $ \to $ & $ \ce{C4} $ & $ + $ & $ \ce{N-}   $ & CBS-QB3    & 238.1 & 239.4 \\
            &            & $ \ce{C4N-} $ & $ \to $ & $ \ce{C4} $ & $ + $ & $ \ce{N-}   $ & CBS-APNO   & 241.8 & 243.0 \\
            &            & $ \ce{C4N-} $ & $ \to $ & $ \ce{C4} $ & $ + $ & $ \ce{N-}   $ & CBS-4M     & 241.8 & 243.0 \\[1ex]
         8b &            & $ \ce{C4N-} $ & $ \to $ & $ \ce{C4-}$ & $ + $ & $ \ce{N}    $ & CBS-QB3    & 141.8 & 142.8 \\
            &            & $ \ce{C4N-} $ & $ \to $ & $ \ce{C4-}$ & $ + $ & $ \ce{N}    $ & CBS-APNO   & 142.5 & 143.3 \\
            &            & $ \ce{C4N-} $ & $ \to $ & $ \ce{C4-}$ & $ + $ & $ \ce{N}    $ & CBS-4M     & 144.6 & 145.8 \\
    \hline
  \end{tabular}
\end{table*}

\begin{table*}
  \centering
  \caption{Dissociation of neutral \ce{C2N}, \ce{C3N}, and \ce{C5N} chains already detected in space. 
    Enthalpies of reaction at zero (subscript 0)
    and room temperature (subscript RT) computed by several CBS protocols.\cite{g16} All values (in kcal/mol)
    refer to the electronic ground states.}
  \label{table:dissociation-c3n-c5n_si}
  \begin{tabular}{rclclclrrr}
    \hline
        No. & Species  &               & Reaction&              &       &                  &  Method & $\Delta_{r} H_{0}^{0}$ & $\Delta_{r} H_{RT}^{0}$ \\
    \hline
         9a  & \ce{C2N} & $ \ce{C2N} $ & $ \to $ & $  \ce{C} $ & $ + $ & $ \ce{CN}     $   & CBS-QB3  & 113.4 & 114.4 \\
             &          & $ \ce{C2N} $ & $ \to $ & $  \ce{C} $ & $ + $ & $ \ce{CN}     $   & CBS-APNO & 113.1 &  114.1 \\
             &          & $ \ce{C2N} $ & $ \to $ & $  \ce{C} $ & $ + $ & $ \ce{CN}     $   & CBS-4M   & 116.6 & 117.7 \\[1ex]
         9b  &          & $ \ce{C2N} $ & $ \to $ & $  \ce{C2}$ & $ + $ & $ \ce{N}     $   & CBS-QB3  & 145.8 & 146.8 \\
             &          & $ \ce{C2N} $ & $ \to $ & $  \ce{C2}$ & $ + $ & $ \ce{N}     $   & CBS-APNO & 149.3 & 148.7 \\
             &          & $ \ce{C2N} $ & $ \to $ & $  \ce{C2}$ & $ + $ & $ \ce{N}     $   & CBS-4M   & 147.5 &  148.7 \\[1ex]
        10a  & \ce{C3N} & $ \ce{C3N} $ & $ \to $ & $  \ce{C} $ & $ + $ & $ \ce{C2N}    $   & CBS-QB3  & 156.8 & 157.9 \\
             &          & $ \ce{C3N} $ & $ \to $ & $  \ce{C} $ & $ + $ & $ \ce{C2N}    $   & CBS-APNO & 158.8 & 159.7 \\
             &          & $ \ce{C3N} $ & $ \to $ & $  \ce{C} $ & $ + $ & $ \ce{C2N}    $   & CBS-4M   & 158.0 & 159.0 \\[1ex]
        10b  &          & $ \ce{C3N} $ & $ \to $ & $  \ce{C2}$ & $ + $ & $ \ce{CN}    $   & CBS-QB3  & 126.0 & 127.1 \\
             &          & $ \ce{C3N} $ & $ \to $ & $  \ce{C2}$ & $ + $ & $ \ce{CN}    $   & CBS-APNO & 131.0 & 132.0 \\
             &          & $ \ce{C3N} $ & $ \to $ & $  \ce{C2}$ & $ + $ & $ \ce{CN}    $   & CBS-4M   & 133.2 &  134.4 \\[1ex]
        10c  &          & $ \ce{C3N} $ & $ \to $ & $  \ce{C3}$ & $ + $ & $ \ce{N}     $   & CBS-QB3  & 132.6 & 134.0 \\
             &          & $ \ce{C3N} $ & $ \to $ & $  \ce{C3}$ & $ + $ & $ \ce{N}     $   & CBS-APNO & 133.9 & 135.2 \\
             &          & $ \ce{C3N} $ & $ \to $ & $  \ce{C3}$ & $ + $ & $ \ce{N}     $   & CBS-4M   & 136.7 & 138.0 \\[1ex]
        12a  & \ce{C5N} & $ \ce{C5N} $ & $ \to $ & $   \ce{C}$ & $ + $ & $ \ce{C4N}   $   & CBS-QB3  & 144.1 & 145.7 \\
             &          & $ \ce{C5N} $ & $ \to $ & $   \ce{C}$ & $ + $ & $ \ce{C4N}   $   & CBS-APNO & 147.2 & 148.1 \\
             &          & $ \ce{C5N} $ & $ \to $ & $   \ce{C}$ & $ + $ & $ \ce{C4N}   $   & CBS-4M   & 147.3 & 148.2 \\[1ex]
        12b  &          & $ \ce{C5N} $ & $ \to $ & $  \ce{C2}$ & $ + $ & $ \ce{C3N}   $   & CBS-QB3  & 139.2 & 140.6 \\
             &          & $ \ce{C5N} $ & $ \to $ & $  \ce{C2}$ & $ + $ & $ \ce{C3N}   $   & CBS-APNO & 143.7 & 144.3 \\
             &          & $ \ce{C5N} $ & $ \to $ & $  \ce{C2}$ & $ + $ & $ \ce{C3N}   $   & CBS-4M   & 144.6 & 145.5 \\[1ex]
        12c  &          & $ \ce{C5N} $ & $ \to $ & $  \ce{C3}$ & $ + $ & $ \ce{C2N}   $   & CBS-QB3  & 126.0 & 127.8 \\
             &          & $ \ce{C5N} $ & $ \to $ & $  \ce{C3}$ & $ + $ & $ \ce{C2N}   $   & CBS-APNO & 128.3 & 129.3 \\
             &          & $ \ce{C5N} $ & $ \to $ & $  \ce{C3}$ & $ + $ & $ \ce{C2N}   $   & CBS-4M   & 133.7 &  134.8 \\[1ex]
        12d  &          & $ \ce{C5N} $ & $ \to $ & $  \ce{C4}$ & $ + $ & $ \ce{CN}   $   & CBS-QB3   &  126.9 & 128.6 \\
             &          & $ \ce{C5N} $ & $ \to $ & $  \ce{C4}$ & $ + $ & $ \ce{CN}   $   & CBS-APNO  & 128.0 & 128.7 \\
             &          & $ \ce{C5N} $ & $ \to $ & $  \ce{C4}$ & $ + $ & $ \ce{CN}   $   & CBS-4M    & 131.6 & 132.7 \\[1ex]
        12e  &          & $  \ce{C5N}$ & $ \to $ & $   \ce{C5}$ & $ + $ & $ \ce{N}   $   & CBS-QB3  & 135.8 & 137.3 \\
             &          & $  \ce{C5N}$ & $ \to $ & $   \ce{C5}$ & $ + $ & $ \ce{N}   $   & CBS-APNO & 136.0 & 136.8 \\
             &          & $  \ce{C5N}$ & $ \to $ & $   \ce{C5}$ & $ + $ & $ \ce{N}   $   & CBS-4M   & 142.9 & 143.8 \\
     \hline
  \end{tabular}
\end{table*}

\begin{table*}
  \centering
  \caption{Relevant exchange reactions.
    Enthalpies of reaction at zero (subscript 0)
    and room temperature (subscript RT) computed by several CBS protocols.\cite{g16} All values (in kcal/mol)
    refer to the electronic ground states.}
  \label{table:exchange_1_si}
  \begin{tabular}{llclclrllrr}
    \hline
        No. &  &          &           & Reaction  &       &        &          &  Method & $\Delta_{r} H_{0}^{0}$ & $\Delta_{r} H_{RT}^{0}$ \\
    \hline
        13   & $ \ce{C5}$ & $ + $ & $  \ce{N} $  & $ \to $ & $  \ce{C}$ & $ + $ & $ \ce{C4N}      $   & CBS-QB3  &    8.3 &    8.4 \\
             & $ \ce{C5}$ & $ + $ & $  \ce{N} $  & $ \to $ & $  \ce{C}$ & $ + $ & $ \ce{C4N}      $   & CBS-APNO &   11.2 &   11.3 \\
             & $ \ce{C5}$ & $ + $ & $  \ce{N} $  & $ \to $ & $  \ce{C}$ & $ + $ & $ \ce{C4N}      $   & CBS-4M   &    4.5 &    4.3 \\[1ex]
        14a  & $ \ce{N} $ & $ + $ & $  \ce{C4H-}$ & $ \to $ & $  \ce{C4N}$ & $ + $ & $  \ce{H-}   $   & CBS-QB3  &   23.9 &   24.3 \\
             & $ \ce{N} $ & $ + $ & $  \ce{C4H-}$ & $ \to $ & $  \ce{C4N}$ & $ + $ & $  \ce{H-}   $   & CBS-APNO &   32.2 &  32.9 \\
             & $ \ce{N} $ & $ + $ & $  \ce{C4H-}$ & $ \to $ & $  \ce{C4N}$ & $ + $ & $  \ce{H-}   $   & CBS-4M   &   24.3 &  24.6 \\[1ex]
        14b  & $ \ce{N} $ & $ + $ & $  \ce{C4H-} $ & $ \to $ & $  \ce{C4N-}$ & $ + $ & $  \ce{H}  $   & CBS-QB3  &  -36.0 &  -35.9 \\ 
             & $ \ce{N} $ & $ + $ & $  \ce{C4H-} $ & $ \to $ & $  \ce{C4N-}$ & $ + $ & $  \ce{H}  $   & CBS-APNO &  -36.6 &  -36.1 \\
             & $ \ce{N} $ & $ + $ & $  \ce{C4H-} $ & $ \to $ & $  \ce{C4N-}$ & $ + $ & $  \ce{H}  $   & CBS-4M   &  -41.6 &  -41.3 \\[1ex]
        14c  & $ \ce{N-}$ & $ + $ & $  \ce{C4H} $ & $ \to $ & $  \ce{C4N-}$ & $ + $ & $  \ce{H}   $   & CBS-QB3  & -125.0 & -124.8 \\  
             & $ \ce{N-}$ & $ + $ & $  \ce{C4H} $ & $ \to $ & $  \ce{C4N-}$ & $ + $ & $  \ce{H}   $   & CBS-APNO & -137.1 & -136.9 \\
             & $ \ce{N-}$ & $ + $ & $  \ce{C4H} $ & $ \to $ & $  \ce{C4N-}$ & $ + $ & $  \ce{H}   $   & CBS-4M   & -129.2 & -129.2 \\[1ex]
        14d  & $ \ce{N-}$ & $ + $ & $  \ce{C4H} $ & $ \to $ & $  \ce{C4N}$ & $ + $ & $  \ce{H-}   $   & CBS-QB3  &  -65.1 &  -64.6 \\
             & $ \ce{N-}$ & $ + $ & $  \ce{C4H} $ & $ \to $ & $  \ce{C4N}$ & $ + $ & $  \ce{H-}   $   & CBS-APNO &  -68.3 &  -67.9 \\
             & $ \ce{N-}$ & $ + $ & $  \ce{C4H} $ & $ \to $ & $  \ce{C4N}$ & $ + $ & $  \ce{H-}   $   & CBS-4M   &  -63.3 &  -63.3 \\[1ex]
        15a  & $ \ce{CN}$ & $ + $ & $  \ce{C3H} $ & $ \to $ & $  \ce{H}$ & $ + $ & $  \ce{C4N}    $   & CBS-QB3  &  -20.5 &  -20.4 \\
             & $ \ce{CN}$ & $ + $ & $  \ce{C3H} $ & $ \to $ & $  \ce{H}$ & $ + $ & $  \ce{C4N}    $   & CBS-APNO &  -18.9 &  -18.9 \\
             & $ \ce{CN}$ & $ + $ & $  \ce{C3H} $ & $ \to $ & $  \ce{H}$ & $ + $ & $  \ce{C4N}    $   & CBS-4M   &  -24.8 &  -24.8 \\[1ex]
        15b  & $ \ce{CN-}$ & $ + $ & $  \ce{C3H}$ & $ \to $ & $  \ce{H}$ & $ + $ & $  \ce{C4N-}   $   & CBS-QB3  &   -2.8 &   -3.0 \\ 
             & $ \ce{CN-}$ & $ + $ & $  \ce{C3H}$ & $ \to $ & $  \ce{H}$ & $ + $ & $  \ce{C4N-}   $   & CBS-APNO &   -4.5 &   -4.6 \\
             & $ \ce{CN-}$ & $ + $ & $  \ce{C3H}$ & $ \to $ & $  \ce{H}$ & $ + $ & $  \ce{C4N-}   $   & CBS-4M    &   -10.2 & -10.2 \\[1ex]
        15c  & $ \ce{CN-}$ & $ + $ & $  \ce{C3H}$ & $ \to $ & $  \ce{H-}$ & $ + $ & $  \ce{C4N}   $   & CBS-QB3  &   57.1 &  57.2 \\
             & $ \ce{CN-}$ & $ + $ & $  \ce{C3H}$ & $ \to $ & $  \ce{H-}$ & $ + $ & $  \ce{C4N}   $   & CBS-APNO &   64.4 &  64.4 \\
             & $ \ce{CN-}$ & $ + $ & $  \ce{C3H}$ & $ \to $ & $  \ce{H-}$ & $ + $ & $  \ce{C4N}   $   & CBS-4M   &   55.7 & 55.8 \\[1ex]
        15d  & $ \ce{CN} $ & $ + $ & $  \ce{C3H-}$ & $ \to $ & $  \ce{H}$ & $ + $ & $  \ce{C4N-}  $   & CBS-QB3  &  -50.9 &  -50.8 \\
             & $ \ce{CN} $ & $ + $ & $  \ce{C3H-}$ & $ \to $ & $  \ce{H}$ & $ + $ & $  \ce{C4N-}  $   & CBS-APNO &  -51.9 &  -51.6 \\
            & $ \ce{CN} $ & $ + $ & $  \ce{C3H-}$ & $ \to $ & $  \ce{H}$ & $ + $ & $  \ce{C4N-}  $   & CBS-4M    &  -56.7 &  -56.8 \\[1ex]
        15e  & $ \ce{CN} $ & $ + $ & $  \ce{C3H-}$ & $ \to $ & $  \ce{H-}$ & $ + $ & $  \ce{C4N}  $   & CBS-QB3  &    9.0 &    9.4 \\  
             & $ \ce{CN} $ & $ + $ & $  \ce{C3H-}$ & $ \to $ & $  \ce{H-}$ & $ + $ & $  \ce{C4N}  $   & CBS-APNO &   16.9 &   17.4 \\
             & $ \ce{CN} $ & $ + $ & $  \ce{C3H-}$ & $ \to $ & $  \ce{H-}$ & $ + $ & $  \ce{C4N}  $   & CBS-4M   &    9.2 &    9.2 \\
     \hline
  \end{tabular}
\end{table*}

\begin{table*}
  \centering
  \caption{Relevant exchange reactions.
    Enthalpies of reaction at zero (subscript 0)
    and room temperature (subscript RT) computed by several CBS protocols.\cite{g16} All values (in kcal/mol)
    refer to the electronic ground states.}
  \label{table:exchange_2_si}
  \begin{tabular}{llclclrllrr}
    \hline
        No. &  &          &           & Reaction  &       &        &          &  Method & $\Delta_{r} H_{0}^{0}$ & $\Delta_{r} H_{RT}^{0}$ \\
    \hline
        16a  & $ \ce{CH} $ & $ + $ & $  \ce{C3N} $ & $ \to $ & $  \ce{H}$ & $ + $ & $  \ce{C4N}   $   & CBS-QB3  &  -59.4 &  -59.2 \\
             & $ \ce{CH} $ & $ + $ & $  \ce{C3N} $ & $ \to $ & $  \ce{H}$ & $ + $ & $  \ce{C4N}   $   & CBS-APNO &  -57.4 &  -57.2 \\
             & $ \ce{CH} $ & $ + $ & $  \ce{C3N} $ & $ \to $ & $  \ce{H}$ & $ + $ & $  \ce{C4N}   $   & CBS-4M   &  -58.9 &  -58.9 \\[1ex]
        16b  & $ \ce{CH-}$ & $ + $ & $  \ce{C3N}$ & $ \to $ & $  \ce{H}$ & $ + $ & $  \ce{C4N-}   $   & CBS-QB3  & -105.6 & -105.8 \\
             & $ \ce{CH-}$ & $ + $ & $  \ce{C3N}$ & $ \to $ & $  \ce{H}$ & $ + $ & $  \ce{C4N-}   $   & CBS-APNO & -105.0 & -105.0 \\
             & $ \ce{CH-}$ & $ + $ & $  \ce{C3N}$ & $ \to $ & $  \ce{H}$ & $ + $ & $  \ce{C4N-}   $   & CBS-4M   & -111.0 & -111.1 \\[1ex]
        16c  & $ \ce{CH-}$ & $ + $ & $  \ce{C3N}$ & $ \to $ & $  \ce{H-}$ & $ + $ & $  \ce{C4N}   $   & CBS-QB3  &  -45.8 &  -45.5 \\
             & $ \ce{CH-}$ & $ + $ & $  \ce{C3N}$ & $ \to $ & $  \ce{H-}$ & $ + $ & $  \ce{C4N}   $   & CBS-APNO &  -36.2 &  -36.0 \\
             & $ \ce{CH-}$ & $ + $ & $  \ce{C3N}$ & $ \to $ & $  \ce{H-}$ & $ + $ & $  \ce{C4N}   $   & CBS-4M   &  -45.1 &  -45.2 \\[1ex]
        16d  & $ \ce{CH} $ & $ + $ & $  \ce{C3N-}$ & $ \to $ & $  \ce{H}$ & $ + $ & $  \ce{C4N-}  $   & CBS-QB3  &  -29.1 &  -29.1 \\
             & $ \ce{CH} $ & $ + $ & $  \ce{C3N-}$ & $ \to $ & $  \ce{H}$ & $ + $ & $  \ce{C4N-}  $   & CBS-APNO &  -31.0 &  -30.8 \\
             & $ \ce{CH} $ & $ + $ & $  \ce{C3N-}$ & $ \to $ & $  \ce{H}$ & $ + $ & $  \ce{C4N-}  $   & CBS-4M   &  -36.8 &  -36.9 \\[1ex]  
        17a  & $ \ce{CH} $ & $ + $ & $  \ce{C3N-}$ & $ \to $ & $  \ce{H-}$ & $ + $ & $  \ce{C4N}  $   & CBS-QB3  &   30.8 &  31.1 \\
             & $ \ce{CH} $ & $ + $ & $  \ce{C3N-}$ & $ \to $ & $  \ce{H-}$ & $ + $ & $  \ce{C4N}  $   & CBS-APNO &   37.9 &  38.2 \\
             & $ \ce{CH} $ & $ + $ & $  \ce{C3N-}$ & $ \to $ & $  \ce{H-}$ & $ + $ & $  \ce{C4N}  $   & CBS-4M   &   29.1 &  29.1 \\[1ex]
        17b  & $ \ce{C2H}$ & $ + $ & $  \ce{C2N}$ & $ \to $ & $  \ce{H} $ & $ + $ & $  \ce{C4N}   $   & CBS-QB3  &   -40.4 &  -40.3 \\
             & $ \ce{C2H}$ & $ + $ & $  \ce{C2N}$ & $ \to $ & $  \ce{H} $ & $ + $ & $  \ce{C4N}   $   & CBS-APNO &   -41.0  &  -40.5 \\
             & $ \ce{C2H}$ & $ + $ & $  \ce{C2N}$ & $ \to $ & $  \ce{H} $ & $ + $ & $  \ce{C4N}   $   & CBS-4M   &  -40.4 &  -40.3 \\[1ex]
        17c  & $ \ce{C2H-}$ & $ + $ & $  \ce{C2N}$ & $ \to $ & $ \ce{H} $ & $ + $ & $  \ce{C4N-}  $   & CBS-QB3  &  -44.6 &  -44.5 \\
             & $ \ce{C2H-}$ & $ + $ & $  \ce{C2N}$ & $ \to $ & $ \ce{H} $ & $ + $ & $  \ce{C4N-}  $   & CBS-APNO &  -46.6 & -46.2 \\
             & $ \ce{C2H-}$ & $ + $ & $  \ce{C2N}$ & $ \to $ & $ \ce{H} $ & $ + $ & $  \ce{C4N-}  $   & CBS-4M   &  -51.7  & -51.5 \\[1ex]
        17d  & $ \ce{C2H-}$ & $ + $ & $  \ce{C2N}$ & $ \to $ & $  \ce{H-}$ & $ + $ & $  \ce{C4N}  $   & CBS-QB3  &   15.3  &  15.7 \\
             & $ \ce{C2H-}$ & $ + $ & $  \ce{C2N}$ & $ \to $ & $  \ce{H-}$ & $ + $ & $  \ce{C4N}  $   & CBS-APNO &   22.2 &  22.8 \\
             & $ \ce{C2H-}$ & $ + $ & $  \ce{C2N}$ & $ \to $ & $  \ce{H-}$ & $ + $ & $  \ce{C4N}  $   & CBS-4M   &    14.2 &  14.5 \\[1ex]
        17e  & $ \ce{C2H} $ & $ + $ & $  \ce{C2N-}$ & $ \to $ & $  \ce{H-}$ & $ + $ & $  \ce{C4N} $   & CBS-QB3  &  10.8  & 11.2 \\
             & $ \ce{C2H} $ & $ + $ & $  \ce{C2N-}$ & $ \to $ & $  \ce{H-}$ & $ + $ & $  \ce{C4N} $   & CBS-APNO &   16.8 & 17.5 \\
             & $ \ce{C2H} $ & $ + $ & $  \ce{C2N-}$ & $ \to $ & $  \ce{H-}$ & $ + $ & $  \ce{C4N} $   & CBS-4M   &   15.1  & 15.3 \\[1ex]
        17f  & $ \ce{C2H} $ & $ + $ & $  \ce{C2N-}$ & $ \to $ & $  \ce{H} $ & $ + $ & $  \ce{C4N-}$   & CBS-QB3  &  -49.1  &  -49.0 \\
             & $ \ce{C2H} $ & $ + $ & $  \ce{C2N-}$ & $ \to $ & $  \ce{H} $ & $ + $ & $  \ce{C4N-}$   & CBS-APNO &   -52.0 &  -51.5 \\
             & $ \ce{C2H} $ & $ + $ & $  \ce{C2N-}$ & $ \to $ & $  \ce{H} $ & $ + $ & $  \ce{C4N-}$   & CBS-4M   &   -50.8 &  -50.6 \\[1ex]
     \hline
  \end{tabular}
\end{table*}

\begin{table*}
  \centering
  \caption{Relevant exchange reactions.
    Enthalpies of reaction at zero (subscript 0)
    and room temperature (subscript RT) computed by several CBS protocols.\cite{g16} All values (in kcal/mol)
    refer to the electronic ground states.}
  \label{table:exchange_3_si}
  \begin{tabular}{llclclrllrr}
    \hline
        No. &  &          &           & Reaction  &       &        &          &  Method & $\Delta_{r} H_{0}^{0}$ & $\Delta_{r} H_{RT}^{0}$ \\
    \hline
        18   & $ \ce{NC2N} $ & $ + $ & $   \ce{C2}  $ & $ \to $ & $  \ce{N} $ & $ + $ & $  \ce{C4N}  $   & CBS-QB3  & 48.6 & 49.0 \\
             & $ \ce{NC2N} $ & $ + $ & $   \ce{C2}  $ & $ \to $ & $  \ce{N} $ & $ + $ & $  \ce{C4N}  $   & CBS-APNO & 45.2 & 45.6 \\
             & $ \ce{NC2N} $ & $ + $ & $   \ce{C2}  $ & $ \to $ & $  \ce{N} $ & $ + $ & $  \ce{C4N}  $   & CBS-4M   & 41.4 & 41.5 \\[1ex]
        19   & $ \ce{NC2N} $ & $ + $ & $   \ce{C2N}  $ & $ \to $ & $  \ce{N2} $ & $ + $ & $  \ce{C4N}  $   & CBS-QB3  & -29.8 & -29.3 \\
             & $ \ce{NC2N} $ & $ + $ & $   \ce{C2N}  $ & $ \to $ & $  \ce{N2} $ & $ + $ & $  \ce{C4N}  $   & CBS-APNO & -29.4 & -28.9 \\
             & $ \ce{NC2N} $ & $ + $ & $   \ce{C2N}  $ & $ \to $ & $  \ce{N2} $ & $ + $ & $  \ce{C4N}  $   & CBS-4M   & -31.7 & -31.4 \\[1ex]
        20   & $ \ce{NC2N} $ & $ + $ & $   \ce{C2H}  $ & $ \to $ & $  \ce{NH} $ & $ + $ & $  \ce{C4N}  $   & CBS-QB3  & 82.3 & 82.7 \\
             & $ \ce{NC2N} $ & $ + $ & $   \ce{C2H}  $ & $ \to $ & $  \ce{NH} $ & $ + $ & $  \ce{C4N}  $   & CBS-APNO & 81.4 & 82.2 \\
             & $ \ce{NC2N} $ & $ + $ & $   \ce{C2H}  $ & $ \to $ & $  \ce{NH} $ & $ + $ & $  \ce{C4N}  $   & CBS-4M   & 78.0 & 78.4 \\
     \hline
  \end{tabular}
\end{table*}

\begin{table}
  \centering
  \caption{Adiabatic electron affinities and ionization potentials (in eV) 
           of \ce{C4N} and \ce{C6N} computed with various CBS protocols.
           Notice that, out of these protocols, the CBS-QB3 $EA$-estimates are the closest to the experimental values 
           $EA_{\ce{C2N}} = 2.74890 \pm 0.00010$\,eV, $EA_{\ce{C4N}} = 3.1113 \pm 0.00010$\,eV and $EA_{\ce{C4N}} = 3.3715 \pm 0.00010$\,eV \cite{Neumark:09}.}
  \label{table:ea-ip_cbs-c2n-c4n-c6n}
  \begin{tabular}{lrrrrrr}
    \hline
  Method   & $EA_{\ce{C2N}}$ &  $IP_{\ce{C2N}}$ & $EA_{\ce{C4N}}$ &  $IP_{\ce{C4N}}$ & $EA_{\ce{C6N}}$ & $IP_{\ce{C6N}}$ \\
    \hline
  CBS-QB3  & 2.7615 & 10.8166 & 3.1351          &  9.6913 & 3.4804 & 8.9994 \\
  CBS-APNO & 2.7728 & 10.8178 & 3.2506          &  9.6332 & 3.5648 & 8.9491 \\
  CBS-4M   & 3.0115 & 11.1315 & 3.4596          & 10.0462 & 3.7693 & 9.5614 \\
    \hline
  \end{tabular}
\end{table}

\end{document}